\documentclass[journal]{IEEEtran}
\usepackage{epsfig,graphicx,subfigure,psfrag,amsmath,cases,epstopdf,cite}
\usepackage{latexsym,amssymb,amsmath,epsfig,subfigure,algorithm,mathtools,bm,relsize}
\usepackage{algorithmic}
\usepackage{color}
\usepackage{url}
\usepackage{scrtime}
\usepackage{array}
\usepackage{float}
\usepackage{tcolorbox}
\usepackage{epstopdf}
\usepackage{multirow}

\newtheorem{Thm}{Theorem}
\newtheorem{Lem}{Lemma}
\newtheorem{Cor}{Corollary}

\DeclareMathOperator{\Tr}{\mathrm{Tr}}

\DeclareMathOperator{\maxo}{\mathrm{maximize}}

\DeclareMathAlphabet\mathbfcal{OMS}{cmsy}{b}{n}

\newcommand{\definedas}{\overset{\underset{\Delta}{}}{=}}
\newcolumntype{L}{>{\arraybackslash\raggedright}m{2cm}}
\makeatletter

\newcommand{\Rmnum}[1]{\expandafter\@slowromancap\romannumeral #1@}
\makeatother

\author{Zhiqiang Wei,~\IEEEmembership{Member,~IEEE,} Xianghao Yu,~\IEEEmembership{Member,~IEEE,}  \\Derrick Wing Kwan Ng,~\IEEEmembership{Fellow,~IEEE,}  and Robert Schober,~\IEEEmembership{Fellow,~IEEE}
\thanks{Zhiqiang Wei is supported by funding from Alexander von Humboldt Foundation. D. W. K. Ng is supported by funding from the UNSW Digital Grid Futures Institute, UNSW, Sydney, under a cross-disciplinary fund scheme and by the Australian Research Council's Discovery Project (DP210102169). Robert Schober's work was supported in part by the German Science Foundation through projects SCHO 831/12-1 and SFB 1483 - Project-ID 442419336, EmpkinS. (\textit{Corresponding author: Xianghao Yu.})

Zhiqiang Wei and Robert Schober are with the Institute for Digital Communications (IDC), Friedrich-Alexander University Erlangen-Nuremberg, Germany (email: zhiqiang.wei; robert.schober@fau.de).

Xianghao Yu is with Department of Electronic and Computer Engineering, the Hong Kong University of Science and Technology, Hong Kong (email: eexyu@ust.hk).

Derrick Wing Kwan Ng is with the School of Electrical Engineering and Telecommunications, the University of New South Wales, Australia (email: w.k.ng@unsw.edu.au).}}

\title{Resource Allocation for Simultaneous Wireless Information and Power Transfer Systems: A Tutorial Overview}

\begin{document}
\maketitle
\begin{abstract}
Over the last decade, simultaneous wireless information and power transfer (SWIPT) has become a practical and promising solution for connecting and recharging battery-limited devices, thanks to significant advances in low-power electronics technology and wireless communications techniques.
To realize the promised potentials, advanced resource allocation design plays a decisive role for revealing, understanding, and exploiting the intrinsic rate-energy tradeoff capitalizing on the dual use of radio frequency (RF) signals for wireless charging and communication.
In this paper, we provide a comprehensive tutorial overview of SWIPT from the perspective of resource allocation design.
The fundamental concepts, system architectures, and RF energy harvesting (EH) models are introduced.
In particular, three commonly adopted EH models, namely the linear EH model, the nonlinear saturation EH model, and the nonlinear circuit-based EH model are characterized and discussed.
Then, for a typical wireless system setup, we establish a generalized resource allocation design framework which subsumes conventional resource allocation design problems as special cases.
Subsequently, we elaborate on relevant tools from optimization theory and exploit them for solving representative resource allocation design problems for SWIPT systems with and without perfect channel state information (CSI) available at the transmitter, respectively.
The associated technical challenges and insights are also highlighted.
Furthermore, we discuss several promising and exciting future research directions for resource allocation design for SWIPT systems intertwined with cutting-edge communication technologies, such as intelligent reflecting surfaces, unmanned aerial vehicle, mobile edge computing, federated learning, and machine learning.
\end{abstract}


\section{Introduction}
\subsection{Overview of Wireless Power Transfer}
The advancements of wireless technologies over the past decades have triggered a massive growth in the
number of wireless devices and sensors, such as smartphones, tablets, earbuds, wearable sensors, and wireless environmental monitoring sensors, fueling the rapid development of the Internet-of-Things (IoT) \cite{7123563} and massive machine type communications (mMTC) \cite{7565189}.
Nowadays, these widely deployed wireless devices are often battery-powered while battery capacities are usually limited which creates a serious performance bottleneck in realizing ubiquitous wireless communication networks.
In particular, the regular manual battery replacement and wired recharging can be costly, cumbersome, dangerous, or even not possible, e.g., for biomedical implants, which significantly limits the lifetime of wireless networks and increases the network operational costs.
In this regard, a viable solution for extending the lifetime of wireless devices is to allow them to scavenge energy from diverse renewable energy sources, e.g., solar, wind, geothermal, and vibration.
However, energy harvesting (EH) from these resources has its own limitations.
In fact, natural energy sources are usually intermittent and uncontrollable and thus EH from natural sources for powering communication may lead to unstable communication services\cite{Bruno2019Review,DerrickOFDMAHybridBS}.
As an alternative, near-field contactless charging technologies, including inductive and magnetic resonant coupling, enable on-demand energy supply, while the range of energy transmission is quite limited \cite{KimInductiveCoupling,Kurs83}.
Moreover, laser-based wireless charging is also a potential technology to achieve efficient power delivery over long distances\cite{summerer2009concepts}.
Unfortunately, this technique requires a line-of-sight (LoS) link and accurate laser beam alignment between transceivers, which may not always be possible in practice.

In contrast, radio frequency (RF) wireless power transfer (WPT), where wireless devices harvest energy from ambient electromagnetic (EM) waves, is a promising technology to provide a stable and controllable wireless energy supply for battery-limited wireless devices.
Moreover, WPT\footnote{In this paper, WPT refers to far-field EM-based wireless power transfer.} is applicable in non-LoS (NLoS)
environments and allows for flexibility in transceiver deployment.
Indeed, thanks to the continuous advancement of low-power electronics technology, the harvested energy is sufficient to support various types of emerging wireless devices and applications, such as radio
frequency identification (RFID) tags and wearable sensors.
Furthermore, since radio signals can be used for transferring both information and power, simultaneously or asynchronously, it is expected that the integration and joint design of WPT and wireless communication systems can enable perpetual sustainability of wireless networks.
Depending on how WPT is combined with communications, several general system architectures can be distinguished\cite{ZhangRuiModel,SuzhiMag,QingQingEEWPC,NguyenWPR,YinghuiWPBC,FengWPMEC,PsomasMEC,XiaoyanMEC,LuyueMEC}:
\begin{itemize}
	\item Simultaneous wireless information and power transfer (SWIPT) systems convey both information and power at the same time and in the same frequency band using the same radio signals \cite{ZhangRuiModel}.
	\item Wireless powered communication (WPC) systems deliver wireless energy to power communication devices\cite{SuzhiMag,QingQingEEWPC}.
	Then, the energy harvested by these devices is utilized for transferring their information to information decoding receivers.
	\item In wireless powered relaying (WPR) systems, a source node delivers both information and power to an energy-limited relay node via SWIPT\cite{NguyenWPR}.
	The relay node harvests energy and decodes information from the received signals.
	In turn, the wirelessly-powered relay assist the end-to-end communications by forwarding the source information to the destination node.
	\item Wireless powered backscatter communication (WPBC) systems convey energy to a wireless tag device\cite{YinghuiWPBC}.
	Then, the wirelessly-powered tag reflects and modulates the incoming RF signal for communication with a tag reader.
	
	\item {Wireless powered mobile edge computing (WPMEC) systems energize resources-constrained mobile edge computing (MEC) devices via WPT \cite{FengWPMEC,PsomasMEC}.
	Following the harvest-then-offload protocol, the MEC devices can offload computation-intensive tasks to the edge cloud or servers \cite{XiaoyanMEC,LuyueMEC} exploiting the harvested energy.}
\end{itemize}

In this paper, we focus on resource allocation design for SWIPT systems as this system architecture serves as a fundamental building block for the other types of WPT-based systems.
Besides, as the name suggests, SWIPT implies the dual use of radio signals, which leads to an essential and non-trivial tradeoff between wireless information transfer (WIT) and WPT, known as rate-energy tradeoff.
To unlock the potential of SWIPT, resource allocation plays a decisive role in optimizing this tradeoff to achieve certain design goals.
However, resource allocation design for SWIPT systems is challenging as the dual use of radio signals in the same link introduces severe coupling between WIT and WPT.
%
%
Moreover, focusing on SWIPT systems facilitates the investigation of concrete resource allocation design methodologies.
These methodologies can then serve as a fundamental platform, which can be extended to other WPT-based system architectures with proper modifications.
%
%
We refer interested readers for more details on resource allocation design for the other four system architectures to \cite{Morsi2020,SuzhiMag,QingQingEEWPC,NguyenWPR,YinghuiWPBC,FengWPMEC,PsomasMEC,XiaoyanMEC,LuyueMEC}.
\subsection{Fundamentals of SWIPT Systems}
\begin{figure}[t]
	\center{\includegraphics[width=2.5in]{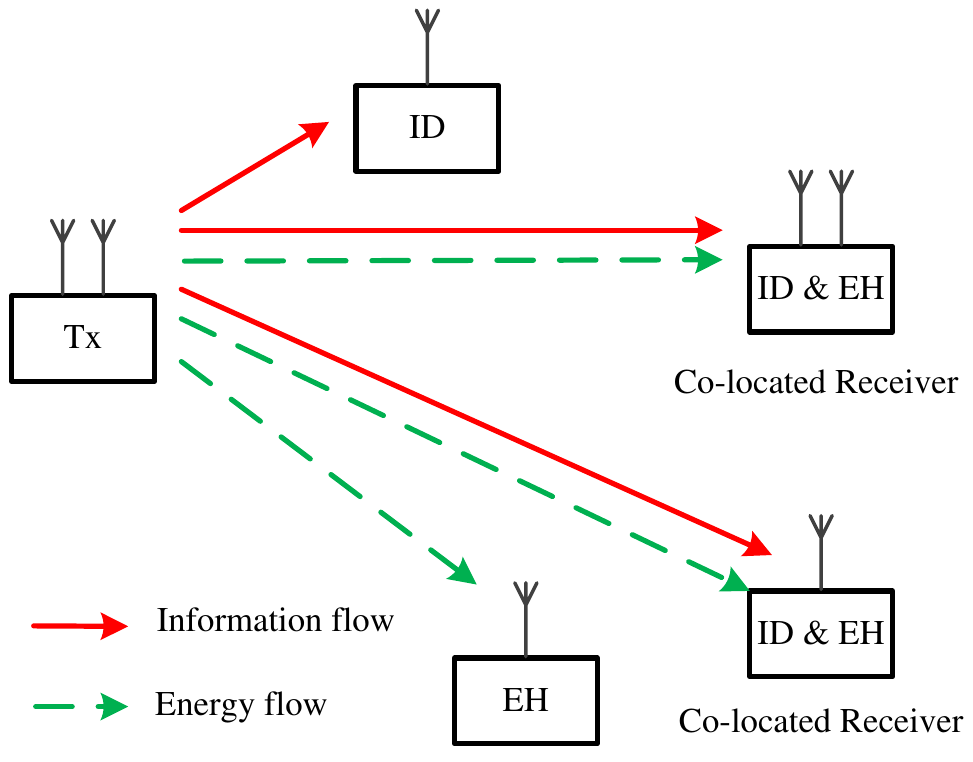}}\vspace{-2mm}
	\caption{A SWIPT system with separated and co-located ID and EH Rxs.}\vspace{-5mm}
	\label{fig:SWIPTSystemModel}
\end{figure}

In a SWIPT system, as shown in Fig. \ref{fig:SWIPTSystemModel}, energy harvesting receivers (EH Rxs) harvest RF energy and information decoding receivers (ID Rxs) decode the information contained in the signals emitted by the transmitter (Tx).
The EH Rxs and the ID Rxs may be physically
separated or co-located in the same user equipment depending on the  application\cite{ZhangRuiModel}.
For separated Rxs, ID and EH are performed based on different received signals at physically separated sites.
In contrast, a co-located Rx comprises both ID and EH hardware modules, which need to share the received RF signals at the analog front-end.
Unfortunately, practical circuits for EH cannot extract information \cite{Bruno2019Review} and vice versa.
As a compromise solution, several Rx architectures have been proposed\cite{ZhangRuiModel} for exploiting the received RF signals for EH and ID:
\begin{itemize}
	\item Time-switching (TS): TS co-located Rxs switch in time between the ID and EH hardware modules, which are synchronized with the Tx. The TS co-located Rx architecture requires the optimization of the TS sequences to achieve a desired rate-energy tradeoff.
	\item Power-splitting (PS): PS co-located Rxs split the power of the received signal into two power streams, where one stream with PS ratio $\rho$ is fed to the ID Rx and the other stream with PS ratio $1-\rho$ is fed to the EH Rx. 
	%
	%
	%

	\item Antenna-switching (AS): When the co-located Rxs are equipped with multiple antennas, an AS co-located Rx architecture can be adopted, where a subset of the antennas are selected for  EH only, while the remaining antennas are solely used for ID.
	Determining the optimal AS policy is critical to balance between wireless charging and the beamforming gain for ID.
\end{itemize}

We note that these three co-located Rx architectures can be combined in different ways, depending on the specific system setups and hardware constraints.
For example, the PS and AS co-located Rx architectures can be combined by equipping a power splitter at each receive antenna.
Not surprisingly, sharing the received signals at the co-located Rx incurs a tradeoff between information transfer and EH.
%
%
Moreover, the signal sharing policy at co-located Rxs has to be jointly designed with the resource allocation at the Tx to achieve a favorable rate-energy tradeoff.

\subsection{Resource Allocation Design for SWIPT Systems}
Resource allocation is a key concept for improving wireless system performance by making the best use of limited resources based on the available system information.
Resource allocation design has been extensively studied in the context of conventional data communication networks \cite{RogerOFDM1999,DerrickOFDMA2012,WeiNOMA7934461}.
In general, wireless system resources are power, the available bandwidth and time, as well as space if multiple antennas are deployed.
The system information exploited for resource allocation design includes channel state information (CSI), queue state information, and quality-of-service (QoS) requirements.
On the other hand, different system design objectives can be set, e.g., maximizing the system sum-rate, maximizing the system energy efficiency, or minimizing the system power consumption, depending on the specific application scenario.
However, resource allocation design for SWIPT systems differs from that in conventional wireless systems in several aspects. 
First, compared to ID Rxs, the mathematical EH models needed to capture the input-output characteristic of practical EH circuits at the EH Rxs are substantially more complicated, which imposes a great challenge for resource allocation design.
Second, energy-based performance metrics, such as the harvested energy and WPT efficiency, are as important as data rate-based metrics.
Third, in conventional data communication systems, co-channel interference is harmful and limits the system performance.
Hence, it should be suppressed or mitigated during resource allocation design.
In contrast, in SWIPT systems, strong interference is a vital energy source for energy-limited receivers and could be harvested to facilitate WPT.
%
%

In the following, we briefly overview seminal works on resource allocation design for SWIPT systems.
Different SWIPT system architectures were first proposed in \cite{ZhangRuiModel} for a multiple-input multiple-output (MIMO) broadcasting system.
Corresponding resource allocation design problems were formulated for maximization of the achievable rate-energy region\cite{ZhangRuiModel}.
Based on the framework in \cite{ZhangRuiModel}, the authors in \cite{LiangTS,LiangPS,ZhouOPS} developed schemes for optimal time/antenna-switching and/or optimal power splitting to achieve various tradeoffs between WIT and WPT.
Furthermore, extending SWIPT to multi-user communication systems is desirable as the property of one-to-many charging can be exploited for increasing the lifetime of a large number of wireless devices, e.g., wireless sensors in IoT networks.
In SWIPT-based multi-user communication systems, the resource allocation can be designed to exploit the multi-user diversity originating from the independent channel fading of different users to optimize the system performance taking into account certain constraints.
Examples of such designs include maximizing the weighted sum-power transferred to EH Rxs under minimum signal-to-interference-plus-noise ratio (SINR) requirements for ID Rx\cite{XuSWIPT} or minimizing the total transmit power under SINR and EH constraints\cite{ShiQingtwc}.
Moreover, compared with narrowband systems, applying SWIPT techniques in multi-carrier communications offers the opportunity to exploit frequency diversity as the frequency domain offers extra degrees of freedom (DoF) for flexible and efficient resource allocation design.
For example, orthogonal frequency-division multiple access (OFDMA) has been applied in SWIPT-based multi-user communication systems\cite{KaibinTSP,JR:WIPT_fullpaper_OFDMA,ZhouTWC}, where ID Rxs retrieve their information from the received signals on subcarriers allocated to them, while EH Rxs harvest the energy from the signals received on all subcarriers.
Besides, a multi-objective optimization (MOO) framework can be adopted to handle the conflicting system design goals of providing communication services while guaranteeing EH performance\cite{JR:MOOP_SWIPT,MengLiSWIPTMOO}.

All the above studies assumed that perfect CSI is available at the Tx, which is challenging to acquire in practice due to channel estimation errors, feedback delays, and quantization noises.
In practice, imperfect CSI at the transmitter (CSIT) usually leads to substantial performance degradation and system outages.
In SWIPT systems, the resulting channel uncertainty affects the performance of both WPT and WIT, and consequently system outages may happen more frequently compared to conventional communication systems.
In fact, SWIPT systems generally require a joint transceiver design and the coupling between WIT and WPT makes the resource allocation design more sensitive to CSIT errors.
Thus, robust resource allocation design is critical for achieving high SWIPT system performance in the presence of channel uncertainty\cite{GaofengAverage,KhandakeOutage,BinbinOutageWorst,XiangRobust,Xin_WorstCase,XinruiWorst}.
For instance, the authors in \cite{XiangRobust} maximized the worst-case harvested direct current (DC) power for the EH Rxs while guaranteeing a minimum data rate for the ID Rxs for all possible channel realizations.
When the channel uncertainty is not taken into account for resource allocation design, frequent violations of the minimum required data rate constraint of the ID Rxs occur\cite{XiangRobust}.
In contrast, the authors' robust design guarantees the performance of the SWIPT system even in the presence of channel uncertainty, at the expense of sophisticated resource allocation.

Furthermore, integrating other cutting-edge technologies, such as full-duplex \cite{LengFDSWIPT,HuFDswipt}, coordinated multipoint (CoMP)\cite{5706317}, physical layer security \cite{JR:rui_zhang_secrecy,NgSecureSWIPT}, and relaying technologies\cite{IoannisSWIPT,HimalJSAC}, into SWIPT systems is expected to further improve the system performance while introducing new resource allocation design challenges.
For instance, by leveraging macro-diversity, CoMP \cite{5706317} can extend the service areas of both WIT and WPT without increasing the overall transmit power\cite{TangCoMPSWIPT}.
However, the power loss in delivering power from the central processor to distributed remote radio heads and the corresponding backhaul capacity limitation for conveying data have to be considered for resource allocation design\cite{7037480}.
Interested readers may refer to \cite{LengFDSWIPT,HuFDswipt,JR:rui_zhang_secrecy,NgSecureSWIPT,IoannisSWIPT,HimalJSAC} and the references therein for more details.

In the early stages of research on resource allocation design for SWIPT systems, most of the works assumed a simple linear EH model.
Yet, from the micro-electronics literature\cite{CN:EH_measurement_2,JR:EH_measurement_1}, it is well-known that the EH circuits converting the received RF energy into electrical energy exhibit strong nonlinearities, especially for high RF input powers.
Therefore, adopting a linear EH model for resource allocation design may introduce a model mismatch, which degrades the system performance\cite{JR:Rania_nonlinear_circuit_based_TCOM,JR:non_linear_model,JR:Elena_TCOM,book:Kwan_power_transfer}.
The authors in \cite{JR:non_linear_model} firstly identified the implications of the nonlinearity of EH circuits for SWIPT system design and proposed a nonlinear saturation EH model, which leads to tractable problem formulations for resource allocation design.
In particular, the nonlinear saturation EH model provides a relationship between the average received RF power and the average harvested DC power for a given input distribution.
%
Most recently, several works \cite{JR:Rania_nonlinear_circuit_based_TCOM,BrunoTSP} proposed nonlinear circuit-based EH models, which characterize the relationship between the instantaneous received RF power and the instantaneous harvested DC power.
These nonlinear circuit-based EH models allow the optimization of the input distribution whereas both the linear EH model and the nonlinear saturation EH model do not.
In general, improving the EH circuit modeling accuracy can improve SWIPT system performance while it generally reduces the tractability of resource allocation design.
A comprehensive treatment of the relation between EH modeling accuracy and the resulting resource allocation design is not available in the literature, yet.

\subsection{Objective and Organization}
The main objective of this paper is to review and discuss resource allocation design for SWIPT systems.
There have been several overview papers on SWIPT\cite{KrikidisSWIPT,NiyatoEH,ZengWPT,Bruno2019Review,clerckx2021wireless}.
The early works in \cite{KrikidisSWIPT,NiyatoEH} relied on the simple linear EH model, whose practicality has been questioned since that time.
%
%
The authors of \cite{ZengWPT} provided an overview of the main communication and signal processing techniques for SWIPT systems for both the linear EH model and the nonlinear EH model.
However, the role of resource allocation for the joint design of the WPT and WIT subsystems was not highlighted.
Also, the authors of \cite{Bruno2019Review} investigated how the receiver architecture and the modeling of the energy harvesters affect the rate-energy tradeoff in SWIPT systems.
Most recently, the authors of \cite{clerckx2021wireless} discussed possible applications of future wireless-powered networks, including computing, sensing, and edge learning.
Nevertheless, the existing tutorial-style overview papers on SWIPT \cite{KrikidisSWIPT,NiyatoEH,ZengWPT,Bruno2019Review,clerckx2021wireless} did not focus on resource allocation design, which motivates this paper.

The remainder of this paper is organized as follows. 
Section II introduces the EH models developed for resource allocation in SWIPT systems.
Section III proposes a general resource allocation design framework for SWIPT systems and outlines corresponding solution methodologies.
Section IV presents three robust resource allocation design approaches to combat the channel uncertainty in SWIPT systems.
Potential future research directions are discussed in Section V, and Section VI concludes this paper.

\section{Energy Harvesting Models}

\begin{figure}[t]
	\center{
		\includegraphics[width=3.4in]{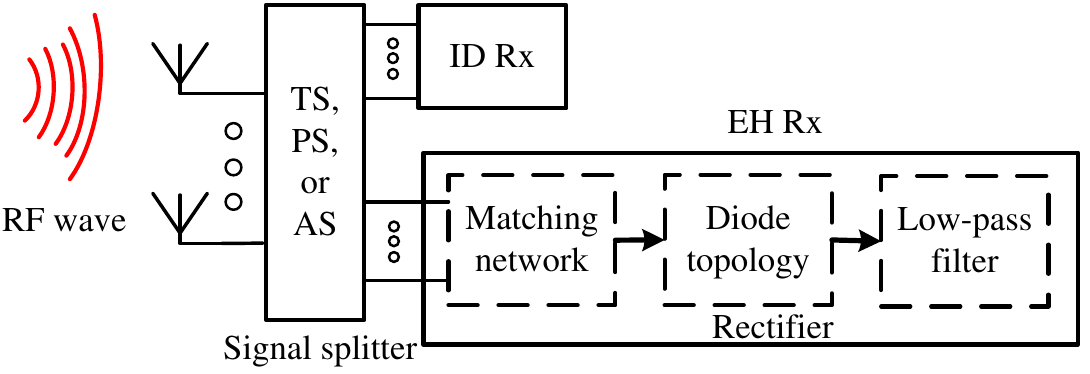}}\vspace{-2mm}
	\caption{Block diagram of typical co-located ID and EH Rxs.}\vspace{-5mm}
	\label{fig:block_diagram_EH}
\end{figure}

To facilitate resource allocation design for spectrally-efficient and energy-efficient SWIPT-based systems, it is necessary to characterize the input-output relationship of RF-based EH circuits via a suitable model\cite{Bruno2019Review}. 
{A practical RF-based EH circuit, known as a \textit{rectenna}, comprises an antenna and a rectifier \cite{JR:Energy_harvesting_circuit,JR:EH_measurement_1}, see Fig. \ref{fig:block_diagram_EH}. 
The rectifying circuit typically includes a matching network, diodes, and a low-pass filter, which converts the received RF energy into electrical energy \cite{JR:Energy_harvesting_circuit,JR:EH_measurement_1}.
Subsequently, the scavenged DC power can be stored in a battery for future use, e.g., for data transmission or signal processing. 
We note that the load resistance representing, e.g., an energy storage module, is part of the low-pass filter and is not explicitly shown in Fig. \ref{fig:block_diagram_EH}.}
To quantify the performance of an RF-based EH circuit, its RF-to-DC energy conversion efficiency has to be accurately modeled
and analyzed. 
In the following, we present three commonly adopted EH models striking a balance between modeling accuracy and modeling tractability.

\subsection{Linear EH Model}
In the conventional linear EH model, e.g., \cite{JR:WIPT_fullpaper_OFDMA,JR:rui_zhang_secrecy,JR:SWIPT_antennas,CN:OFDM_Kwan},  the average harvested power at the EH receiver, $P_{\mathrm{out} }^{\mathrm{Linear}}$, is modeled by the following linear equation:
\begin{equation}\label{eqn:linear_EH_model}
P_{\mathrm{out} }^{\mathrm{Linear}}=\eta  P^{\mathrm{EH}}_{\mathrm{Rx} },
\end{equation}\noindent
where $P^{\mathrm{EH}}_{\mathrm{Rx} }$ denotes the average power of the input RF signals for EH and $0<\eta \leq1$ is the constant power conversion efficiency. Its value can be obtained by curve fitting based on measured harvested DC power data. 
According to \eqref{eqn:linear_EH_model}, the harvested power at the EH Rx is linearly and directly proportional to the received RF power. 
This linear model was commonly adopted in the early works on SWIPT, e.g., \cite{JR:EE_SWIPT_Massive_MIMO,Ding2014}, as it facilitates simple resource allocation design and performance analysis. 
Yet, this model is not accurate as the output power of the EH circuit is not bounded even for large RF input powers.

\subsection{Nonlinear Circuit-based EH Model}
Practical RF-based EH circuits inevitably introduce nonlinearities for the end-to-end WPT due to the nonlinear components required for converting the RF signal to DC power\cite{CN:EH_measurement_2,JR:EH_measurement_1}. 
%
%
Considering the single-diode EH circuit shown in \cite[Fig.~2]{JR:Rania_nonlinear_circuit_based_TCOM}, the authors proposed an approximate closed-form expression for the output DC power with respect to (w.r.t.) the instantaneous received RF power assuming a sinusoidal input excitation signal\footnote{As shown in \cite[Eq.~(1)]{JR:Rania_nonlinear_circuit_based_TCOM}, the received RF signal is typically narrowband as the bandwidth is usually much smaller than the carrier frequency.
Therefore, a sinusoidal input excitation signal is sufficient for characterizing the EH circuit. 
Moreover, despite the assumed fixed sinusoidal input excitation signal, \eqref{eqn:output_power_circuit_basedV3} enables waveform design via optimization of the distribution of the input symbols $x$.}.
In particular, the instantaneous harvested DC power is given by \cite{JR:Rania_nonlinear_circuit_based_TCOM}
\begin{equation}\label{eqn:output_power_circuit_basedV3}
P_{\rm out}^{\mathrm{Cir}}(\beta)=\min\hspace{-1mm} \left(\hspace{-1mm}\left[\frac{1}{\varsigma}W_0\hspace{-1mm}\left(\varsigma e^\varsigma I_0(\beta)\right)\hspace{-1mm}-\hspace{-1mm}1\right]^2 \hspace{-1mm}I^2_{\rm s} R_{\mathrm{L}}, \frac{B_{\rm v}^2}{4 R_{\rm L}}\right),
\end{equation}
where $\beta\!\definedas\! \sqrt{2}B \left|h_{\mathrm{E}}\right| \left|x\right|$, $x\in \mathbb{C}$ denotes the transmitted complex symbol, and $h_{\mathrm{E}}\in \mathbb{C}$ is the channel gain from the Tx to the EH Rx.
Constants $B$ and $\varsigma$ are parameters that depend on the elements of the EH circuit.
Besides, $I_{\rm s}$ is the diode's reverse bias saturation current, $R_{\rm L}$ is the load resistance, and $B_{\mathrm{v}}$ is the diode reverse breakdown voltage.
Moreover, $W_0\left(\cdot\right)$ is the principal branch
of the LambertW function\cite{corless1996lambertw} and $I_0\left(\cdot\right)$ is the zeroth order modified Bessel function of the first kind.
The LambertW and Bessel functions make resource allocation design based on the expression for the harvested power in \eqref{eqn:output_power_circuit_basedV3} challenging.

In \cite{JR:Rania_nonlinear_circuit_based_TCOM}, it has been demonstrated that the characteristic of the model in \eqref{eqn:output_power_circuit_basedV3} closely matches the harvested DC power obtained from Advanced Design System (ADS) circuit simulations \cite{ADS}.
The output DC power of the rectifying circuit is highly nonlinear w.r.t. its RF input power due to the reverse breakdown of the diode.
In fact, for exceedingly large EH circuit input powers, the harvested DC power in \eqref{eqn:output_power_circuit_basedV3} saturates at $P_{\rm out}^{\mathrm{Cir}}(\beta)=\frac{B_{\rm v}^2}{4 R_{\rm L}}$\cite{Georgiadis_WPT_book_2016}. 
The nonlinear circuit-based EH model in \eqref{eqn:output_power_circuit_basedV3} characterizes the instantaneous input-output relationship of the energy harvester.
The average harvested DC power is given by ${\mathcal E}_{x}\{P_{\rm out}^{\mathrm{Cir}}(\beta)\}$, where the expectation ${\mathcal E}_{x}\{\cdot\}$ is taken w.r.t. the distribution of input symbol $x$.
As such, the circuit-based EH model allows the optimization of the input distribution of the transmit symbols to achieve a certain design objective, e.g., maximizing the average harvested DC power ${\mathcal E}_{x}\{P_{\rm out}^{\mathrm{Cir}}(\beta)\}$.

\begin{figure}[t]
	\centering
	\includegraphics[width=3.0in]{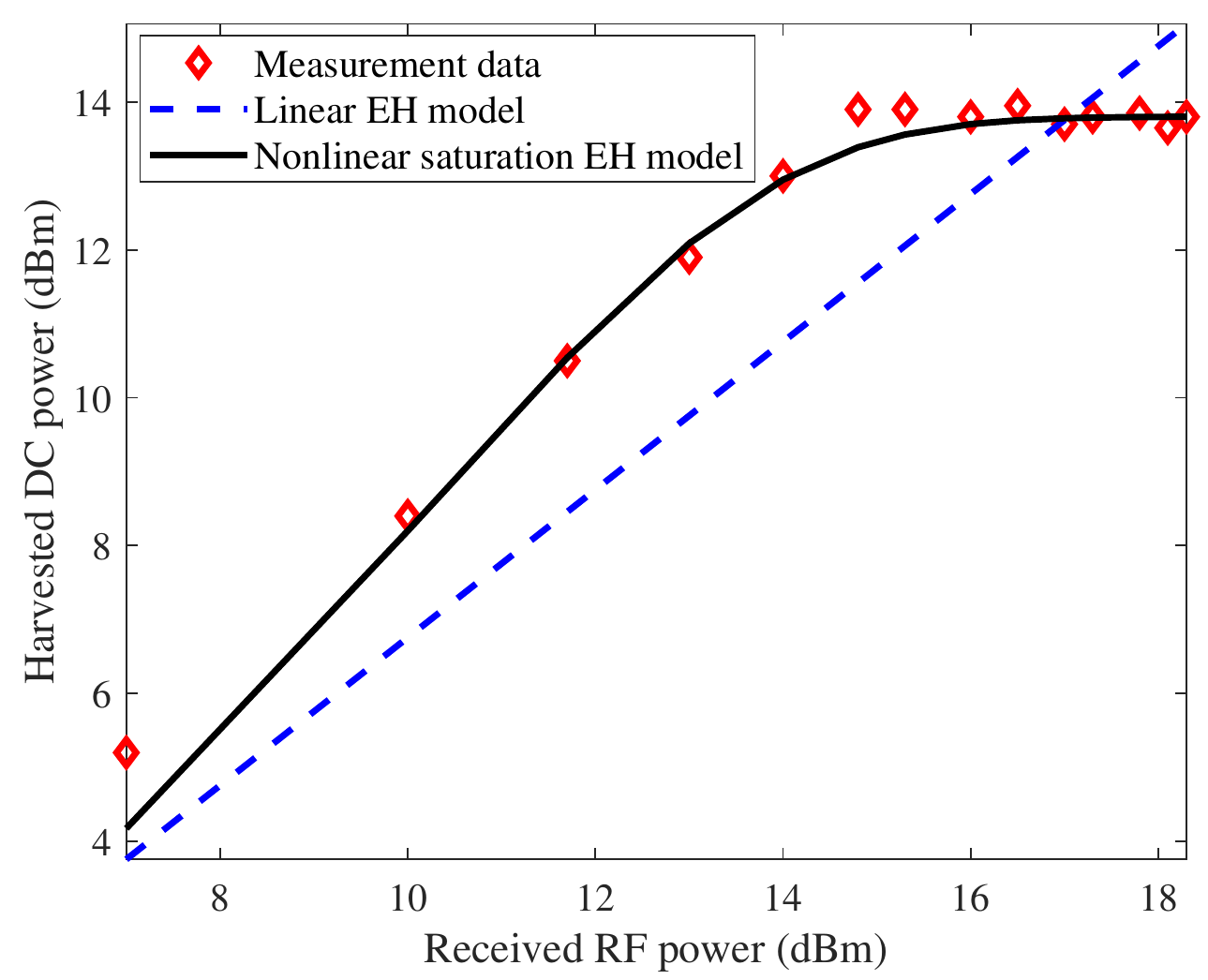}\vspace{-2mm}
	\caption{A comparison between measurement data from \cite{CN:EH_measurement_2}, the average harvested DC power for the nonlinear saturated model in \eqref{eqn:EH_non_linear} and the conventional linear EH model in \eqref{eqn:linear_EH_model}. The model parameters are obtained by curve fitting via least square algorithm with $P^{\mathrm{Sat}}=0.024$ Watt, $b=0.014$, $a=150$, and $\eta = 0.474$.}\vspace{-5mm} \label{fig:comparsion_EH}
\end{figure}

\subsection{Nonlinear Saturation EH Model}

Although the above circuit-based model can accurately capture the nonlinear input-output relationship of a practical EH circuit, the corresponding analytical expressions make resource allocation design very challenging.
Indeed, such a tailor-made method relies on specific implementation details of the EH circuits (e.g., the circuit schematic in \cite[Fig.~2]{JR:Rania_nonlinear_circuit_based_TCOM}) and the resulting mathematical expressions are different for different types of EH circuits. 
%
%
To isolate the system model from specific implementation details,  a parametric nonlinear EH model was proposed in \cite{JR:non_linear_model,JR:Elena_TCOM,book:Kwan_power_transfer} to  facilitate resource allocation algorithm design by striking a balance between modeling accuracy and tractability. 
In particular, for an arbitrary received RF signal with average power $P^{\mathrm{EH}}_{\mathrm{Rx}}$,  the average harvested DC power, $P_{\mathrm{out} }^{\mathrm{Sat}}$, is modeled as:
\begin{equation}\label{eqn:EH_non_linear}
P_{\mathrm{out} }^{\mathrm{Sat}}=
\frac{[\Psi
	- P^{\mathrm{Sat}}\Omega ]}{1-\Omega },\, \Omega =\frac{1}{1+\exp(a b )},
\end{equation}
where $\Psi=\frac{P^{\mathrm{Sat}}}{1+\exp\Big(-a (P^{\mathrm{EH}}_{\mathrm{Rx} }-b )\Big)}$ is a logistic (sigmoid) function and its input $P^{\mathrm{EH}}_{\mathrm{Rx} }$ represents the average received RF power for EH.
Parameter $P^{\mathrm{Sat}}$ is a positive constant representing the maximum available power at the output of the energy harvester when the EH circuit is saturated due to an exceedingly large input RF power.
Constant $a$ denotes the nonlinear charging rate w.r.t. the input power and constant $b $ is related to the minimum turn-on voltage of the EH circuit modeling the circuit sensitivity.
These three parameters jointly determine the shape of the logistic function which depends on the physical characteristics of the RF EH circuit and their values can be estimated by applying standard curve fitting algorithms.
Moreover, parameters $a$ and $b$ depend on the excitation waveform and have to be determined for a given input distribution.
More importantly, the saturation model in  \eqref{eqn:EH_non_linear} is a continuous quasi-convex function which is generally more tractable than the circuit-based EH model for resource allocation design, cf. \cite{JR:WPCN_nonlinear_Elena,CN:Elena_non_linear_scheduling}.

In Fig.~\ref{fig:comparsion_EH}, we show an example for curve fitting based on the measurement data in \cite{CN:EH_measurement_2} for both the linear EH model and the nonlinear saturation EH model. 
As can be observed, the parametric nonlinear model closely matches the experimental results provided in \cite{CN:EH_measurement_2}  for the RF power harvested by a practical EH circuit.  
In contrast, the conventional linear RF EH model fails to capture the nonlinear nature of practical EH circuits, especially in the high received RF power regime.
{We note that the received RF power that leads to saturation depends on the EH circuit used.
	While the circuit considered in \cite{CN:EH_measurement_2} causes saturation for received RF powers exceeding $P^{\mathrm{EH}}_{\mathrm{RX}} = 14.5$ dBm, the circuits reported in \cite{VarastehWPT} and \cite{BrunoTSP,BrunoBeneficial} saturate already for $P^{\mathrm{EH}}_{\mathrm{RX}} = -5$ dBm and $P^{\mathrm{EH}}_{\mathrm{RX}} = 0$ dBm, respectively.}

\begin{table*}[t]
	\caption{Comparison between considered RF-based EH models. }\label{tab:EH_models_comparison}\vspace{-2mm} \centering
	\scriptsize
	\begin{tabular}{c|c|c|c}\hline
		& Linear EH Model \eqref{eqn:linear_EH_model}&  Nonlinear Circuit-based EH Model  \eqref{eqn:output_power_circuit_basedV3}&  Nonlinear Saturation EH Model \eqref{eqn:EH_non_linear}\\
		\hline
		Based on  & Curve fitting of a linear function & EH circuit schematic& Curve fitting of a logistic function \\\hline
		Waveform/distribution design possible?  &$\times $ & \checkmark & $\times $   \\
		Models DC
		power
		saturation?   &$\times $ & \checkmark & $\checkmark $   \\
		Closed-form for analysis  & $\checkmark $& $\times $& $\checkmark $   \\
		Resource allocation difficulty  & Easy & Difficult & Medium   \\
		\hline
	\end{tabular}
	\vspace{-2mm}
\end{table*} 

To summarize, Table \ref{tab:EH_models_comparison} provides a comparison of the three considered RF-based EH models. 
In general, both the linear EH model and the nonlinear saturation EH model are more tractable for resource allocation design due to their simple input-output relationship.
In particular, the nonlinear saturation EH model can accurately capture the receiver sensitivity and output DC power saturation of practical RF-based EH circuits. 
However, since the curve fitting for finding the model parameters has to be performed for a given input distribution, this model does not allow the optimization of the input distribution.
In contrast, although the circuit-based model generally leads to more complicated expressions for the harvested power, it offers the possibility to jointly optimize the resource allocation and the input distribution for improving the system performance.
Indeed, depending on the setting, all models discussed in this section may be acceptable approximations of the behavior of a practical energy harvester.
The system designer has to decide which model provides the best tradeoff between resource allocation design complexity and performance for a given application.

\subsection{Other Nonlinear EH Models}
In addition to the nonlinear EH models discussed in Sections II-B and II-C, the authors in \cite{Waveform_design_WPT_Clerckx_2016} proposed a tractable Taylor expansion-based EH model accounting for the nonlinear diode characteristics for a multisine excitation signal.
In particular, this model expresses the harvested DC power in terms of a time average of a polynomial of the input RF signal.
Yet, compared with the nonlinear circuit-based EH model, the model in \cite{Waveform_design_WPT_Clerckx_2016} assumes a perfect matching network between the antenna and the rectifier and ignores the reverse-bias breakdown mode of the rectifying diode.
%
%
Moreover, although the EH models in Sections II-B, II-C, and \cite{Waveform_design_WPT_Clerckx_2016} capture the nonlinearity of EH circuits, they still have limitations. 
%
%
For instance, since a non-zero time is required for ramping up/down the voltage across the reactive elements of realistic EH circuits before reaching the steady state, EH circuits are not memoryless \cite{PaulART}. 
One possible approach for handling this issue is to model the dynamic of the EH circuit by a Markov decision process (MDP), e.g., \cite{JR:Energy_harvesting_circiut_memory_Robert}. 
On the other hand, the imperfection of hardware components introduces further nonlinearities to the input-output relationship of practical EH circuits \cite{PaulART,JR:Energy_harvesting_circiut_memory_Robert}.
Since an accurate analytical model for capturing all nonlinear effects is generally not tractable, a machine learning (ML)-based approach to adaptively model the nonlinear characteristics of practical EH circuits can be adopted \cite{JR:Energy_harvesting_circiut_memory_Robert}.
Nevertheless, such an ML approach models the EH circuit as a black box which generally offers limited insights for SWIPT system design.

\section{Resource Allocation Design for SWIPT Systems}
In this section, we introduce a representative SWIPT system model and establish a general resource allocation design framework.
Solution methodologies for addressing the formulated problem are introduced for the EH models considered in Section II. 
We first investigate the resource allocation design based on the linear EH model in \eqref{eqn:linear_EH_model} and the nonlinear saturation EH model in \eqref{eqn:EH_non_linear}, where we assume a Gaussian input distribution.
Then, we extend our consideration to the nonlinear circuit-based EH model in \eqref{eqn:output_power_circuit_basedV3}.
Finally, a simulation example is presented to illustrate the effectiveness of the introduced methodologies and to unveil insights for SWIPT system deigns.

\subsection{System Model}
We consider a multi-user multiple-input single-output (MISO) SWIPT system, where a Tx equipped with $N_{\mathrm{t}}$ antennas serves $K$ single-antenna users.
We assume each user is equipped with a co-located PS Rx.
{This is because the rate-energy region of SWIPT systems employing PS Rxs is convex and subsumes that of SWIPT systems employing TS Rxs for both the linear EH model and the nonlinear saturation EH model \cite{BrunoTSP,Bruno2019Review}.}
In fact, TS Rxs can be mimicked by forcing the PS ratio to 1 and 0 across time.
Note that fixing the PS ratio to $1$ and $0$ can also emulate the case of separated Rxs.

The baseband transmitted signal is given by
\begin{equation}\label{SWIPTTxSignal}
	\mathbf{x} = \sum_{k=1}^{K} \mathbf{w}_k s_k + \mathbf{v},
\end{equation}
where $s_k \sim \mathcal{CN}(0,1)$, $k \in \mathcal{K} = \{1,\ldots,K\}$, denotes the modulated symbol intended for user $k$ and $\mathbf{w}_k \in \mathbb{C}^{N_{\mathrm{t}}\times 1}$ is the corresponding WIT beamforming vector.
Here, $\mathbb{C}^{A\times B}$ stands for the set of $A\times B$ complex matrices and $\mathcal{CN}(\mu,\sigma^2)$ denotes the circularly symmetric complex Gaussian (CSCG) distribution with mean $\mu$ and variance $\sigma^2$.
Vector $\mathbf{v} \in \mathbb{C}^{N_{\mathrm{t}}\times 1}$ is an energy signal that is modeled as a complex pseudo-random sequence with covariance matrix\footnote{Theoretically, a deterministic energy signal is as effective as a random one in terms of WPT \cite{Bruno2019Review}. Yet, the latter can be easily shaped to satisfy potential constraints on the spectrum mask.} $\mathbf{V}={\cal E}\{\mathbf{v}\mathbf{v}^{\mathrm{H}}\}$, where $\left(\cdot\right)^{\mathrm{H}}$ stands for the Hermitian transpose of a vector or matrix.
Let us denote the vector characterizing the baseband equivalent frequency flat fading channel from the Tx to user $k$ as $\mathbf{h}_k\in{\mathbb{C}^{N_{\mathrm{t}} \times 1}}$.
The baseband received signal at user $k$ is given by
\begin{equation}
	y_k = \mathbf{h}^{\mathrm{H}}_k \left(\sum_{k'=1}^{K} \mathbf{w}_{k'} s_{k'}+ \mathbf{v}\right)  + n^{\mathrm{A}}_k,
\end{equation}
where $n^{\mathrm{A}}_k \sim \mathcal{CN}(0,\sigma_{\mathrm{A}}^2)$ denotes the additive white Gaussian noise (AWGN) at the receive antennas with power $\sigma_{\mathrm{A}}^2$.
For now, we assume that the CSI is perfectly known at the Tx and the users.
The extension to the case of imperfect CSI will be discussed in Section IV.
Ignoring the antenna noise power\cite{Bruno2019Review} as its contribution to the harvested power is negligible, the received RF power at user $k$ is equal to the equivalent baseband signal power, which is given by 
\begin{align}
\hspace{-2mm}P_{k,\mathrm{Rx}} = {\cal E}\{\left|y_k\right|^2\} 
 = \sum_{k'=1}^{K}\Tr\left( \mathbf{W}_{k'} \mathbf{H}_k\right) + \Tr\left( \mathbf{V}\mathbf{H}_k\right),
\end{align}
where $\Tr\left( \cdot\right)$ is the trace of a matrix, $\mathbf{W}_{k'} = \mathbf{w}_{k'}\mathbf{w}^{\mathrm{H}}_{k'}$, and $\mathbf{H}_k = \mathbf{h}_{k}\mathbf{h}^{\mathrm{H}}_{k}$.
User $k$ splits its received signal at the analog RF front-end with a PS ratio $\left(1-\rho_{k}\right)$ for EH, i.e., $P^{\mathrm{EH}}_{k,\mathrm{Rx}} = \left(1-\rho_{k}\right) P_{k,\mathrm{Rx}}$.
Note that both the desired signal and the inter-user interference (IUI) contribute to $P^{\mathrm{EH}}_{k,\mathrm{Rx}}$.
For the linear EH model, the harvested DC power at user $k$ is  given by
\begin{equation}\label{LinearEHModel}
	P_{k,\mathrm{out} }= \eta  P^{\mathrm{EH}}_{k,\mathrm{Rx} }.
\end{equation}
In contrast, for the nonlinear saturation EH model, the harvested DC power at user $k$ is given by
\begin{equation}\label{NonlinearEHModelSWIPT}
P_{k,\mathrm{out} }=\frac{[\Psi_k
	- P^{\mathrm{Sat}}\Omega ]}{1-\Omega },
\end{equation}
where $\Psi_k= \frac{P^{\mathrm{Sat}}}{1+\exp\Big(-a (P^{\mathrm{EH}}_{k,\mathrm{Rx} }-b )\Big)}$.
Besides, user $k$ decodes its information based on the other power stream with a PS ratio $\rho_{k}$ and the corresponding achievable rate is given by
\begin{equation}\label{AchievableRate}
	R_k = \log_2\hspace{-1mm}\left(\hspace{-1mm}1+\frac{\rho_{k}\Tr\left(\mathbf{W}_{k} \mathbf{H}_k\right)}{\rho_{k}\left(\hspace{-1mm}\Tr\hspace{-0.5mm}\left(\sum\limits_{k'\neq k}^{K} \hspace{-1mm}\mathbf{W}_{k'} \mathbf{H}_k \hspace{-1mm}+ \hspace{-1mm} \mathbf{V}\mathbf{H}_k\hspace{-1mm}\right) \hspace{-1mm}+\hspace{-1mm} \sigma_{\mathrm{A}}^2\hspace{-1mm}\right) \hspace{-1mm}+\hspace{-1mm} \sigma_{\mathrm{P}}^2}\hspace{-1mm}\right),
\end{equation}
where $\sigma_{\mathrm{P}}^2$ denotes the power of the AWGN introduced by the overall signal processing including the power splitting and the RF-to-baseband signal conversion at the ID Rx.
{Furthermore, we define the total power consumption minus the harvested DC power of SWIPT systems as 
	\begin{equation}\label{PowerModel}
	P_{\mathrm{Sys}} = P_{\mathrm{C}} + {\xi \mathrm{Tr} \left(\sum_{k=1}^{K} \mathbf{W}_{k} + \mathbf{V}\right) - \sum_{k=1}^{K} P_{k,\mathrm{out} }},
	\end{equation}
	where $P_{\mathrm{C}}$ comprises the constant circuit power consumption of both the Tx and the $K$ users.
	Here, $\xi \mathrm{Tr}\left(\sum\nolimits_{k=1}^{K} \mathbf{W}_{k} + \mathbf{V}\right)$ is the power
	dissipation of the power amplifier of the Tx and $\frac{1}{\xi}$ with $\xi \ge 1$ is the power amplifier efficiency.
	Although $P_{k,\mathrm{out} }$ might be much smaller compared with the first and second terms in \eqref{PowerModel}, $P_{\mathrm{Sys}}$ is physically meaningful as it represents the net power consumption of the SWIPT system\cite{JR:WIPT_fullpaper_OFDMA,ShiTSP}.
	%
	}
	%

Next, we define two important utility functions for the considered system, i.e., the WIT and WPT efficiencies, which are given by 
\begin{align}
	\mathcal{U}_{\mathrm{WIT}}^{\mathrm{Eff}}\left(\rho_{k},\hspace{-0.5mm}\mathbf{W}_k,\hspace{-0.5mm}\mathbf{V}\right) &= \frac{R_{\mathrm{WS}}\hspace{-0.5mm}\left(\rho_{k},\hspace{-0.5mm}\mathbf{W}_k,\hspace{-0.5mm}\mathbf{V}\right)}{P_{\mathrm{D}}\hspace{-0.5mm}\left(\mathbf{W}_k,\hspace{-0.5mm}\mathbf{V}\right)\hspace{-1mm}-\hspace{-1mm}P_{\mathrm{EH}}\hspace{-0.5mm}\left(\rho_{k},\hspace{-0.5mm}\mathbf{W}_k,\hspace{-0.5mm}\mathbf{V}\right)} \;\;\text{and}\;\;\label{ObjFunctionI}\\
	\mathcal{U}_{\mathrm{WPT}}^{\mathrm{Eff}}\left(\rho_{k},\hspace{-0.5mm}\mathbf{W}_k,\hspace{-0.5mm}\mathbf{V}\right) &= \frac{P_{\mathrm{EH}}\left(\rho_{k},\hspace{-0.5mm}\mathbf{W}_k,\hspace{-0.5mm}\mathbf{V}\right)}{P_{\mathrm{D}}\left(\mathbf{W}_k,\hspace{-0.5mm}\mathbf{V}\right)},\label{ObjFunctionII}
\end{align}
respectively, where $R_{\mathrm{WS}}\left(\rho_{k},\mathbf{W}_k,\mathbf{V}\right) = \sum_{k=1}^{K} \alpha_k R_k$ represents the weighted system sum-rate, $P_{\mathrm{D}}\left(\mathbf{W}_k,\mathbf{V}\right) = P_{\mathrm{C}} + \xi \Tr \left(\sum_{k=1}^{K} \mathbf{W}_{k} + \mathbf{V}\right)$ denotes the power dissipation required for wireless information and power transfer, and $P_{\mathrm{EH}}\left(\rho_{k},\mathbf{W}_k, \mathbf{V}\right) = \sum_{k=1}^{K} P_{k,\mathrm{out} }$ is the total power harvested by all users.
Here, $\alpha_k \ge 0$, $\forall k$, are non-negative weights which account for the priorities of different users and their values can be set to facilitate fairness in resource allocation.
The system WIT efficiency  $\mathcal{U}_{\mathrm{WIT}}^{\mathrm{Eff}}$ denotes the number of bits delivered while consuming one joule of net energy\cite{ZhiqiangChapter}.
The system WPT efficiency $\mathcal{U}_{\mathrm{WPT}}^{\mathrm{Eff}}$ represents the amount of harvested power while consuming one Watt of system power.
The former utility function is often adopted for the conventional communication-centric resource allocation designs, e.g., \cite{CN:OFDM_Kwan,JR:EE_SWIPT_Massive_MIMO,JR:WPC_Rui_Zhang}, while the latter one is suitable for power-centric designs \cite{QingqingWuSWIPT,BoaventuraDC}.
Besides, the two utility functions in \eqref{ObjFunctionI} and \eqref{ObjFunctionII} can be jointly considered in a MOO framework\cite{JR:MOOP_SWIPT,JR:Yan_MOOP}.
In general, MOO aims to provide a set of Pareto optimal resource allocation policies.
In particular, a resource allocation policy is Pareto optimal if there is no other policy that improves at least one of the objectives without detriment to the other objectives\cite{JR:MOOP_SWIPT}. 
There are various methods to handle MOO programming (MOOP) problems\cite{JR:MOOP_SWIPT}.
The crux of MOO methods is to convert MOOP to single-objective optimization programming (SOOP) via some parametric transformation such that the Pareto optimal set can be found by solving the SOOP problem.
For example, one can adopt the weighted Tchebycheff method to investigate the tradeoff between the WIT and WPT efficiencies for SWIPT systems as in\cite{JR:MOOP_SWIPT}.
Without loss of generality, we consider SOOP in the following for illustration.

\subsection{Problem Formulation}
One possible design goal for resource allocation in SWIPT systems is the maximization of the WIT efficiency.
This leads to the following optimization problem:
\begin{center}
\begin{tcolorbox}[title = {$\mathcal{P}_{\mathrm{WIT}}$:  WIT Efficiency Maximization}]
	\vspace{-5mm}
	\begin{align}\label{RAProblemFormulation}
	\underset{\begin{subarray}{c}
		0\le\rho_{k}\le 1,\\
		\mathbf{W}_k, \mathbf{V} \in \mathbb{H}^{N_{\mathrm{t}} \times N_{\mathrm{t}}}
		\end{subarray}}{\mathrm{maximize}} &&&\hspace{-0mm} \mathcal{U}_{\mathrm{WIT}}^{\mathrm{Eff}}\left(\rho_{k},\mathbf{W}_k,\mathbf{V}\right)
	\\
	\mathrm{s. t.}&&&\hspace{-0mm}\mbox{C1:}\,\sum_{k=1}^{K}\Tr \left( \mathbf{W}_{k}\right) + \Tr \left( \mathbf{V}\right)\le P_{\max},\notag\\
	&&&\hspace{-0mm}\mbox{C2:}\,R_{k}\ge R_{k,\mathrm{req}},\quad \forall k,\notag\\
	&&&\hspace{-0mm}\mbox{C3:}\,P_{k,\mathrm{out} } \ge P_{k,\mathrm{req} },\quad\forall k,\notag\\
	&&&\hspace{-0mm}\mbox{C4:}\,\mathcal{U}_{\mathrm{WPT}}^{\mathrm{Eff}} \left(\rho_{k},\mathbf{W}_k,\mathbf{V}\right) \ge \mathrm{WPT}^{\mathrm{Eff}}_{\mathrm{req}},\notag\\
	&&&\hspace{-0mm}\mbox{C5:}\,\mathrm{Rank} \left(\mathbf{W}_k\right) \le 1, \quad\forall k,\notag\\
	&&&\hspace{-0mm}\mbox{C6:}\,\mathbf{W}_k, \mathbf{V} \succeq \mathbf{0}, \quad\forall k,\notag
	\end{align}
	\vspace{-8mm}\par\noindent
\end{tcolorbox}
\end{center}
\noindent where $\mathbb{H}^{A \times A}$ stands for the set of $A \times A$ complex Hermitian matrices.
Constraint C1 limits the average transmit power to the available power budget $P_{\max}$ of the Tx.
Constraint C2 is an individual QoS constraint for user $k$ and constant $R_{k,\mathrm{req}}$ is the corresponding minimum required data rate.
Constraint C3 is the individual EH constraint for user $k$ and $P_{k,\mathrm{req} }$ is the corresponding minimum required harvested DC power.
Constraint C4 guarantees that the WPT efficiency does not fall below a given threshold $\mathrm{WPT}^{\mathrm{Eff}}_{\mathrm{req}}$.
Constraint C5 jointly with $\mathbf{W}_k \in \mathbb{H}^{N_{\mathrm{t}} \times N_{\mathrm{t}}}$ is imposed to
guarantee that $\mathbf{W}_{k} = \mathbf{w}_{k}\mathbf{w}^{\mathrm{H}}_{k}$ holds after optimization.
Constraint C6 ensures that $\mathbf{W}_{k}$ and $\mathbf{V}$ are positive semidefinite matrices.

{Problem formulation $\mathcal{P}_{\mathrm{WIT}}$ intends to make the WIT as efficient as possible under constraints on the WPT efficiency and minimum requirements for the harvested power and the data rate.  
$\mathcal{P}_{\mathrm{WIT}}$ is of practical interest when the circuit power consumptions for WIT and WPT are non-negligible or even overwhelming compared to the transmit power, which might be the case, e.g., for SWIPT systems operating in the millimeter wave frequency bands \cite{BinSWIPTmmWave}.
}
%
%
The formulated problem $\mathcal{P}_{\mathrm{WIT}}$ allows not only the investigation of the rate-energy tradeoff \cite{Bruno2019Review} but also of the tradeoff between the WPT and WIT efficiencies.
Note that one can swap the objective function with constraint C4 when the WPT efficiency is more critical for achieving the design objectives of the 
overall system.
The problem solving methodology detailed in Section III-C can be adopted to address both problems.

Additionally, the problem  in \eqref{RAProblemFormulation} provides a general unifying framework which subsumes existing resource allocation designs as special cases\cite{JR:WPC_Rui_Zhang,JR:WPCN_nonlinear_Elena,XuSWIPT}.
{For example, by omitting $P_{\mathrm{D}}\left(\mathbf{W}_k, \mathbf{V}\right)$ and $P_{\mathrm{EH}}\left(\rho_{k},\mathbf{W}_k, \mathbf{V}\right)$ in the objective function of \eqref{RAProblemFormulation}, the resulting problem formulation becomes the conventional weighted sum-rate maximization problem \cite{JR:WPC_Rui_Zhang}}:
\begin{center}
	\begin{tcolorbox}[title = {$\mathcal{P}_{\mathrm{Rate}}$:  Rate Maximization}]
	\vspace{-5mm}
\begin{align}\label{RAProblemFormulationRate}
\underset{\begin{subarray}{c}
	0\le\rho_{k}\le 1,\\
	\mathbf{W}_k, \mathbf{V} \in \mathbb{H}^{N_{\mathrm{t}} \times N_{\mathrm{t}}}
	\end{subarray}}{\mathrm{maximize}} R_{\mathrm{WS}}\left(\rho_{k},\mathbf{W}_k,\mathbf{V}\right)\;\;
\mathrm{s.t.}\;\;\mbox{C1-C6}.
\end{align}
\vspace{-5mm}\par\noindent
\end{tcolorbox}
\end{center}
%
%
%
%
{$\mathcal{P}_{\mathrm{Rate}}$ is suitable for optimization of systems with demanding requirements on the data rate, such as SWIPT systems operating in mobile cellular networks.}
{On the other hand, when there is a stringent constraint on the power consumption of the Tx, such as in unmanned aerial vehicle (UAV)-enabled SWIPT systems \cite{XieUAVSWIPT},  $R_{\mathrm{WS}}\left(\rho_{k},\mathbf{W}_k, \mathbf{V}\right)$ and $P_{\mathrm{EH}}\left(\rho_{k},\mathbf{W}_k, \mathbf{V}\right)$ can be removed from the objective function of \eqref{RAProblemFormulation} such that the problem degenerates to a power minimization problem \cite{JR:WPCN_nonlinear_Elena}:}
\begin{center}
	\begin{tcolorbox}[title = {$\mathcal{P}_{\mathrm{Power}}$:  Power Minimization}]
	\vspace{-5mm}
\begin{align}\label{RAProblemFormulationPower}
\underset{\begin{subarray}{c}
	0\le\rho_{k}\le 1,\\
	\mathbf{W}_k, \mathbf{V} \in \mathbb{H}^{N_{\mathrm{t}} \times N_{\mathrm{t}}}
	\end{subarray}}{\mathrm{minimize}} P_{\mathrm{D}}\left(\mathbf{W}_k, \mathbf{V}\right)\;\;
\mathrm{s. t.}\;\;\mbox{C1-C6}.
\end{align}
\vspace{-5mm}\par\noindent
\end{tcolorbox}
\end{center}
{\noindent Furthermore, for applications where wireless charging is the key for prolonging the lifetime of battery-limited users, e.g., IoT systems \cite{ChunshengGIOT} and Internet of Everything (IoE) networks \cite{ShuaifeiIoE}, $R_{\mathrm{WS}}\left(\rho_{k},\mathbf{W}_k, \mathbf{V}\right)$ and $P_{\mathrm{D}}\left(\mathbf{W}_k, \mathbf{V}\right)$ can be omitted in the objective function of \eqref{RAProblemFormulation} to maximize the total harvested power \cite{XuSWIPT}:}
\begin{center}
	\begin{tcolorbox}[title = {$\mathcal{P}_{\mathrm{EH}}$:  Harvested Power Maximization}]
	\vspace{-5mm}
	\begin{align}\label{RAProblemFormulationEH}
	\underset{\begin{subarray}{c}
		0\le\rho_{k}\le 1,\\
		\mathbf{W}_k, \mathbf{V} \in \mathbb{H}^{N_{\mathrm{t}} \times N_{\mathrm{t}}}
		\end{subarray}}{\mathrm{maximize}}  P_{\mathrm{EH}}\left(\rho_{k},\mathbf{W}_k, \mathbf{V}\right)\;\;
\mathrm{s. t.}\;\;\mbox{C1-C6}.
	\end{align}
	\vspace{-5mm}\par\noindent
\end{tcolorbox}
\end{center}

The above formulated problems are typical instances of SWIPT resource allocation design problems.
From an optimization point of view, problems $\mathcal{P}_{\mathrm{Power}}$ and $\mathcal{P}_{\mathrm{EH}}$ are easier to solve compared with $\mathcal{P}_{\mathrm{WIT}}$ and $\mathcal{P}_{\mathrm{Rate}}$ due to their simpler objective functions.
Nevertheless, $\mathcal{P}_{\mathrm{Power}}$ and $\mathcal{P}_{\mathrm{EH}}$ are also non-convex and the non-convexity arises from the coupling between the PS ratio $\rho_k$ and the beamforming matrices $\left(\mathbf{W}_k, \mathbf{V}\right)$ and from the rank-one constraint C5.
$\mathcal{P}_{\mathrm{WIT}}$ and $\mathcal{P}_{\mathrm{Rate}}$ are more challenging to solve due to the additional non-convexity introduced by IUI.
The fractional objective function further contributes to the degree of difficulty involved in solving $\mathcal{P}_{\mathrm{WIT}}$.
{In fact, solving $\mathcal{P}_{\mathrm{WIT}}$, $\mathcal{P}_{\mathrm{Rate}}$, $\mathcal{P}_{\mathrm{Power}}$, and $\mathcal{P}_{\mathrm{EH}}$ is NP-hard \cite{ZhiQuanNPHard}.
Thus, finding the globally optimal solution is challenging and entails prohibitive complexity.}
{Global optimization approaches, such as monotonic optimization \cite{TuyMonotonic,MatthiesenTSP,7812683,ZapponeEEMonotonic} and branch-and-bound (BnB) \cite{konno2000branch,WeiNOMA7934461}, can be used to find the optimal solutions of the formulated problems.
	In particular, monotonic optimization targets a subset of non-convex optimization problems described in terms of monotonic functions and difference of monotonic (d.m.) functions.
	This subsumes a large number of non-convex optimization problems as most non-convex functions, including the fractional functions in \eqref{ObjFunctionI} and \eqref{ObjFunctionII}, can be transformed to d.m. functions, as is detailed in \cite{7812683,ZapponeEEMonotonic}.
	On the other hand, BnB is also a widely adopted approach that successively divides the feasible region (Branch) into subregions and systematically discards non-promising subregions based on lower bounds or upper bounds (Bound) \cite{konno2000branch}.
	This partial enumeration strategy can be used to solve a wide array of global optimization problems, including monotonic optimization problems \cite{Tuy2005Chapter}.
	BnB algorithms converge to globally optimal solutions in a finite number of iterations if the branching operation is consistent and the selection operation is bound improving \cite{WeiNOMA7934461}.
	Yet, BnB usually has a lower speed of convergence compared to monotonic optimization as it cannot exploit the structure of the optimization problem.
	The computational complexity of both methods is generally exponential w.r.t. the number of input variables.
	More details on monotonic and BnB optimization for resource allocation design can be found in \cite{7812683,ZapponeEEMonotonic,konno2000branch,WeiNOMA7934461}.}

{A common approach to handle fractional objective functions is Dinkelbach's method \cite{dinkelbach1967nonlinear,CaiEEUAV}.
With this method, the fractional objective function is equivalently transformed into a subtractive form by introducing an auxiliary parameter.
The resulting subtractive optimization problem is solved in an inner loop and the auxiliary parameter is iteratively updated in an outer loop.
If the resulting inner optimization problem can be solved globally, the corresponding algorithm converges to the globally optimal solution of the original problem.
However, if the transformed optimization problem in subtractive form is also non-convex, usually only a suboptimal solution of the inner optimization problem can be obtained with an affordable computational complexity, e.g., by using successive convex approximation (SCA) \cite{WeiNOMA7934461} or the weighted sum mean square error (WSMSE) method \cite{QingjiangWSMSE}.
In this case, the convergence of Dinkelbach's method cannot be guaranteed.
Moreover, it is well known that Dinkelbach's method can only handle a single fractional objective function\cite{ZhiqiangChapter}.
For instance, $R_{\mathrm{WS}}\left(\rho_{k},\mathbf{W}_k,\mathbf{V}\right)$ in $\mathcal{P}_{\mathrm{WIT}}$ and $\mathcal{P}_{\mathrm{Rate}}$ is a sum of logarithms of fractional functions and thus Dinkelbach's method is not applicable.
Moreover, even for $\mathcal{P}_{\mathrm{Power}}$ and $\mathcal{P}_{\mathrm{EH}}$, the multiplication of $\rho_k$ and $\left(\mathbf{W}_k, \mathbf{V}\right)$ in $P_{k,\mathrm{out} }$ and $R_k$ prevents the application of Dinkelbach's method.
}
Hence, in this paper, we exploit the fractional programming (FP) method recently proposed in \cite{ShenFP} which can handle the fractional/multiplicative functions and even the function of fractional/multiplicative functions more flexibly.
Similar to Dinkelbach's method, FP also introduces auxiliary parameters to decouple the optimization variables and updates the optimization variables and auxiliary parameters iteratively.
However, the adopted \textit{quadratic transformation} \cite{ShenFP,ZhiqiangEENOMA} for FP is more flexibly such that the resultant inner optimization problem is usually convex.
Therefore, the FP method is guaranteed to converge to a stationary point of the original optimization problem and also enjoys a polynomial-time computational complexity.
In the following, we first present a solution methodology for optimization problem $\mathcal{P}_{\mathrm{WIT}}$ based on the quadratic transformation \cite{ShenFP,ZhiqiangEENOMA}, which can handle the severe variable coupling and can be readily used for developing a concrete algorithm for resource allocation.
Then, the solutions for $\mathcal{P}_{\mathrm{Rate}}$, $\mathcal{P}_{\mathrm{Power}}$, and $\mathcal{P}_{\mathrm{EH}}$ can be obtained by omitting the corresponding terms in the algorithm accordingly, since $\mathcal{P}_{\mathrm{WIT}}$ subsumes the other three problems.

\subsection{Solution Methodology for $\mathcal{P}_{\mathrm{WIT}}$}
FP is based on the {quadratic transformation} recently proposed in \cite{ShenFP}, which introduces an auxiliary variable to convert a fractional form function into an equivalent subtractive form, i.e.,
\begin{align}
&\mathop {{\maxo}}\limits_{{\bf{x}} \in {\cal X}} \;\frac{{{f_{{\rm{Obj}}}}({\bf{x}})}}{{{g_{{\rm{Obj}}}}({\bf{x}})}}\;\;
{\rm{s.t.}}\;\frac{{{f_{i,{\rm{Cons}}}}({\bf{x}})}}{{{g_{i,{\rm{Cons}}}}({\bf{x}})}} \ge {f_{i,{\rm{Convex}}}}({\bf{x}}),\; \forall i, \label{FractionalForm}\\[-2mm]
\Leftrightarrow &\underset{\mathbf{x}\in \mathcal{X}, y_{\rm{Obj}},y_{i,{\rm{Cons}}}}{\maxo}\; 2 y_{\rm{Obj}} \sqrt{{f_{{\rm{Obj}}}}(\mathbf{x})} - y_{\rm{Obj}}^2 {g_{{\rm{Obj}}}}(\mathbf{x}) \label{SubtractiveForm}\\
{\rm{s.t.}}\;&2y_{i,{\rm{Cons}}} \sqrt{{f_{i,{\rm{Cons}}}}(\hspace{-0.25mm}\mathbf{x}\hspace{-0.25mm})} \hspace{-1mm}-\hspace{-1mm} y_{i,{\rm{Cons}}}^2 {g_{i,{\rm{Cons}}}}(\hspace{-0.25mm}\mathbf{x}\hspace{-0.25mm}) \hspace{-1mm}\ge\hspace{-1mm} {f_{i,{\rm{Convex}}}}(\hspace{-0.25mm}{\bf{x}}\hspace{-0.25mm}),\; \forall i,\notag
\end{align}
where $y_{\rm{Obj}},y_{i,{\rm{Cons}}} \in \mathbb{R}$ are auxiliary variables and ${f_{i,{\rm{Convex}}}}({\bf{x}})$ is a convex function w.r.t. $\mathbf{x}$.
The proof of the equivalence between \eqref{FractionalForm} and \eqref{SubtractiveForm} is provided in \cite{ShenFP}. 
When ${{f_{{\rm{Obj}}}}({\bf{x}})}$ and ${{f_{i,{\rm{Cons}}}}({\bf{x}})}$ are concave functions w.r.t. $\mathbf{x}$, ${{g_{{\rm{Obj}}}}({\bf{x}})}$ and ${{g_{i,{\rm{Cons}}}}({\bf{x}})}$ are convex functions w.r.t. $\mathbf{x}$, and $\mathcal{X}$ is a convex set, the subtractive functions $2 y_{\rm{Obj}} \sqrt{{f_{{\rm{Obj}}}}(\mathbf{x})} - y_{\rm{Obj}}^2 {g_{{\rm{Obj}}}}(\mathbf{x})$ and $2y_{i,{\rm{Cons}}} \sqrt{{f_{i,{\rm{Cons}}}}(\mathbf{x})} - y_{i,{\rm{Cons}}}^2 {g_{i,{\rm{Cons}}}}(\mathbf{x})$ are concave functions w.r.t. $\mathbf{x}$.
Then, the resulting problem in \eqref{SubtractiveForm} is a convex optimization problem for given $y_{\rm{Obj}}$ and $y_{i,\rm{Cons}}$.
Moreover, for given $\mathbf{x}$, the optimal auxiliary variables are given by
\begin{equation}
	y_{\rm{Obj}} = \frac{\sqrt{f_{\rm{Obj}}(\mathbf{x})}}{g_{\rm{Obj}}(\mathbf{x})}\;\text{and}\;y_{i,\rm{Cons}} = \frac{\sqrt{f_{i,\rm{Cons}}(\mathbf{x})}}{g_{i,\rm{Cons}}(\mathbf{x})},
\end{equation}
respectively.

\begin{figure*}[!t]
	\setcounter{equation}{22}
	\begin{align}\label{SubtractiveFunctions_QP}
	\mathcal{G}_{\mathrm{Obj}}\left(\mathbf{W}_k, \hspace{-0.5mm}\mathbf{V},\hspace{-0.5mm} \gamma_k,\hspace{-0.5mm}P_{k,\mathrm{out} },\hspace{-0.5mm}\beta_{\mathrm{Obj}}\right) &\hspace{-1mm}=\hspace{-1mm}2 \beta_{\mathrm{Obj}} \sqrt{\sum_{k=1}^{K} \hspace{-1mm}\alpha_k \log_2\hspace{-1mm}\left(1\hspace{-1mm}+\hspace{-1mm}\gamma_k\right)} -\beta^2_{\mathrm{Obj}} \left[{P_{\mathrm{D}}\left(\mathbf{W}_k, \mathbf{V}\right) \hspace{-0.5mm}-\hspace{-0.5mm} \sum_{k=1}^{K} P_{k,\mathrm{out} }}\right]\notag\\[-0.5mm]
	\mathcal{G}_{\mathrm{C7}}\left(\mathbf{W}_k,\hspace{-0.5mm}\mathbf{V},\hspace{-0.5mm} \rho_{k},\hspace{-0.5mm}\beta_{k,\mathrm{C7}}\right) &\hspace{-1mm}=\hspace{-1mm}2\beta_{k,\mathrm{C7}} \sqrt{\Tr\left(\mathbf{W}_{k} \mathbf{H}_k\right)} 
	\hspace{-1mm}- \hspace{-1mm}\beta^2_{k,\mathrm{C7}} \hspace{-1mm}\left[{\hspace{-1mm}\left(\hspace{-1mm}\Tr\hspace{-1mm}\left(\sum_{k'\neq k}^{K} \hspace{-1mm}\mathbf{W}_{k'} \mathbf{H}_k \hspace{-1mm}+\hspace{-1mm} \mathbf{V}\mathbf{H}_k\hspace{-1mm}\right) \hspace{-1mm}+\hspace{-1mm} \sigma_{\mathrm{A}}^2\hspace{-1mm}\right) \hspace{-1mm}+\hspace{-1mm} \frac{\sigma_{\mathrm{P}}^2}{\rho_{k}}}\right]\hspace{-0.5mm}\notag\\[-0.5mm]
	\mathcal{G}_{\mathrm{C8}}\left(\mathbf{W}_k,\hspace{-0.5mm}\mathbf{V},\hspace{-0.5mm} P_{k,\mathrm{out} },\hspace{-0.5mm}\beta_{k,\mathrm{C8}}\right) & \hspace{-1mm}= \hspace{-1mm}2\beta_{k,{\mathrm{C8}}}\sqrt{\eta {\Tr\left(\sum_{k'=1}^{K} \mathbf{W}_{k'} \mathbf{H}_k + \mathbf{V}\mathbf{H}_k\right)}} - \beta^2_{k,{\mathrm{C8}}} P_{k,\mathrm{out} }
	\end{align}
	\vspace{-6mm}\par\noindent
	\hrulefill
	\vspace{-5mm}
\end{figure*}

{As a result, an iterative algorithm can be developed to update $\mathbf{x}$ and $\left(y_{\rm{Obj}}, y_{i,\rm{Cons}}\right)$ in an alternating manner.
Note that this algorithm is guaranteed to converge to a suboptimal solution of the original problem in \eqref{FractionalForm} if the transformed problem in \eqref{SubtractiveForm} can be solved globally.
We refer interested readers to \cite{ShenFP,ZhiqiangEENOMA} for a detailed proof of the convergence.
Furthermore, the algorithm has a polynomial-time computational complexity, which is well-suited for real-time implementation.}

In the following, we show how to perform the quadratic transformation to obtain a suboptimal solution of $\mathcal{P}_{\mathrm{WIT}}$ for both the linear EH model and the nonlinear saturation EH model.

\subsubsection{Linear EH model}
In this section, we transform $\mathcal{P}_{\mathrm{WIT}}$ into an equivalent optimization problem, which is suitable for applying the quadratic transformation.
Then, an iterative algorithm is developed to achieve a stationary point of 
$\mathcal{P}_{\mathrm{WIT}}$.
In the objective function of $\mathcal{P}_{\mathrm{WIT}}$, we can observe that the numerator is not a concave function w.r.t. the optimization variables due to the IUI and the denominator is not a convex function w.r.t. the optimization variables due to the coupling between $\rho_k$ and $\left(\mathbf{W}_k, \mathbf{V}\right)$.
To address these challenges, we introduce two auxiliary optimization variables $\gamma_k$ and $P_{k,\mathrm{out} }$ and add two corresponding constraints:
\setcounter{equation}{19}
\begin{align}
	\mbox{C7:}&\,\frac{\Tr\left(\mathbf{W}_{k} \mathbf{H}_k\right)}{\left(\hspace{-0.5mm}\Tr\hspace{-0.5mm}\left(\sum_{k'\neq k}^{K} \hspace{-0.5mm}\mathbf{W}_{k'} \mathbf{H}_k \hspace{-0.5mm}+\hspace{-0.5mm}  \mathbf{V}\mathbf{H}_k\hspace{-0.5mm}\right) \hspace{-0.5mm}+\hspace{-0.5mm} \sigma_{\mathrm{A}}^2\hspace{-0.5mm}\right) \hspace{-0.5mm}+\hspace{-0.5mm} \frac {\sigma_{\mathrm{P}}^2}{\rho_{k}}}\ge \gamma_k \;\text{and}\notag\\
	\mbox{C8:}&\,\eta \frac{\Tr\left(\sum_{k'=1}^{K} \mathbf{W}_{k'} \mathbf{H}_k + \mathbf{V}\mathbf{H}_k\right)}{P_{k,\mathrm{out} }} \ge  \frac{1}{\left(1-\rho_{k}\right)}, \;\forall k.\label{AdditionalConstraints}
\end{align}
It can be verified that constraints C7 and C8 are satisfied with equality at the optimal point.
Then, the problem in \eqref{RAProblemFormulation} can be rewritten as follows:
\begin{center}
	\begin{tcolorbox}[title = {$\mathcal{P}_{\mathrm{WIT}}$:  WIT Efficiency Maximization with Linear EH Model}]
	\vspace{-4mm}
\begin{align}\label{RAProblemFormulationII}
\underset{	\begin{subarray}{c}
	0\le\rho_{k}\le 1, \gamma_k,P_{k,\mathrm{out} } \\
	\mathbf{W}_k, \mathbf{V} \in \mathbb{H}^{N_{\mathrm{t}} \times N_{\mathrm{t}}}
	\end{subarray}}{\mathrm{maximize}} &&& \hspace{-2mm} \frac{\sum_{k=1}^{K} \alpha_k \log_2\left(1+\gamma_k\right)}{P_{\mathrm{D}}\left(\mathbf{W}_k, \mathbf{V}\right)-\sum_{k=1}^{K} P_{k,\mathrm{out} }}
\\
&&&\hspace{-18.8mm}\mathrm{s. t.}\;\;\mbox{C1},\mbox{C3},\mbox{C5-C8}\notag\\
&&&\hspace{-12mm}\mbox{C2:}\,\gamma_k\ge 2^{R_{k,\mathrm{req}}}-1,\; \forall k,\notag\\
&&&\hspace{-12mm}\mbox{C4:}\,\sum_{k=1}^{K} P_{k,\mathrm{out} } \ge  \mathrm{WPT}^{\mathrm{Eff}}_{\mathrm{req}} P_{\mathrm{D}}\left(\mathbf{W}_k, \mathbf{V}\right).\notag
\end{align}
\vspace{-5mm}\par\noindent
\end{tcolorbox}
\end{center}

Adopting the quadratic transformation in \eqref{FractionalForm} and \eqref{SubtractiveForm}, the problem in \eqref{RAProblemFormulationII} can be transformed into the following equivalent optimization problem:
\begin{center}
	\begin{tcolorbox}[title = {$\mathcal{P}_{\mathrm{WIT}}$:  WIT Efficiency Maximization with Linear EH Model}]
	\vspace{-5mm}
\begin{align}\label{RAProblemFormulationIII}
\underset{	\begin{subarray}{c}
	0\le\rho_{k}\le 1, \gamma_k,P_{k,\mathrm{out} } \\
	\mathbf{W}_k, \mathbf{V} \in \mathbb{H}^{N_{\mathrm{t}} \times N_{\mathrm{t}}},\\
	\beta_{\mathrm{Obj}},\beta_{k,\mathrm{C7}},\beta_{k,\mathrm{C8}}
	\end{subarray}}{\mathrm{maximize}} &&&\hspace{-2mm}
\mathcal{G}_{\mathrm{Obj}}\left(\mathbf{W}_k, \mathbf{V},\gamma_k,P_{k,\mathrm{out} },\beta_{\mathrm{Obj}}\right)
\\[-2mm]
&&& \hspace{-22.8mm}\mathrm{s.t.} \;\;\mbox{C1-C5},\,\notag\\
&&&\hspace{-16mm}\mbox{C7:}\,\mathcal{G}_{\mathrm{C7}}\left(\mathbf{W}_k, \mathbf{V}, \rho_{k},\beta_{\mathrm{C7}}\right)\ge \gamma_k, \;\forall k,\notag\\
&&&\hspace{-16mm}\mbox{C8:}\,\mathcal{G}_{\mathrm{C8}}\left(\mathbf{W}_k, \mathbf{V}, P_{k,\mathrm{out} },\beta_{k,\mathrm{C8}}\right) \ge \frac{1}{\left(1-\rho_{k}\right)}, \forall k,\notag
\end{align}
\vspace{-5mm}\par\noindent
\end{tcolorbox}
\end{center}
\noindent where $\beta_{\mathrm{Obj}},\beta_{k,\mathrm{C7}}, \beta_{k,\mathrm{C8}} \in \mathbb{R}$ denote the auxiliary variables corresponding to the objective function, C7, and C8, respectively.
The corresponding subtractive functions $\mathcal{G}_{\mathrm{Obj}}$, $\mathcal{G}_{\mathrm{C7}}$, and $\mathcal{G}_{\mathrm{C8}}$ are given in \eqref{SubtractiveFunctions_QP} at the top of next page.

The resulting subtractive functions in \eqref{SubtractiveFunctions_QP} are concave w.r.t. the optimization variables for given auxiliary variables.
Except for constraint C5, the problem in \eqref{RAProblemFormulationIII} is a convex optimization problem given the auxiliary variables.
On the other hand, given $\left(\mathbf{W}_k, \mathbf{V}, \gamma_k,P_{k,\mathrm{out} }\right)$, the optimal auxiliary variables can be updated as follows
\setcounter{equation}{23}
\begin{align}
	\beta_{\mathrm{Obj}} &=  \frac{\sqrt{\sum_{k=1}^{K} \hspace{-1mm}\alpha_k \log_2\hspace{-1mm}\left(1\hspace{-1mm}+\hspace{-1mm}\gamma_k\right)}}{{P_{\mathrm{D}}\left(\mathbf{W}_k, \mathbf{V}\right) \hspace{-0.5mm}-\hspace{-0.5mm} \sum_{k=1}^{K} P_{k,\mathrm{out} }}}, \notag\\
		\beta_{k,\mathrm{C7}} &= \frac{\sqrt{\Tr\left(\mathbf{W}_{k} \mathbf{H}_k\right)}}{{\hspace{-1mm}\left(\hspace{-1mm}\Tr\hspace{-1mm}\left(\sum_{k'\neq k}^{K} \hspace{-1mm}\mathbf{W}_{k'} \mathbf{H}_k \hspace{-1mm}+\hspace{-1mm} \mathbf{V}\mathbf{H}_k\hspace{-1mm}\right) \hspace{-1mm}+\hspace{-1mm} \sigma_{\mathrm{A}}^2\hspace{-1mm}\right) \hspace{-1mm}+\hspace{-1mm} \frac{\sigma_{\mathrm{P}}^2}{\rho_{k}}}},\forall k,\;\;\text{and}\notag\\
			\beta_{k,{\mathrm{C8}}} & = \frac{\sqrt{\eta {\Tr\left(\sum_{k'=1}^{K} \mathbf{W}_{k'} \mathbf{H}_k + \mathbf{V}\mathbf{H}_k\right)}}}{P_{k,\mathrm{out}}},\forall k. \label{AuXVariableUpdate}
\end{align}

Now, an algorithm can be developed by solving \eqref{RAProblemFormulationIII} for given $\left(\beta_{\mathrm{Obj}},\beta_{k,\mathrm{C7}}, \beta_{k,\mathrm{C8}}\right)$ and updating $\left(\beta_{\mathrm{Obj}},\beta_{k,\mathrm{C7}}, \beta_{k,\mathrm{C8}}\right)$ according to \eqref{AuXVariableUpdate} alternatingly.
{Such an algorithm is guaranteed to converge to a stationary point of the problem in \eqref{RAProblemFormulationIII} if the inner problem, i.e., the problem in \eqref{RAProblemFormulationIII} for given $\left(\beta_{\mathrm{Obj}},\beta_{k,\mathrm{C7}}, \beta_{k,\mathrm{C8}}\right)$, can be solved globally \cite{ShenFP,ZhiqiangEENOMA}.}
To solve \eqref{RAProblemFormulationIII} for given $\left(\beta_{\mathrm{Obj}},\beta_{k,\mathrm{C7}}, \beta_{k,\mathrm{C8}}\right)$, we adopt semidefinite relaxation (SDR) \cite{ZhiQuanTSPM} to
remove constraint C5 from \eqref{RAProblemFormulationIII}.
The resulting problem is a convex semidefinite programming (SDP) problem, which can be solved by numerical convex solvers, such as CVX\cite{book:convex}.
If the solution of the relaxed SDP problem satisfies constraint C5, i.e., $\mathrm{Rank} \left(\mathbf{W}_k\right) \le 1$, it is the optimal solution. 
Now, we study the tightness of the SDP relaxation in the following theorem.
\begin{Thm}\label{Theorem1}
	Assuming that the channel vectors of all users, $\mathbf{h}_k$, $\forall k$, are mutually statistically independent, given $\left(\beta_{\mathrm{Obj}},\beta_{k,\mathrm{C7}}, \beta_{k,\mathrm{C8}}\right)$ and $P_{\max}>0$, the optimal WIT beamforming matrices $\mathbf{W}^*_k$ of the relaxed version of the problem in \eqref{RAProblemFormulationIII} are rank-one, i.e., $\mathrm{Rank} \left(\mathbf{W}^*_k\right) \le 1$, $\forall k$, with probability one. Furthermore, the optimal WPT beamformer is $\mathbf{v}^* = \mathbf{0}$, where $\mathbf{0}$ denotes the vector with all zero entries.
\end{Thm}
\begin{proof}
	See Appendix for a proof of Theorem \ref{Theorem1}.
	%
\end{proof}

According to Theorem \ref{Theorem1}, the SDP relaxation is tight. 
Moreover, when the energy signal cannot be canceled at the Rxs, a dedicated WPT signal $\mathbf{v}$ is not needed for maximizing the WIT efficiency despite the EH constraint.
In fact, it has been demonstrated that, for the linear EH model, the CSCG input distribution is optimal in terms of both WIT and WPT \cite{Bruno2019Review}.
Therefore, the optimal WIT beamformer, $\mathbf{w}^*_k$, is sufficient to maximize the WIT and WPT efficiencies simultaneously in SWIPT systems\footnote{Note that if the energy signal $\mathbf{v}$ is known at the ID Rxs, it can be canceled at user $k$ to further improve the spectral efficiency\cite{XuSWIPT}. In such a case, the energy signal can in fact improve the WPT performance, i.e., a non-zero $\mathbf{v}$ is generally optimal.}.

\subsubsection{Nonlinear Saturation EH Model}
For the nonlinear saturation EH model in \eqref{NonlinearEHModelSWIPT}, constraint C8 in \eqref{AdditionalConstraints} needs to be reformulated accordingly.
Besides, we need to additionally introduce an optimization variable  $P^{\mathrm{EH}}_{k,\mathrm{Rx} }$ to facilitate the quadratic transformation.
In this case, the problem in \eqref{RAProblemFormulation} can be rewritten as follows:
\begin{center}
	\begin{tcolorbox}[title = {$\mathcal{P}_{\mathrm{WIT}}$:  WIT Efficiency Maximization with Nonlinear Saturation EH Model}]
	\vspace{-4mm}
\begin{align}\label{RAProblemFormulationIV}
\underset{	\begin{subarray}{c}
0\le\rho_{k}\le 1, \gamma_k,P_{k,\mathrm{out} } \\\mathbf{W}_k, \mathbf{V} \in \mathbb{H}^{N_{\mathrm{t}} \times N_{\mathrm{t}}}, P^{\mathrm{EH}}_{k,\mathrm{Rx} }
	\end{subarray}}{\mathrm{maximize}} &&&\hspace{-5mm} \frac{\sum_{k=1}^{K} \alpha_k \log_2\left(1+\gamma_k\right)}{P_{\mathrm{D}}\left(\mathbf{W}_k, \mathbf{V}\right)-\sum_{k=1}^{K} P_{k,\mathrm{out} }}
\\
&&&\hspace{-34.8mm}\mathrm{s.t.}\;\;\mbox{C1-C7},\notag\\
&&&\hspace{-28mm}\mbox{C8a:}\, \frac{\sum_{k'=1}^{K}\hspace{-1mm}\Tr\left( \mathbf{W}_{k'} \mathbf{H}_k\right) \hspace{-1mm}+\hspace{-1mm} \Tr\left( \mathbf{V}\mathbf{H}_k\right)}{P^{\mathrm{EH}}_{k,\mathrm{Rx} }} \hspace{-1mm}\ge \hspace{-1mm} \frac{1}{\left(1\hspace{-1mm}-\hspace{-1mm}\rho_{k}\right)}, \;\forall k,\notag\\
&&&\hspace{-28mm}\mbox{C8b:}\,\frac{1}{1\hspace{-1mm}+\hspace{-1mm}\exp\hspace{-0.5mm}\left[-a\left(P^{\mathrm{EH}}_{k,\mathrm{Rx} } \hspace{-1mm}-\hspace{-1mm} b\right)\right]} \hspace{-1mm}\ge\hspace{-1mm} \frac{P_{k,\mathrm{out} }}{P^{\mathrm{Sat}}}\left(1\hspace{-1mm}-\hspace{-1mm}\Omega\right)\hspace{-1mm} +\hspace{-1mm} \Omega,\forall k.\notag
\end{align}
\vspace{-4mm}\par\noindent
\end{tcolorbox}
\end{center}
Now, the quadratic transformation can be used to transform the problem in \eqref{RAProblemFormulationIV} into an equivalent subtractive form by introducing auxiliary variables associated with the objective function, C7, C8a, and C8b.
{Comparing \eqref{RAProblemFormulationII} and \eqref{RAProblemFormulationIV}, we can observe that C8a in \eqref{RAProblemFormulationIV} is equivalent to C8 in \eqref{RAProblemFormulationII} when the RF-to-DC conversion does not cause any energy loss, i.e., $\eta=1$.
	In fact, the difference between \eqref{RAProblemFormulationII} and \eqref{RAProblemFormulationIV} is the additional constraint C8b in \eqref{RAProblemFormulationIV}.
	The similar structure of both problems implies that the SDP relaxation is also tight for \eqref{RAProblemFormulationIV}.
	This is formally stated in the following corollary.
	\begin{Cor}\label{Cor1}
		Assuming that the channel vectors of all users, $\mathbf{h}_k$, $\forall k$, are mutually statistically independent, given the auxiliary variables associated with the objective function, C7, C8a, and C8b, if $P_{\max}>0$, the optimal WIT beamforming matrices $\mathbf{W}^*_k$, $\forall k$, of the problem in \eqref{RAProblemFormulationIV} satisfy $\mathrm{Rank} \left(\mathbf{W}^*_k\right) \le 1$ with probability one. Besides, the optimal WPT beamformer is $\mathbf{v}^* = \mathbf{0}$.
	\end{Cor}
	\begin{proof}
		This corollary can be proved by noting that constraint C8b in \eqref{RAProblemFormulationIV} does not affect the KKT conditions associated with $\left(\mathbf{W}_k, \mathbf{V}\right)$.
		Hence, the proof of Theorem \ref{Theorem1} in the Appendix is also valid for the nonlinear saturation EH model.
	\end{proof}

	Therefore, the SDR approach can be employed to develop a suboptimal resource allocation algorithm by updating the optimization variables in \eqref{RAProblemFormulationIV} and corresponding auxiliary variables iteratively.
	The convergence is guaranteed since given the auxiliary variables, the problem in \eqref{RAProblemFormulationIV} can be solved globally via SDR.}
Moreover, the additional constraint C8b in \eqref{RAProblemFormulationIV} implies that the resource allocation design based on the linear EH model may outperform that based on the nonlinear saturation EH model if the EH circuit was actually linear.
However, in practice, the resource allocation design based on the linear EH model suffers from a model mismatch and the expected system performance is not achievable, as shown in Section III-D.
%

\begin{table*}[t]\caption{System parameters.}\vspace{-2mm}\label{tab:SimulationParametersII} \centering
	\scriptsize
	\begin{tabular}{c|c|c|c}\hline
		Number of users & $K = 3$ & Number of transmit antennas & $N_t = 16$ \\
		Path loss exponent & 2.5 & Transmit antenna gain & 10 dBi \\
		Multipath fading distribution & Rician fading & Rician factor & 2 dB \\
		Antenna noise power & $\sigma_{\mathrm{A}}^2 = -100$ dBm  &  Processing noise  & $\sigma_{\mathrm{P}}^2 = -80$ dBm \\
		Minimum harvested DC power & $P_{k,\mathrm{req} } = 1 \mu \text{W}$  &  Minimum WPT efficiency & $\mathrm{WPT}^{\mathrm{Eff}}_{\mathrm{req}} = 10^{-5} \sim 10^{-4}$ \\
		Minimum rate requirement &$R_{k,\mathrm{req}} = 1$ bit/s/Hz& Maximum transmit power & $P_{\max} = 30$ dBm\\ 
		Circuit power consumption &$P_{\mathrm{C}} = 1$ W & Power amplifier parameter  & ${\xi} = 1.5$ \\
		Outage probability&$\kappa_{R_k}=\kappa_{P_k}= \kappa_{\mathrm{EH}} =0.001\sim 0.1$ &CSI error parameter &$\frac{\sigma^2_{{h}_k}}{\left|{h}_k\right|^2} =\frac{e^2_{h_k}}{\left|{h}_k\right|^2} = 0.01\sim 0.1$ \\\hline
		EH model parameters & \multicolumn{3}{c}{$\eta = 0.474$, $P^{\mathrm{Sat}}=0.024$ Watt, $b=0.014$, and $a=150$}  \\
		\hline
	\end{tabular}
	\vspace{-5mm}
\end{table*} 

\subsubsection{Nonlinear Circuit-based EH Model}
Adopting a nonlinear circuit-based EH model requires the joint design of the input distribution and the resource allocation.
The optimal input distribution that maximizes the WIT performance under an EH constraint is unique, discrete, and finite \cite{JR:Rania_nonlinear_circuit_based_TCOM}.
However, finding the optimal input distribution to maximize a general utility function, such as \eqref{ObjFunctionI}, typically results in an intractable problem.
%
%
Also, the coupling of the input distribution and the resource allocation variables imposes a challenge for optimization.

In the following, we propose an iterative suboptimal design approach, where the input distribution and the resource allocation are designed in an alternating manner.
In particular, in the $i$-th iteration, given the resource allocation policy $\left(\mathbf{W}^i_k, \mathbf{V}^i, \rho^i_{k}\right)$, the optimal input distribution enjoying the highest WIT efficiency is numerically determined.
Motivated by the fact that the optimal input distribution for WIT is the zero-mean CSCG distribution and the optimal input distribution for WPT has an on-off characteristic\cite{Bruno2019Review,JR:Rania_nonlinear_circuit_based_TCOM}, we propose the following channel input:
\begin{equation}
s_k = \sqrt{\alpha^i_{k,\mathrm{I}}}s_{k,\mathrm{I}} + \sqrt{\alpha^i_{k,\mathrm{E}}} s_{k,\mathrm{E}},
\end{equation}
where $s_{k,\mathrm{I}}$ denotes the information-bearing symbols for user $k$ following a zero-mean CSCG distribution and $s_{k,\mathrm{E}}$ denotes the energy-bearing symbols for user $k$ following an on-off distribution.
Here, $\alpha^i_{k,\mathrm{I}},\alpha^i_{k,\mathrm{E}} \ge0$ represent the powers allocated for WIT and WPT to user $k$ in the $i$-th iteration, respectively.
To facilitate the joint design, we assume all user devices employ the same EH circuits and $\alpha^i_{k,\mathrm{I}} = \alpha^i_{k',\mathrm{I}}$ and $\alpha^i_{k,\mathrm{E}} = \alpha^i_{k',\mathrm{E}}$, $\forall k \neq k'$.
Assuming the impulsive signal $s_{k,\mathrm{E}}$ can be canceled at the Rx via successive interference cancellation (SIC), the achievable rate is given by \eqref{AchievableRate} if only the CSCG signal is used for WIT.
Given $\left(\mathbf{W}^i_k, \mathbf{V}^i, \rho^i_{k}\right)$, we can employ the nonlinear circuit-based EH model in \eqref{eqn:output_power_circuit_basedV3} to simulate the harvested DC power and numerically select the optimal power allocation, $\left(\alpha^i_{k,\mathrm{I}},\alpha^i_{k,\mathrm{E}}\right)$, which yields the highest WIT efficiency.
Then, the model parameters of the nonlinear saturation EH model in \eqref{eqn:EH_non_linear} can be updated by curve fitting based on the harvested DC power obtained from the nonlinear circuit-based EH model for the optimized input distribution.
The resource allocation for the $\left(i+1\right)$-th iteration is designed based on the selected input distribution and the updated nonlinear saturation EH model, where problem formulation \eqref{RAProblemFormulationIV} and the corresponding solution methodologies  are applicable.

\subsection{Simulation Result for an Exemplary SWIPT System}

\begin{figure}[t]
	\centering
	\includegraphics[width=3.0in]{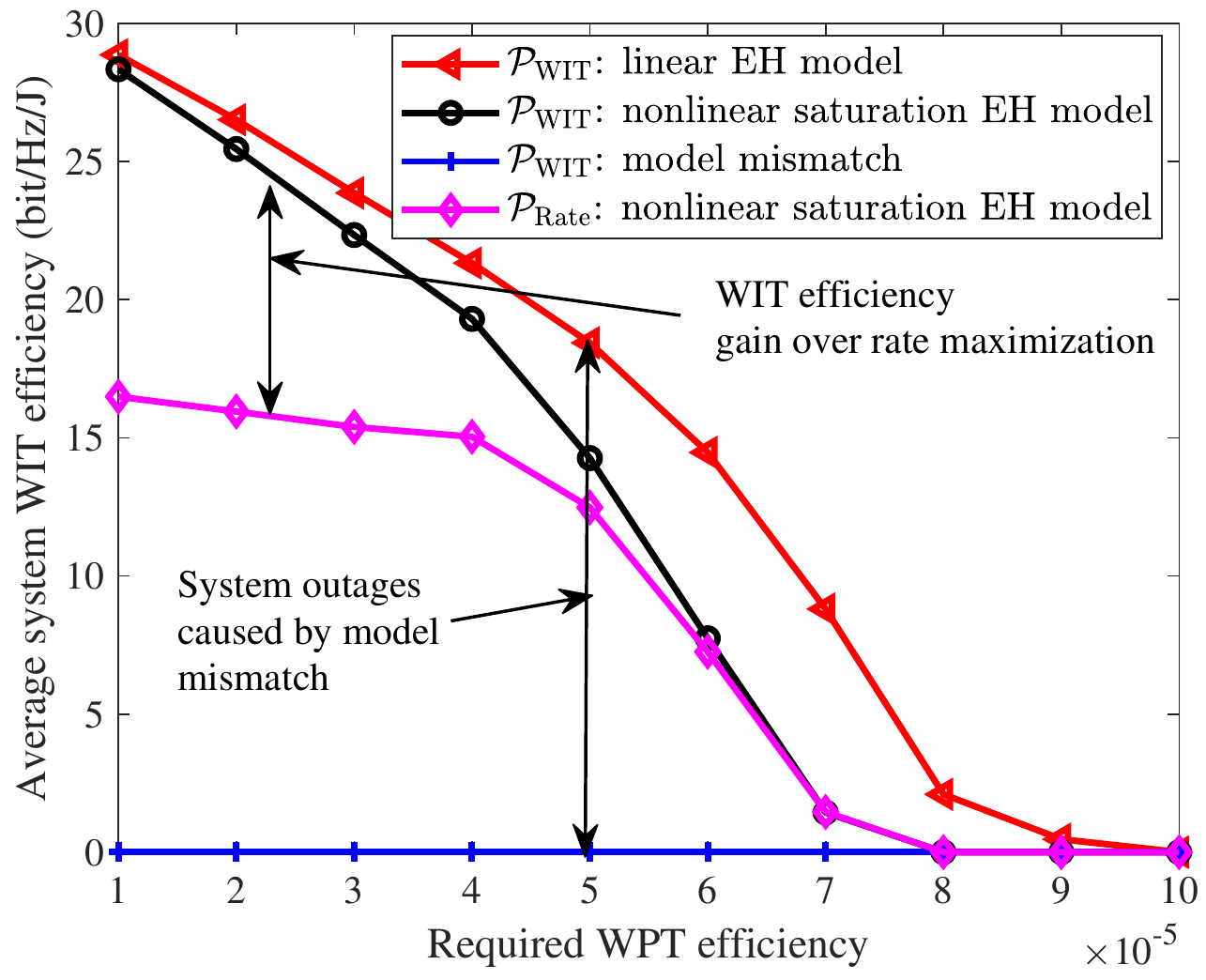}\vspace{-2mm}
	\caption{Average system WIT efficiency (bit/J) versus the minimum required WPT efficiency.} \label{fig:EEVSEH}\vspace{-5mm}
\end{figure}

{In this section, we provide a simulation result for an exemplary SWIPT system	 to reveal some insights for resource allocation design.
	The average system WIT efficiency $\mathcal{U}_{\mathrm{WIT}}^{\mathrm{Eff}}$ is evaluated for different minimum required WPT efficiencies $\mathrm{WPT}^{\mathrm{Eff}}_{\mathrm{req}}$ in Fig. \ref{fig:EEVSEH}.
	We set $\mathbf{v} = \mathbf{0}$ and adopt maximum ratio transmission (MRT) precoding for the initialization of the resource allocation algorithm described in Sections III-C1 and III-C2, i.e., $\mathbf{w}_k = \mathbf{h}_k$, $\forall k$.
	All simulation parameters are provided in Table \ref{tab:SimulationParametersII}.
	Here, the EH model parameters are obtained by curve fitting based on the measurement data in \cite{CN:EH_measurement_2}, cf. Fig. \ref{fig:comparsion_EH}.
	We solve the WIT maximization problem $\mathcal{P}_{\mathrm{WIT}}$ for both the linear EH model and the nonlinear saturation EH model in \eqref{RAProblemFormulationII} and \eqref{RAProblemFormulationIV}, respectively.
	As the required WPT efficiency increases, we observe from Fig. \ref{fig:EEVSEH} that the system WIT efficiencies for both EH models decrease since more resources have to be dedicated to WPT.
	We can further observe that the system WIT efficiency for the linear EH model is higher than that for the nonlinear saturation EH model since the latter model usually leads to an additional constraint for resource allocation  compared to the former one, as shown in \eqref{RAProblemFormulationII} and \eqref{RAProblemFormulationIV}.
	However, linear RF-to-DC conversion is not realizable due to the inevitable nonlinearity of EH circuits.
	To show the impact of the resulting resource allocation mismatch, we solve the WIT maximization problem $\mathcal{P}_{\mathrm{WIT}}$ based on the linear EH model and evaluate the resulting system performance for the nonlinear saturation EH model (``$\mathcal{P}_{\mathrm{WIT}}$: model mismatch'').
	Note that we set the system WIT efficiency to zero when any QoS or EH constraint cannot be satisfied to account for the corresponding penalty.
	Since at least one of the QoS and EH constraints in \eqref{RAProblemFormulationII} is active at the optimal point, the resource allocation policy designed based on the linear EH model inevitably leads to system outages when employed in a practical SWIPT system with nonlinear EH.
	Furthermore, for comparison, we solve the rate maximization problem $\mathcal{P}_{\mathrm{Rate}}$ based on the nonlinear saturation EH model in \eqref{RAProblemFormulationRate}.
	We observe that maximizing the system sum-rate results in a lower WIT efficiency in the low WPT efficiency regime while it achieves almost the optimal WIT efficiency in the high WPT efficiency regime.
	In fact, as indicated in \eqref{ObjFunctionII}, to achieve a high WPT efficiency, a low transmit power and a high PS ratio for EH are needed, and vice versa.
	Therefore, in the low WPT efficiency regime, maximizing the system sum-rate causes the SWIPT system to operate in the high power regime and thus it decreases the WIT efficiency due to the diminishing return in spectral efficiency when allocating more transmit power.
	In contrast, in the high WPT efficiency regime, maximizing the system sum-rate cannot exhaust the power budget at the Tx as 
	only a small transmit power can be afforded due to the stringent WPT efficiency constraint.
	In the low power regime, the spectral efficiency increases almost linearly w.r.t. the transmit power and thus transmitting all the allowable power is the most energy-efficient option.
	
\section{Resource Allocation Design for SWIPT Systems with Imperfect CSIT}
The resource allocation design problems for SWIPT systems formulated above are based on the assumption of perfect CSIT.
However, in the presence of CSIT errors, resource allocation design for SWIPT systems suffers from substantial performance degradation, which calls for robust designs.
In this section, we introduce three different approaches to robust resource allocation design, namely average robust design\cite{GaofengAverage}, outage-constrained robust design\cite{KhandakeOutage,BinbinOutageWorst}, and worst-case robust design\cite{XiangRobust,Xin_WorstCase,XinruiWorst}.
We first discuss how to incorporate channel uncertainty in the problem formulation to make the resource allocation robust.
Then, corresponding solution methodologies are discussed and  examples for outage-constrained and worst-case robust designs are provided for illustration.
\subsection{CSIT Error Model}
To capture the imperfection of CSIT, we model the channel from the Tx to user $k$ as 
\begin{equation}\label{ImperfectCSITModel}
\mathbf{h}_k = \hat{\mathbf{h}}_k + \Delta \mathbf{h}_k, \forall k,
\end{equation}
where $\hat{\mathbf{h}}_k$ denotes the estimate of $\mathbf{h}_k$ and $\Delta \mathbf{h}_k$ is the corresponding CSIT error.
In the literature, there are generally two approaches for modeling CSIT errors\cite{GaofengAverage,KhandakeOutage,BinbinOutageWorst,XiangRobust,Xin_WorstCase,XinruiWorst}:
\begin{itemize}
	\item Statistical CSIT Error Model\cite{GaofengAverage,KhandakeOutage,BinbinOutageWorst}: 
	\begin{equation}\label{StatisCSIeRROR}
		\Delta \mathbf{h}_k \in \mathcal{CN}\left(\mathbf{0},\sigma^2_{{h}_k}\mathbf{I}_{N_{\mathrm{t}}}\right),
	\end{equation}
	where $\sigma^2_{{h}_k}$ is the variance of the channel estimation error of user $k$ and $\mathcal{CN}(\boldsymbol{\mu},\boldsymbol{\Sigma})$ denotes the  CSCG distribution with mean $\boldsymbol{\mu}$ and variance $\boldsymbol{\Sigma}$.
	\item Bounded CSIT Error Model\cite{BinbinOutageWorst,XiangRobust,Xin_WorstCase,XinruiWorst}:
	\begin{equation}\label{bOUNDCSIeRROR}
		\left\|\Delta \mathbf{h}_k\right\| \le e_{h_k},
	\end{equation}
	where $e_{h_k}$ denotes the maximum value of the norm of the CSIT error of user $k$.
\end{itemize}

{The statistical CSIT error model assumes the CSIT errors, which may be caused by a noising estimation process, can be approximated by a Gaussian random variable \cite{GaofengAverage,KhandakeOutage,BinbinOutageWorst}.
	In contrast, for the bounded CSIT error model, the CSIT error is assumed to be within a known norm-bounded set, without any assumption on its distribution \cite{BinbinOutageWorst,XiangRobust,Xin_WorstCase,XinruiWorst}.
	One typical application scenario of the bounded CSIT error model is to capture  CSI errors resulting from quantization.
	For robust resource allocation design based on the bounded CSIT error model, certain constraints have to be satisfied for all possible errors within the uncertainty set, which ultimately leads to a worst-case design.
	On the other hand, with the statistical CSIT error model, the resource allocation can be made robust w.r.t. the average system performance or performance outages, which leads to average robust designs and outage-constrained robust designs, respectively.
	These different approaches will be discussed in detail in the following.}

\subsection{Average Robust Design}
This design methodology aims to maximize the average system performance with a constraint on the average QoS based on the statistical CSIT error model\cite{GaofengAverage}.
For instance, the average robust design problem corresponding to the problem in \eqref{RAProblemFormulationRate} can be formulated as follows:
\begin{center}
	\begin{tcolorbox}[title = {$\mathcal{P}^{\mathrm{Avg}}_{\mathrm{Rate}}$:  Average Robust Design for Rate Maximization}]
	\vspace{-5mm}
	\begin{align}\label{RAProblemFormulationRateAverage}
	\underset{\begin{subarray}{c}
		0\le\rho_{k}\le 1,\\
		\mathbf{W}_k, \mathbf{V} \in \mathbb{H}^{N_{\mathrm{t}} \times N_{\mathrm{t}}}
		\end{subarray}}{\mathrm{maximize}} &&&\hspace{-0mm} {\mathcal E}_{\Delta \mathbf{h}_k} \{R_{\mathrm{WS}}\left(\rho_{k},\mathbf{W}_k,\mathbf{V}\right)|\hat{\mathbf{h}}_k\}
	\\
	&&&\hspace{-29.8mm}\mathrm{s. t.}\;\;\mbox{C1},\,\mbox{C5},\, \mbox{C6}\notag\\
	&&&\hspace{-23mm}\overline{\mbox{C2}}:\,{\mathcal E}_{\Delta \mathbf{h}_k}\{R_{k}|\hat{\mathbf{h}}_k\}\ge R_{k,\mathrm{req}},\; \forall k,\notag\\
	&&&\hspace{-23mm}\overline{\mbox{C3}}:\,{\mathcal E}_{\Delta \mathbf{h}_k}\left\{P_{k,\mathrm{out} }|\hat{\mathbf{h}}_k\right\} \ge P_{k,\mathrm{req} },\;\forall k,\notag\\
	&&&\hspace{-23mm}\overline{\mbox{C4}}:\,{\mathcal E}_{\Delta \mathbf{h}_k}\left\{\sum_{k=1}^{K}P_{k,\mathrm{out} } |\hat{\mathbf{h}}_k\right\} \ge P_{\mathrm{D}}\left(\mathbf{W}_k, \mathbf{V}\right) \mathrm{WPT}^{\mathrm{Eff}}_{\mathrm{req}},\notag
	\end{align}
	\vspace{-4mm}\par\noindent
\end{tcolorbox}
\end{center}
\noindent where the expectation is taken over the CSIT error distribution in \eqref{StatisCSIeRROR}.
Note that this kind of problem formulation does not penalize instantaneous QoS outages, as long as the desired average performance is achieved.
However, it is usually challenging to obtain a closed-form expression for the average performance metric, such as the system sum-rate in \eqref{RAProblemFormulationRateAverage} or the WIT efficiency in \eqref{RAProblemFormulation}, which limits the applicability of the average robust design methodology.
One possible approach to resolve this issue is the Monte Carlo method\cite{Birge2011introduction}, where the average system performance metric in the objective function and the constraints are approximated by the corresponding empirical mean via random sampling of the channel based on the CSIT error distribution\cite{Birge2011introduction}.
For instance, the average system sum-rate in \eqref{RAProblemFormulationRateAverage} can be approximated by ${\mathcal E}_{\Delta \mathbf{h}_k} \{R_{\mathrm{WS}}\left(\rho_{k},\mathbf{W}_k,\mathbf{V}\right)|\hat{\mathbf{h}}_k\} \approx \frac{1}{I} \sum_{i=1}^{I} R_{\mathrm{WS}}\left(\rho_{k},\mathbf{W}_k,\mathbf{V}\right)|_{{\mathbf{h}}_k = \hat{\mathbf{h}}_k+{\Delta \mathbf{h}^i_k}}$, where ${\Delta \mathbf{h}^i_k}$ denotes the $i$-th sample of the CSIT error according to \eqref{StatisCSIeRROR} and $I$ is the total number of samples.
Yet, this approach increases the computational complexity for solving \eqref{RAProblemFormulationRateAverage}.

\subsection{Outage-constrained Robust Design}
This design approach optimizes a probabilistic or deterministic objective function while guaranteeing the probabilities with which the QoS and EH requirements are satisfied \cite{KhandakeOutage,BinbinOutageWorst}.
Outage-constrained robust designs are suitable for application scenarios which can tolerate system performance outages while require a limit to the frequency of their occurrence.
For instance, for problem \eqref{RAProblemFormulationPower}, we may minimize the system power consumption under probabilistic QoS and EH constraints:
\begin{center}
	\begin{tcolorbox}[title = {$\mathcal{P}^{\mathrm{Outage}}_{\mathrm{Power}}$:  Outage-constrained Robust Design for Power Minimization}]
		\vspace{-5mm}
		\begin{align}\label{RAProblemFormulationPowerOutage}
		\underset{\begin{subarray}{c}
			0\le\rho_{k}\le 1,\\
			\mathbf{W}_k, \mathbf{V} \in \mathbb{H}^{N_{\mathrm{t}} \times N_{\mathrm{t}}}
			\end{subarray}}{\mathrm{minimize}} &&&\hspace{-0mm} P_{\mathrm{D}}\left(\mathbf{W}_k, \mathbf{V}\right)
		\\
	&&&\hspace{-26.8mm}\mathrm{s. t.}\;\;\mbox{C1, C5, C6,}\,\notag\\
&\hspace{-2mm}&&\hspace{-20mm}\widetilde{\mbox{C2}}\hspace{-1mm}:\hspace{-0.5mm}{\mathrm{Pr}}_{\Delta \mathbf{h}_k}\hspace{-1mm}\left\{R_{k}\ge R_{k,\mathrm{req}}|\hat{\mathbf{h}}_k\right\}\ge 1\hspace{-1mm}-\hspace{-1mm}\kappa_{R_k},\quad \forall k,\notag\\
&\hspace{-2mm}&&\hspace{-20mm}\widetilde{\mbox{C3}}\hspace{-1mm}:\hspace{-0.5mm}{\mathrm{Pr}}_{\Delta \mathbf{h}_k}\hspace{-1mm}\left\{P_{k,\mathrm{out} }\ge P_{k,\mathrm{req} }|\hat{\mathbf{h}}_k\right\} \ge 1\hspace{-1mm}-\hspace{-1mm}\kappa_{P_k} ,\quad\forall k,\notag\\
&\hspace{-2mm}&&\hspace{-20mm}\widetilde{\mbox{C4}}\hspace{-1mm}:\hspace{-0.5mm}{\mathrm{Pr}}_{\Delta \mathbf{h}_k}\hspace{-1mm}\left\{\mathcal{U}_{\mathrm{WPT}}^{\mathrm{Eff}} \hspace{-1mm}\left(\rho_{k},\hspace{-1mm}\mathbf{W}_k,\hspace{-1mm}\mathbf{V}\right)\hspace{-1mm}\ge\hspace{-1mm} \mathrm{WPT}^{\mathrm{Eff}}_{\mathrm{req}}|\hat{\mathbf{h}}_k\right\}\hspace{-1mm}\ge\hspace{-1mm}1\hspace{-1mm}-\hspace{-1mm}\kappa_{\mathrm{EH}} .\notag
		\end{align}
	\vspace{-6mm}\par\noindent
	\end{tcolorbox}
\end{center}
	Constraints $\widetilde{\mbox{C2}}$-$\widetilde{\mbox{C4}}$ ensure that the probabilities of satisfying the QoS and EH requirements are larger than given corresponding thresholds.
	Here, $0 \le \kappa_{R_k}, \kappa_{P_k}, \kappa_{\mathrm{EH}} <1$ denote the maximum tolerable outage probabilities for given constraints on the target data rate, harvested power, and WPT efficiency, respectively.
	The probabilistic constraints in $\widetilde{\mbox{C2}}$-$\widetilde{\mbox{C4}}$ do not admit simple closed-form expressions, which is typical for this kind of problem formulation and imposes a challenge for robust resource allocation design\cite{Kun-Yu2014}.
	Fortunately, since the probabilistic constraints $\widetilde{\mbox{C2}}$-$\widetilde{\mbox{C4}}$ can be transformed into complex Gaussian quadratic forms, the problem can be reformulated to obtain another type of robust formulation, namely a safe convex approximation of the original problem \eqref{RAProblemFormulationPowerOutage}\cite{Kun-Yu2014}. 
	This convex approximation approach is based on the large deviation inequality for {complex Gaussian quadratic forms}, i.e., the Berstein-type inequality, which bounds the probability that a sum of random variables deviates from its mean \cite{Kun-Yu2014}. To begin, let us recall the following lemma.
	\begin{Lem}[Bernstein-type inequality \cite{Kun-Yu2014}]\label{bernstein}
		Consider the following random variable $f(\mathbf{t})=\mathbf{t}^{\mathrm H}\mathbf{Q}\mathbf{t}+2\Re\{\mathbf{t}^{\mathrm H}\mathbf{u}\}$, where $\mathbf{t}\sim\mathcal{CN}(\mathbf{0},\mathbf{I}_M)$, $\mathbf{Q}\in \mathbb{H}^{M\times M}$, and $\mathbf{u}\in \mathbb{C}^{M\times 1}$. For all $\kappa >0$, the following inequality holds:
		\begin{equation}\label{BernsteinPro}
			\textrm{Pr}\left\{f(\mathbf{t})\geq \Upsilon\left(\kappa\right)\right\}\geq 1 - e^{-\kappa},
		\end{equation}
		where $\Upsilon\left(\kappa\right) = \textrm{Tr}\left( \mathbf{Q}\right)-\sqrt{2\kappa}\sqrt{\|\mathbf{Q}\|_{\mathrm{F}}^2+2\|\mathbf{u}\|^2}-\kappa \lambda^+(\mathbf{Q})$, $\|\cdot\|_{\mathrm{F}}^2$ denotes the matrix Frobenius norm, $\lambda^+(\mathbf{Q}) = \max\{\lambda_{\mathrm{max}}(-\mathbf{Q}),0\}$, and $\lambda_{\mathrm{max}}(\cdot)$ denotes the maximum eigenvalue of a matrix.
		Since $\Upsilon\left(\kappa\right)$ is monotonically decreasing, the Bernstein-type inequality in \eqref{BernsteinPro} can be rewritten as follows
		\begin{equation}\label{BernsteinProII}
			\textrm{Pr}\left\{f(\mathbf{t}) + g \geq 0 \right\}\geq 1 - e^{-\Upsilon^{-1}\left(-g\right)},
		\end{equation}
		where $\Upsilon^{-1}\left(\cdot\right)$ denotes the inverse function of $\Upsilon\left(\cdot\right)$.
	\end{Lem}

	The Bernstein-type inequality provides a lower bound for $\textrm{Pr}\left\{f(\mathbf{t}) + g \geq 0 \right\}$ and $e^{-\Upsilon^{-1}\left(-g\right)} \le \tau$ implies $\textrm{Pr}\left\{f(\mathbf{t}) + g \geq 0 \right\} \ge 1-\tau$, where $0\le \tau <1$ is a constant.
	{Due to the intractability of the probabilities in $\widetilde{\mbox{C2}}$-$\widetilde{\mbox{C4}}$ in \eqref{RAProblemFormulationPowerOutage}, it is generally a formidable challenge to analyze the tightness of Bernstein-type inequalities.
	However, a comparative analysis has been provided in \cite{Kun-Yu2014}, which showed that the Bernstein-type inequality is tighter compared with two other bounds, i.e., the sphere bound and the decomposition-based large deviation inequality \cite{Kun-Yu2014}.}
	
	In the following, we discuss how to handle constraint $\widetilde{\mbox{C2}}$ in \eqref{RAProblemFormulationPowerOutage} based on the Bernstein-type inequality introduced in Lemma \ref{bernstein}.
	
	In particular, substituting \eqref{ImperfectCSITModel} for $R_k$ in $\widetilde{\mbox{C2}}$, we have
	\begin{align}\label{C2Bernstein}
		\hspace{-2mm}&{\mathrm{Pr}}_{\Delta \mathbf{h}_k}\{R_{k}\ge R_{k,\mathrm{req}}|\hat{\mathbf{h}}_k\} \ge 1-\kappa_{R_k} \notag\\
		\hspace{-2mm}\Leftrightarrow&{\mathrm{Pr}}_{\mathbf{t}_k}\{{\mathbf{t}^{\mathrm{H}}_k}\mathbf{Q}_k{\mathbf{t}_k} + 2 \Re\{{\mathbf{t}^{\mathrm{H}}_k}\mathbf{u}_k\} + g_k \ge 0\} \ge 1-\kappa_{R_k},
	\end{align}
	where ${\mathbf{t}_k} = \frac{{\Delta \mathbf{h}_k}}{\sigma_{{h}_k}}\sim\mathcal{CN}(\mathbf{0},\mathbf{I}_{N_{\mathrm{t}}})$, $\mathbf{Q}_k = \mathbf{W}_k - \left(2^{R_{k,\mathrm{req}}}-1\right)\left(\sum_{k'\neq k}^{K} \mathbf{W}_{k'} + \mathbf{V}\right)$, $\mathbf{u}_k = \frac{\mathbf{Q}^{\mathrm{H}}_k\hat{\mathbf{h}}_k}{{\sigma_{{h}_k}}}$, and $g_k =\frac{{\hat{\mathbf{h}}^{\mathrm{H}}_k}\mathbf{Q}_k{ \hat{\mathbf{h}}_k} -\left(2^{R_{k,\mathrm{req}}}-1\right)\left(\sigma^2_{\mathrm{A}} + \frac{\sigma^2_{\mathrm{P}}}{\rho_{k}}\right)}{{\sigma^2_{{h}_k}}}$.
	According to \eqref{BernsteinProII}, \eqref{C2Bernstein} is always satisfied if the following inequality holds
	\begin{align}\label{C2Bernstein_II}
		&e^{-\Upsilon^{-1}\left(-g_k\right)} \le \kappa_{R_k} \notag\\
		\Leftrightarrow& \textrm{Tr}\left( \mathbf{Q}_k\right)-\sqrt{2\ln\left(\frac{1}{\kappa_{R_k}}\right)
		}\sqrt{\|\mathbf{Q}_k\|_{\mathrm{F}}^2+2\|\mathbf{u}_k\|^2}\notag\\
	&+\ln{\left(\kappa_{R_k}\right)} \lambda^+(\mathbf{Q}_k) + g_k \ge 0.
	\end{align}
	By introducing suitable slack variables, the constraint in \eqref{C2Bernstein_II} can be transformed into the following linear matrix inequality (LMI) and second-order cone (SOC) constraints:
	\begin{align}\label{C2BernsteinIII}
		&\textrm{Tr}\left( \mathbf{Q}_k\right) - \sqrt{2\ln\left(\frac{1}{\kappa_{R_k}}\right)}d_k + \ln\left(\kappa_{R_k}\right) z_k + g_k \ge 0,\notag\\
		&\sqrt{\|\mathbf{Q}_k\|_{\mathrm{F}}^2\hspace{-1mm}+\hspace{-1mm}2\|\mathbf{u}_k\|^2} \le d_k,\\
		&z_k \mathbf{I}_{N_{\mathrm{t}}} + \mathbf{Q}_k \succeq \mathbf{0},\;\text{and}\notag\\
		&z_k \ge 0,\notag
	\end{align}
		which can be handled by convex optimization approaches.
		In other words, the constraints in \eqref{C2BernsteinIII} impose a convex restriction for the probabilistic constraint $\widetilde{\mbox{C2}}$ in \eqref{RAProblemFormulationPowerOutage}.
	Applying a similar transformation procedure to constraints $\widetilde{\mbox{C3}}$ and $\widetilde{\mbox{C4}}$ in \eqref{RAProblemFormulationPowerOutage}, the resulting problem can be directly handled by the methodologies introduced in Section III-C.
	
\subsection{Worst-case Robust Design}
This approach is based on the bounded CSIT error model in \eqref{bOUNDCSIeRROR} and optimizes the worst-case system performance under worst-case QoS constraints\cite{Xin_WorstCase,XinruiWorst}.
	For example, the problem in \eqref{RAProblemFormulationEH} can be reformulated as follows:
\begin{center}
	\begin{tcolorbox}[title = {$\mathcal{P}^{\mathrm{Worst}}_{\mathrm{EH}}$:  Worst-case Robust Design for Harvested Power Maximization}]
		\vspace{-5mm}
		\begin{align}\label{RAProblemFormulationEHImperfectCSIT}
		\underset{\begin{subarray}{c}
			0\le\rho_{k}\le 1,\\
			\mathbf{W}_k, \mathbf{V} \in \mathbb{H}^{N_{\mathrm{t}} \times N_{\mathrm{t}}}
			\end{subarray}}{\mathrm{maximize}} &&&\hspace{-0mm}\underset{\Delta \mathbf{h}_k, \forall k}{\min} \;\; C\left(\rho_{k},\mathbf{W}_k, \mathbf{V}\right)
		\\
	&&&\hspace{-26.8mm}\mathrm{s.t.}\;\;\mbox{C1, C5, C6,}\,\notag\\
&&&\hspace{-20mm}\widehat{\mbox{C2}}:\,\underset{\Delta \mathbf{h}_k}{\min} \; R_{k}\ge R_{k,\mathrm{req}},\quad \forall k,\notag\\
&&&\hspace{-20mm}\widehat{\mbox{C3}}:\,\underset{\Delta \mathbf{h}_k}{\min} \;P_{k,\mathrm{out} } \ge P_{k,\mathrm{req} },\quad\forall k,\notag\\
&&&\hspace{-20mm}\widehat{\mbox{C4}}:\,\underset{\Delta \mathbf{h}_k, \forall k}{\min} \;\sum_{k=1}^{K}P_{k,\mathrm{out} } \ge P_{\mathrm{D}}\left(\mathbf{W}_k, \mathbf{V}\right) \mathrm{WPT}^{\mathrm{Eff}}_{\mathrm{req}},\notag
		\end{align}
		\vspace{-4mm}\par\noindent
	\end{tcolorbox}
\end{center}
	\noindent where $\Delta \mathbf{h}_k$ follows \eqref{bOUNDCSIeRROR} and the $\min$ in the objective function and constraints $\widehat{\mbox{C2}}$-$\widehat{\mbox{C4}}$ is over the CSIT error which leads to a guaranteed worst-case performance.
	Due to the imperfect CSIT, there are infinitely many possibilities for the objective function and constraints $\widehat{\mbox{C2}}$-$\widehat{\mbox{C4}}$ in \eqref{RAProblemFormulationEHImperfectCSIT}.
	This obstacle can often be circumvented by exploiting a complex Gaussian quadratic form inequality\cite{book:convex}.
	To this end, the implication in the following lemma is useful to obtain tractable restrictions for $\widehat{\mbox{C2}}$-$\widehat{\mbox{C4}}$:
	\begin{Lem}[S-Procedure \cite{book:convex}]\label{Lemma1}
		Let a function $f_m(\mathbf{t}),m\in\{1,2\},\mathbf{t}\in \mathbb{C}^{N\times 1},$ be defined as 
		\begin{equation}
		\label{eqn:S-proc-function}f_m(\mathbf{t})=\mathbf{t}^{\mathrm{H}}\mathbf{A}_m\mathbf{t}+2 \hspace*{0mm}\Re\hspace*{0mm} \{\mathbf{b}_m^{\mathrm{H}}\mathbf{t}\}+c_m,
		\end{equation}
		where $\mathbf{A}_m\in\mathbb{H}^N$, $\mathbf{b}_m\in\mathbb{C}^{N\times 1}$, and $c_m\in\mathbb{R}^{1\times 1}$. Then, the implication $f_1(\mathbf{t})\le 0\Rightarrow f_2(\mathbf{t})\le 0$  holds if and only if there exists a variable $\epsilon\ge 0$ such that
		\begin{equation}\label{eqn:S-proc-LMI}\epsilon
		\begin{bmatrix}
		\mathbf{A}_1 & \mathbf{b}_1          \\
		\mathbf{b}_1^{\mathrm{H}} & c_1           \\
		\end{bmatrix} -\begin{bmatrix}
		\mathbf{A}_2 & \mathbf{b}_2          \\
		\mathbf{b}_2^{\mathrm{H}} & c_2           \\
		\end{bmatrix}          \succeq \mathbf{0} ,
		\end{equation}
		provided that there exists a point $\mathbf{\hat{t}}$ such that $f_k(\mathbf{\hat{t}})<0$.
	\end{Lem}

	Note that the worst-case robust design approach is preferable for mission-critical applications where the QoS and EH constraints cannot be violated even in the presence of CSIT uncertainty.
	In the following, we show exemplarily how to handle constraint $\widehat{\mbox{C2}}$ in \eqref{RAProblemFormulationEHImperfectCSIT} employing the S-Procedure in Lemma \ref{Lemma1}.
	In particular, we define the optimization variable:
	\begin{equation}
	\widehat{\gamma}_k = \underset{\Delta \mathbf{h}_k}{\min}\; \frac{\rho_{k}\Tr\left(\mathbf{W}_{k} \mathbf{H}_k\right)}{\rho_{k}\left(\hspace{-1mm}\Tr\hspace{-0.5mm}\left(\sum\limits_{k'\neq k}^{K} \hspace{-1mm}\mathbf{W}_{k'} \mathbf{H}_k \hspace{-1mm}+ \hspace{-1mm} \mathbf{V}\mathbf{H}_k\hspace{-1mm}\right) \hspace{-1mm}+\hspace{-1mm} \sigma_{\mathrm{A}}^2\hspace{-1mm}\right) \hspace{-1mm}+\hspace{-1mm} \sigma_{\mathrm{P}}^2},\forall k,
	\end{equation}
	such that constraint $\widehat{\mbox{C2}}$ in \eqref{RAProblemFormulationEHImperfectCSIT} can be rewritten as:
	\begin{align}
		\widehat{\mbox{C2a}}:&\,\widehat{\gamma}_k\ge 2^{R_{k,\mathrm{req}}}-1,\; \forall k,\notag\\
		\widehat{\mbox{C2b}}:&\,\underset{\Delta \mathbf{h}_k}{\min} f_2\left(\Delta \mathbf{h}_k\right)\le 0,
	\end{align} 
	where $f_2\left(\Delta \mathbf{h}_k\right) = {\Delta \mathbf{h}^{\mathrm{H}}_k} {\mathbf{\Omega }}_k {\Delta \mathbf{h}_k} + 2\Re\{{ \hat{\bf{ h}}_k^{\rm{H}}{{\bf{\Omega }}_k}}{\Delta \mathbf{h}_k}\} + y_k$, ${{\bf{\Omega }}_k} = \left(\sum_{k'\neq k}^{K} \mathbf{W}_{k'} + \mathbf{V}\right) - \frac{\mathbf{W}_{k}}{\widehat{\gamma}_k}$, and $g_k = { \hat{\bf{ h}}_k^{\rm{H}}{{\bf{\Omega }}_k}{{\hat{\bf{ h}}}_k} + \sigma _{\rm{A}}^2 + \frac{\sigma _{\rm{P}}^2}{\rho_{k}}}$.
	According to Lemma \ref{Lemma1}, we define $f_1\left(\Delta \mathbf{h}_k\right) = \left\|\Delta \mathbf{h}_k\right\|^2 = \Delta \mathbf{h}^{\mathrm{H}}_k \Delta \mathbf{h}_k\le e_{h_k}$ and $f_2\left(\Delta \mathbf{h}_k\right)\le 0$ holds if and only if
	\begin{equation}\label{C2LMI}
	\overline{\widehat{\mbox{C2b}}}: \;\;\boldsymbol{\Xi} = \left[ \hspace{-0.5mm}{\begin{array}{*{20}{c}}
		{{\epsilon _{k}}{\bf{I}}\hspace{-0.5mm} -\hspace{-0.5mm} {{\bf{\Omega }}_k}}&{ - {\bf{\Omega }}_k^{\rm{H}}{{\hat{\bf{ h}}}_k}}\\
		{ - {{\hat{\bf{ h}}}^{\rm{H}}_k}{\bf{\Omega }}_k}&{ - {e_{h_k}}{\epsilon _{k}} \hspace{-0.5mm}-\hspace{-0.5mm} g_k}
		\end{array}} \hspace{-0.5mm}\right]\hspace{-0.5mm} \succeq \hspace{-0.5mm}{\bf{0}},
	\end{equation}
	where ${\epsilon _{k}}\ge 0$.
	We note that to maximize the total power harvested by all users, constraint $\overline{\widehat{\mbox{C2b}}}$ will be satisfied with equality at the optimal point, i.e., $\widehat{\gamma}_k =  2^{R_{k,\mathrm{req}}}-1$.
	Introducing an auxiliary variable $\beta_{\overline{\widehat{\mathrm{C2b}}}} \ge \frac{1}{\rho_{k}}$, the inequality in \eqref{C2LMI} can be transformed into an LMI.
	Applying the transformation procedure above to constraints $\widehat{\mbox{C3}}$ and $\widehat{\mbox{C4}}$ in \eqref{RAProblemFormulationEHImperfectCSIT}, the resulting problem can be handled by the quadratic transformation and SDR approaches discussed in Section III-C.

\subsection{{Complexity Analysis and Implementation Details}}
{In this section, we analyze the computational complexity of the introduced solution methodology and discuss some practical aspects arising in its implementation.
	
	The proposed algorithms are based on the quadratic transformation and SDR.
	Since the auxiliary variables are updated based on closed-form expressions, e.g., \eqref{AuXVariableUpdate}, the computational complexity of the proposed optimization methods is dominated by the SDP required for solving the transformed problems in subtractive form for given auxiliary variables.
	It is well-known that SDP has a polynomial worst-case computational complexity \cite{VandenbergheSDP}.
	In particular, when the interior-point method is employed, the worst-case complexity of SDP is 
	$\mathcal{O} \left(\sqrt{N_{\mathrm{Var}}} \left(N_{\mathrm{Cons}}N^2_{\mathrm{Var}}\right) \log \left(\frac{1}{\epsilon_{\mathrm{SDP}}}\right)\right)$ \cite{HaotianIPM}, where the big-O notation $\mathcal{O}\left(\cdot\right)$ describes the order of computational complexity, $N_{\mathrm{Var}}$ denotes the number of optimization variables, $N_{\mathrm{Cons}}$ denotes the number of constraints, and ${\epsilon_{\mathrm{SDP}}} >0 $ specifies the accuracy of SDP.
	Considering this, the worst-case computational complexity for solving \eqref{RAProblemFormulationIII} is $\mathcal{O} \left(\hspace{-1mm}\sqrt{3K\hspace{-1mm}+\hspace{-1mm}(1\hspace{-1mm} +\hspace{-1mm} K) N^2_{\mathrm{t}}} \hspace{-1mm}\left(\hspace{-0.5mm}(4K\hspace{-1mm}+\hspace{-1mm}2)(3K\hspace{-1mm}+\hspace{-1mm}(1 \hspace{-1mm}+\hspace{-1mm}K) N^2_{\mathrm{t}})^2\right) \hspace{-0.5mm}\log\hspace{-0.5mm} \left(\hspace{-0.5mm}\frac{1}{\epsilon_{\mathrm{SDP}}}\hspace{-0.5mm}\right)\hspace{-0.5mm}\right)$.
	As can be observed, the computational complexity of the proposed algorithm scales with $N^5_{\mathrm{t}}$ and $K^{3.5}$.
	%
	
	To implement the proposed resource allocation design, all users first have to inform the Tx of the SWIPT system about their minimum required data rates and their minimum required EH powers via dedicated feedback links.
	Then, the Tx has to acquire the CSI through uplink training by exploiting the channel reciprocity in time division duplexing (TDD) systems or through downlink training and feedback in frequency division duplexing (FDD) systems\cite{HongxiangCSI}.
	Compared with conventional wireless Rxs, EH Rxs are usually more severely power-limited and thus the required uplink training or feedback leads to higher channel estimation errors.
	Therefore, it is imperative to take the CSIT errors into account for the design of robust resource allocation algorithms, as explained earlier in this section.
	The resource allocation design is computed at the Tx in a centralized manner.
	The obtained beamforming vectors are used to generate the transmit signal, while the obtained PS ratios are informed to all users via dedicated control links for splitting the received RF signal for EH and ID.}

\subsection{Simulation Result for an Exemplary SWIPT System}
	\begin{figure}[t]
		\centering
		\includegraphics[width=3.0 in]{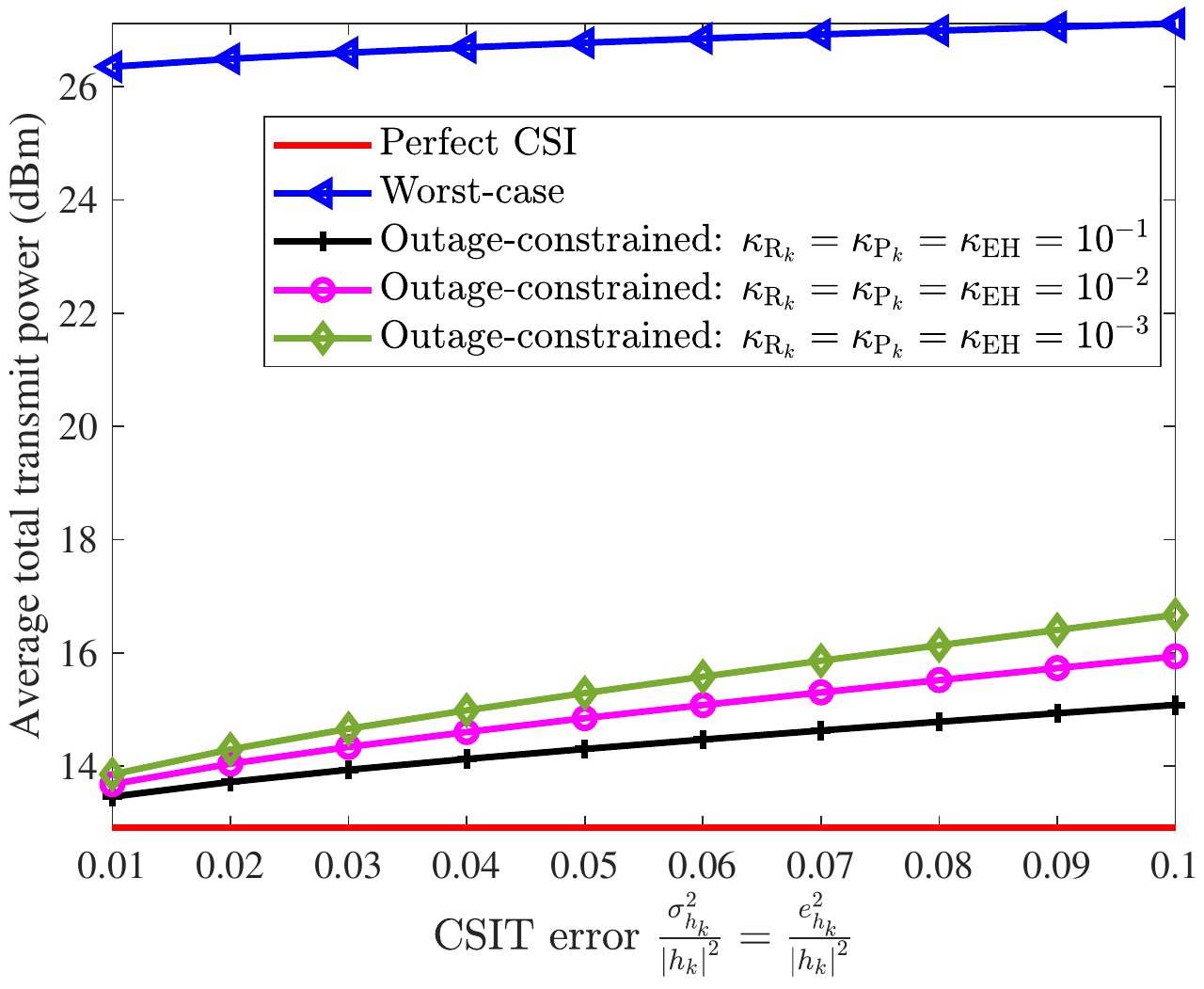}\vspace{-2mm}
		\caption{Average total transmit power versus the CSIT error with worst-case and outage-constrained robust resource allocation designs.}\vspace{-5mm} \label{fig:TxPowerVSCSIError}
	\end{figure}

	To illustrate the effectiveness of the proposed robust designs, we use the worst-case and outage-constrained robust design methods to minimize the total transmit power, i.e., the problem in \eqref{RAProblemFormulationPowerOutage}, considering a nonlinear saturation EH model.
	All simulation parameters are taken from Table \ref{tab:SimulationParametersII}.
	Fig. \ref{fig:TxPowerVSCSIError} shows the total transmit power versus the CSIT error uncertainty $\frac{\sigma^2_{{h}_k}}{\left|{h}_k\right|^2} =\frac{e^2_{h_k}}{\left|{h}_k\right|^2}$.
	We observe that both robust designs consume more power than the system with perfect CSIT as they have to account for the channel uncertainty for resource allocation.
	As expected, for both robust designs, more power is needed as the CSIT becomes more uncertain.
	Moreover, the worst-case robust design consumes significantly more power than the outage-constrained robust design.
	In fact, the worst-case design does not tolerate any violation of the QoS and EH constraints and thus it requires significantly more power.
	Compared with the outage-constrained robust design, the nonlinearly coupled variables of the bounds needed for the worst-case robust design, such as \eqref{C2LMI}, impose further restrictions, which ultimately leads to a higher transmit power.
	For the outage-constrained robust design, more power is needed when the outage probabilities, i.e., $\kappa_{\mathrm{R}_k}$, $\kappa_{\mathrm{P}_k}$, and $\kappa_{\mathrm{EH}}$, become smaller, as the QoS and EH outage constraints become more stringent.
	
\section{Future Research Directions}
In this section, we discuss some potential research directions for resource allocation design in SWIPT systems, including the joint waveform and resource allocation design, intelligent reflecting surface (IRS)-assisted SWIPT systems, UAV-enabled SWIPT systems, SWIPT-enabled MEC systems, and the role of ML.

\subsection{Joint Waveform and Resource Allocation Design}
While resource allocation design for SWIPT systems has been extensively studied in the literature\cite{LiangTS,LiangPS,ZhouOPS,LengFDSWIPT,HuFDswipt,JR:rui_zhang_secrecy,NgSecureSWIPT,IoannisSWIPT,HimalJSAC}, joint waveform and resource allocation design for the nonlinear circuit-based EH model is still an open problem.
As mentioned in Section III-C, the optimal waveform for maximizing the total harvested power is usually deterministic \cite{Bruno2019Review} or follows a discrete finite distribution \cite{JR:Rania_nonlinear_circuit_based_TCOM}, which is different from the widely-adopted CSCG signal for conventional WIT.
In fact, it is expected that exploiting the nonlinearity can enlarge the rate-energy tradeoff region\cite{Bruno2019Review}.
Firstly, how to design the waveform for optimization of a generalized system utility function is still unknown.
Secondly, for resource allocation design, closed-form expressions for the  achievable rate and the harvested DC power are needed, but are usually not available or cumbersome when the signals depart from the conventional CSCG distribution.
Thirdly, for waveform design,  essentially an optimal input distribution has to be found while conventional resource allocation designs are parametric in nature.
Therefore, the coupling of resource allocation and waveform design calls for a general parametric input distribution, which can lead to a tractable performance characterization for SWIPT systems.
This would allow the joint design of the waveform and the resource allocation for improving the system performance.
Moreover, waveform optimization for multi-carrier SWIPT systems is even more challenging.
In particular, WPT has to be carried out based on the time-domain waveform in passband, while for WIT, the adopted signal on each subcarrier is relevant.
{In addition, frequency-selective fading provides frequency diversity, which also needs to be taken into account for joint resource allocation and waveform optimization.
	For instance, power allocation in the frequency domain affects both the system sum-rate and the WIT efficiency \cite{book:david_wirelss_com}.
	On the other hand, power allocation across subcarriers might be an effective approach to alter the distribution of the time-domain signal waveform which can be exploited to improve the WPT efficiency.}
	{Moreover, if multiple antennas are available at the EH Rx, the energy combining can be performed either at the RF or the DC level\cite{ShanpuMIMOWPT}.
	Besides, the Tx/Rx beamforming, the waveform, and the resource allocation should be jointly optimized to strike a balance between WPT and WIT, particularly if nonlinear EH models are employed\cite{ShanpuWaveformMIMO}.
	However, this area of research is still largely unexplored.}

\subsection{IRS-assisted SWIPT}
Recently, IRSs have been proposed to enable the
establishment of a programmable radio environment and have attracted extensive attention from the wireless research community\cite{di2019smart,yu2021smart}.
By exploiting a large number of low-cost passive elements that can reflect the signals with adjustable amplitudes and phase shifts\cite{di2019smart,yu2021smart}, IRSs have the potential to establish favorable communication links for both ID and EH Rxs that would otherwise be blocked or in deep fades\cite{QingqingWuSWIPT}.
Owing to their capability to reconfigure the channel and their limited power consumption, IRSs are a promising candidate for enabling intelligent and energy-efficient SWIPT \cite{hu2021robust}. 
However, to fully unleash the potentials of IRSs in improving SWIPT system performance, the reflection coefficients of the IRSs have to be jointly designed with the WIT and WPT beamforming at the transmitter, which introduces new challenges for resource allocation design.
Besides, the passivity of the IRS elements introduces additional highly non-convex constraints, e.g., unit modulus constraints or binary constraints, which makes the resource allocation design in IRS-assisted SWIPT systems more challenging\cite{9133130}.
Moreover, if the nonlinear circuit-based EH model is adopted, the amalgamation of SWIPT and IRS calls for joint waveform, resource allocation, and passive beamforming design, which is still an open problem.

\subsection{UAV-enabled SWIPT}
{One critical issue of SWIPT systems is the limited WPT efficiency due to the severe wireless propagation loss, especially when the EH Rxs are far from the Tx.
	Besides, since the EH Rxs close to the Tx can harvest significantly higher amounts of power than remote EH Rxs, fairness issues arise for resource allocation in SWIPT systems.
	With the recent advancement of UAV manufacturing technologies and the resulting substantial cost reductions, low-altitude UAVs become a potential mobile platform for providing highly efficient and fair SWIPT for massive EH Rxs distributed over a large area, e.g., UAV-enabled IoT systems \cite{9136595,9174765}.
	In particular, UAV-enabled SWIPT provides an effective approach to combat channel fading as it can establish a LoS link to the EH Rxs with high probability \cite{ZhiqiangUAV}.
	Moreover, benefiting from its high maneuverability, the UAV's trajectory
	can be designed to adapt to the actual signal propagation environment and EH Rx distribution, which
	provides additional design DoFs for resource allocation in SWIPT systems.
	Nevertheless, several critical issues have to be taken into account for resource allocation.
	Firstly, the UAV's trajectory affects the performances of both WIT and WPT significantly and thus has to be jointly designed with the optimal resource allocation.
	For instance, when considering the nonlinear saturation EH model, the UAV does not need to fly closer to EH Rxs whose received RF power is already in the saturation regime.
	Moreover, when serving a large number of EH Rxs over a large area, it is impractical for the UAV to cruise over the devices one-by-one due to its limited flight duration \cite{XieUAVSWIPT}.
	As a remedy, device grouping can be performed such that the UAV's trajectory can be designed based on the distribution of the formed EH Rx clusters.
	Deploying multiple UAVs for SWIPT is also a viable option to address this issue.
	However, how to efficiently coordinate WIT, WPT, and trajectory design is a new problem to be tackled.}

\subsection{Mobile Edge Computing and Federated Learning}
The IoE era represents the next wave of the wireless revolution \cite{7123563}, and will connect billions of devices, objects, and machines\cite{ShuaifeiIoE}. 
This trend is driving a paradigm shift in wireless communications, from ``connecting humans'' to ``connecting things''. 
To realize this novel paradigm, a technological breakthrough towards a new generation of \emph{low-energy} mobile devices with \emph{improved processing capability} is required. 
To prolong the battery life and to minimize the processing delay, partial computation tasks that are supposed to be processed locally at mobile devices are offloaded to nearby servers located at the edge of the wireless networks \cite{8642372}. 
In such MEC systems, SWIPT can work concurrently to further increase the lifetime of the batteries of the mobile devices and relieve the burden of power-limited wireless systems.
The integration of MEC and SWIPT in modern wireless systems has received much attention from both industry and academia \cite{9140412}. 
In particular, proper resource allocation algorithms are needed to jointly design the offloading decision, beamforming, and covariance matrix of the transmitted energy signals for transmit power minimization, EH energy maximization, rate maximization, or task completion time minimization.
The resulting optimization problems are typically non-convex mixed integer nonlinear programs (MINLPs) and efficient solutions have to be developed.

Originating from a similar idea as MEC, federated learning (FL) has been proposed where the computation abilities of the mobile users are typically specialized for training a deep learning model. 
Without directly offloading the raw data of the clients to the edge server, FL is a promising approach for training a deep learning model while preserving data privacy. 
In particular, during the training process of FL, the mobile users need to  train a learning model locally and transmit the model information to the edge server to build consensus \cite{pmlr-v54-mcmahan17a}.
However, periodically training a deep learning model and transmitting the bulky model is computation- and communication-intensive.
Therefore, equipping mobile users with the ability of EH for prolonging the battery usage is a promising approach to enable FL in future intelligent wireless networks \cite{9374105}.
In this regard, an important task in SWIPT-based FL is to optimize the portion of harvested energy allocated to communication with the edge server and local computation, respectively.
In addition, the integration of SWIPT and conventional energy-saving techniques in FL, e.g., model compression, adaptive transmission, and hierarchical FL \cite{qiao2021communication}, needs further investigation.

\begin{figure*}
	\begin{align}
	\mathcal{L} &= \left( {\beta _{{\rm{Obj}}}^2\xi  + {\alpha _{{\rm{C1}}}} + {\alpha _{{\rm{C4}}}}{\rm{EH}}_{{\rm{req}}}^{{\rm{Eff}}}} \right){\rm{Tr}}\left( {\sum\limits_{k = 1}^K {{{\bf{W}}_k}}  + {\bf{V}}} \right) - \sum\limits_{k = 1}^K {{\rm{Tr}}\left( {{\boldsymbol{\Lambda} _{k,{\rm{C6}}}}{{\bf{W}}_k}} \right)}  - {\rm{Tr}}\left( {{\boldsymbol{\Lambda} _{{\bf{v}},{\rm{C6}}}}{\bf{V}}} \right)\label{LagrangianFunction}\\
	& - \sum\limits_{k = 1}^K {{\alpha _{k{\rm{,C7}}}}\hspace{-1mm}\left[ {2{\beta _{k,{\rm{C7}}}}\sqrt {\Tr\left( {{{\bf{W}}_k}{{\bf{H}}_k}} \right)}  \hspace{-1mm}-\hspace{-1mm} \beta _{k,{\rm{C7}}}^2\hspace{-1mm}\Tr\hspace{-1mm}\left(\hspace{-0.5mm} {\sum\limits_{k' \ne k}^K {} \hspace{-1mm}{{\bf{W}}_{k'}}{{\bf{H}}_k}\hspace{-1mm} +\hspace{-1mm} {\bf{V}}{{\bf{H}}_k}} \hspace{-1mm}\right)}\hspace{-1mm} \right]} \hspace{-1mm} - \sum\limits_{k = 1}^K {{\alpha _{k{\rm{,C8}}}}\hspace{-1mm}\left[ {2{\beta _{k,{\rm{C8}}}}\sqrt {\eta \Tr \hspace{-1mm}\left( {\sum\limits_{k' = 1}^K {{{\bf{W}}_{k'}}} {{\bf{H}}_k} \hspace{-1mm}+\hspace{-1mm} {\bf{V}}{{\bf{H}}_k}} \hspace{-1mm}\right)} } \right]} \hspace{-1mm}+ \hspace{-1mm}\vartheta  \notag
	\end{align}
	\vspace{-6mm}\par\noindent
\hrulefill
\vspace{-5mm}
\end{figure*}

\subsection{Machine Learning-based Design}
Effective resource allocation design is critical for realizing efficient utilization of the  available resources within a complex environment. 
Conventional design methodologies based on mathematical optimization may not be directly applicable to large-scale SWIPT networks as the complexity of the optimal designs often scales exponentially with the network size. 
As a remedy, ML is a promising tool as it can be used to optimize large-scale systems without relying on analytical models\cite{qiao2021communication,7792374}.
{Recently, several papers have exploited ML for solving communication-centric resource allocation design problems in conventional wireless networks \cite{MengyuanLearning,MatthiesenLearning,LiangLearning}.} 
In fact, deep learning networks can be leveraged to characterize the complicated mapping relations between given wireless channel conditions and the corresponding resource allocation for SWIPT systems\cite{8626195}.
%
The samples/labels for learning are typically generated by conventional optimal iterative algorithms, e.g., monotonic optimization \cite{7812683}, which may be computationally challenging.
However, as a large number of  samples are typically needed for training an accurate model, how to reduce the sample size while maintaining inference accuracy is a crucial task.
Furthermore, ML can be applied to characterize complicated nonlinear EH models \cite{6485022}. 
Specifically, while different EH models have been proposed to characterize energy harvesters, there still exists a tradeoff between accuracy and tractability in the context of optimization. 
If a proper ML network model can be found to represent energy harvesters, e.g., neural network-based long short-term memory (LSTM)\cite{Hochreiter1997long}, efficient optimization of SWIPT systems will become feasible.
{Deep reinforcement learning (DRL), which is a combination of reinforcement learning and deep neural networks, is a useful tool for resource allocation design for SWIPT systems which have casual information and exhibit state transition features, e.g., SWIPT systems with limited energy buffers at the Tx or EH Rxs\cite{AndreaLearning,ManLearning}.
In particular, when applying SWIPT to dynamic systems, such as UAV networks, the resource allocation design problem can be modeled as an MDP and DRL can be used for maximizing the long-term system reward.
However, in the presence of a large number of EH nodes, DRL requires extensive signaling to feedback the global system state to all EH nodes.
In this case, a multi-agent reinforcement learning (MARL) approach can facilitate the learning of the online resource allocation policy in a distributed fashion\cite{SharmaDRL}.
These aspects are interesting directions for future research.
}

\section{Conclusions}
This paper provided a tutorial overview on resource allocation design for SWIPT systems.
First, the recent literature on SWIPT systems was reviewed with an emphasis on Rx architectures, EH models, and resource allocation design.
Three widely-adopted EH models, namely the linear EH model, the nonlinear saturation EH model, and the nonlinear circuit-based EH model, were characterized and their significant impact on resource allocation design was highlighted.
For a typical SWIPT downlink system, we established a generalized resource allocation design framework which subsumes existing designs as special cases.
Then, focusing on the WIT efficiency maximization problem for both the linear EH model and the nonlinear saturation EH model, we developed an efficient suboptimal solution employing quadratic transformation, which can handle the severe variable coupling typical for SWIPT and can be readily used for developing a concrete resource allocation algorithm.
Besides, the joint design of the input distribution and the resource allocation for the  nonlinear circuit-based EH model was discussed.
To reduce the severe performance degradation suffered by SWIPT systems in the case of channel uncertainty, three robust resource allocation design methodologies were considered.
Simulation results for exemplary SWIPT systems demonstrated the effectiveness of the proposed solution methodologies and provided several insights for SWIPT system design: 1) resource allocation design for SWIPT systems based on the linear EH model leads to inevitable system outages in practice due to the EH model mismatch; 2) a resource allocation maximizing the system sum-rate can achieve the same performance as maximizing the WIT efficiency when the required system WPT efficiency is sufficiently high; 3) robust resource allocation design in the presence of CSIT errors requires significantly higher transmit powers than the case with perfect CSIT, in particular for worst-case robust designs.
Last but not the least, future research directions for resource allocation in SWIPT systems were highlighted, including joint waveform and resource allocation design, IRS-assisted SWIPT systems, UAV-enabled SWIPT systems, applications in emerging MEC systems, and the role of ML.

\section*{Appendix - Proof of Theorem \ref{Theorem1}}
{For given $\left(\beta_{\mathrm{Obj}},\beta_{k,\mathrm{C7}},  \beta_{k,\mathrm{C8}}\right)$, the SDP relaxed problem in \eqref{RAProblemFormulationIII} is convex w.r.t. the remaining optimization variables and satisfies Slater's constraint qualification \cite{book:convex}. Therefore, strong duality holds and solving the dual problem is equivalent to solving the primal problem \cite{book:convex}. 
In the following, we prove Theorem \ref{Theorem1} by exploiting the Karush-Kuhn-Tucker (KKT) conditions of \eqref{RAProblemFormulationIII}.
To start with, the Lagrangian function of the primal problem in \eqref{RAProblemFormulationIII} is given by \eqref{LagrangianFunction}, shown at the top of this page,
where ${{\alpha _{{\rm{C1}}}}},{{\alpha _{{\rm{C4}}}}}, {{\alpha _{k,{\rm{C7}}}}}, {{\alpha _{k,{\rm{C8}}}}} \ge 0$ are the Lagrange multipliers associated with constraints C1, C4, C7, and C8, respectively, 
$\boldsymbol{\Lambda}_{k,{\rm{C6}}},\boldsymbol{\Lambda}_{{\bf{v}},{\rm{C6}}} \in \mathbb{C}^{N_{\mathrm{t}} \times N_{\mathrm{t}}}$ are the 
Lagrange multiplier matrices for 
the positive semidefinite constraint C6 for matrices $\mathbf{W}_k$ and $\mathbf{V}$, respectively, and $\vartheta$ denotes the collection of terms that are independent of $\left(\mathbf{W}_k, \mathbf{V}\right)$.
The dual problem of \eqref{RAProblemFormulationIII} is given by
\begin{equation}\label{DualProblem}
\underset{	\begin{subarray}{c}
	{{\alpha _{{\rm{C1}}}}},{{\alpha _{{\rm{C4}}}}}, {{\alpha _{k,{\rm{C7}}}}}, {{\alpha _{k,{\rm{C8}}}}} \ge 0 \\
	\boldsymbol{\Lambda}_{k,{\rm{C6}}},\boldsymbol{\Lambda}_{{\bf{v}},{\rm{C6}}} \succeq \mathbf{0} \end{subarray}}{\mathrm{maximize}} \hspace{5mm}\underset{	\begin{subarray}{c}
	0\le\rho_{k}\le 1, \gamma_k,P_{k,\mathrm{out} } \\
	\mathbf{W}_k, \mathbf{V} \in \mathbb{H}^{N_{\mathrm{t}} \times N_{\mathrm{t}}}
	\end{subarray}}{\mathrm{inf}} \mathcal{L}.
\end{equation}

The KKT conditions for the optimal $\left(\mathbf{W}^*_k, \mathbf{V}^*\right)$ are obtained as follows:
\begin{align}\label{kktConditions}
\mbox{K1}:& \;	{{\alpha^* _{{\rm{C1}}}}},{{\alpha^* _{{\rm{C4}}}}} \ge 0, {{\alpha^* _{k,{\rm{C7}}}}}, {{\alpha^* _{k,{\rm{C8}}}}} > 0, \notag\\[-0.5mm]
\mbox{K2}: &\;	\boldsymbol{\Lambda}^*_{k,{\rm{C6}}},\boldsymbol{\Lambda}^*_{{\bf{v}},{\rm{C6}}} \succeq \mathbf{0}, \notag\\[-0.5mm]
\mbox{K3}:&\; \Tr\left(\boldsymbol{\Lambda}^*_{k,{\rm{C6}}}\mathbf{W}^*_k \right) =0, \Tr\left({{\boldsymbol{\Lambda}^* _{{\bf{v}},{\rm{C6}}}}{\bf{V}}^*}\right) =0, \notag\\[-0.5mm]
\mbox{K4}:&\; {\nabla _{{{\bf{W}}^*_k}}} \mathcal{L} = \mathbf{0},\;\text{and} \notag\\[-0.5mm]
\mbox{K5}:&\; {\nabla _{{{\bf{V}^*}}}} \mathcal{L} = \mathbf{0} ,
\end{align}
where ${\nabla _{{{\bf{W}}^*_k}}} \mathcal{L}$ and ${\nabla _{{{\bf{V}}^*}}} \mathcal{L}$ denote the gradients of the Lagrangian function $\mathcal{L}$ w.r.t. matrices $\mathbf{W}^*_k$ and $\mathbf{V}^*$, respectively.
By examining K4 and K5 in \eqref{kktConditions}, we have
\begin{align}
\hspace{-2mm}{\nabla _{{{\bf{W}}^*_k}}} \mathcal{L}& = {\boldsymbol{\Lambda}^* _{{\bf{v}},{\rm{C6}}}} \hspace{-0.5mm}-\hspace{-0.5mm} \boldsymbol{\Lambda}^*_{k,{\rm{C6}}} \hspace{-0.5mm}-\hspace{-0.5mm} {\alpha^* _{k{\rm{,C7}}}}\beta _{k,{\rm{C7}}}^2{{\bf{H}}_k} \notag\\
&- \hspace{-0.5mm}{\alpha^* _{k{\rm{,C7}}}}{\beta _{k,{\rm{C7}}}}\frac{{{{\bf{H}}_k}}}{{\sqrt {\Tr\left( {{{\bf{W}}^*_k}{{\bf{H}}_k}} \right)} }} \;\text{and}\label{GridentLW}\\
\hspace{-2mm}\boldsymbol{\Lambda}^*_{{\bf{v}},{\rm{C6}}} & = \hspace{-1mm}\left(\hspace{-1mm} {\beta _{{\rm{Obj}}}^2\xi  \hspace{-1mm}+\hspace{-1mm} {\alpha _{{\rm{C1}}}} \hspace{-1mm}+ \hspace{-1mm}{\alpha _{{\rm{C4}}}}{\rm{EH}}_{{\rm{req}}}^{{\rm{Eff}}}} \right) \hspace{-1mm}\mathbf{I}_{N_{\mathrm{t}}} \hspace{-1mm}+\hspace{-1mm} \sum_{k'=1}^{K}\hspace{-1mm}{\alpha _{k'{\rm{,C7}}}}\beta _{k',{\rm{C7}}}^2{{\bf{H}}_{k'}} \notag\\
&\hspace{-5mm}- \hspace{-1mm}\sum_{k'=1}^{K} \hspace{-1mm}{\alpha _{k'{\rm{,C8}}}}\beta _{k',{\rm{C8}}}\frac{\eta{\bf{H}}_{k'}}{\sqrt{\eta\hspace{-0.5mm} \Tr\hspace{-0.5mm} \left( {\sum\limits_{k'' = 1}^K\hspace{-1mm} {{{\bf{W}}^*_{k''}}} {{\bf{H}}_{k'}} \hspace{-1mm}+\hspace{-1mm} {\bf{V}^*}{{\bf{H}}_{k'}}} \hspace{-1mm}\right)}},\label{GridentLV}
\end{align}
respectively. 
Moreover, based on \cite[Proposition~3.1]{XuSWIPT}, $\boldsymbol{\Lambda}^*_{{\bf{v}},{\rm{C6}}}$ is a full-rank matrix with probability one, i.e., $\mathrm{Rank}\left(\boldsymbol{\Lambda}^*_{{\bf{v}},{\rm{C6}}}\right) = N_{\mathrm{t}}$, when the channels, ${{\bf{h}}_k}$, $\forall k$, are mutually statistically independent.
Therefore, from the complementary slackness condition in K3, we obtain $\mathbf{v}^* = \mathbf{0}$.

Furthermore, by exploiting K4 and \eqref{GridentLW} and a basic inequality for the ranks of matrices, we have
\begin{align}
\mathrm{Rank}\left(\boldsymbol{\Lambda}^*_{{\bf{v}},{\rm{C6}}}\right) &\le \mathrm{Rank}\left(\boldsymbol{\Lambda}^*_{k,{\rm{C6}}}\right) \notag\\
& + \mathrm{Rank}\left(\hspace{-1mm}\left( {\alpha _{k,{\rm{C7}}}^*\beta _{k,{\rm{C7}}}^2{\rm{ + }}\frac{{\alpha _{k,{\rm{C7}}}^*{\beta _{k,{\rm{C7}}}}}}{{\sqrt {{\rm{Tr}}\left( {{\bf{W}}_k^*{{\bf{H}}_k}} \right)\hspace{-1mm}} }}} \right)\hspace{-1mm}{{\bf{H}}_k}\hspace{-1mm}\right)\notag\\
&\Rightarrow \mathrm{Rank}\left(\boldsymbol{\Lambda}^*_{k,{\rm{C6}}}\right) \ge N_{\mathrm{t}}-1.
\end{align}
Moreover, ${\bf{W}}_k^* \neq \mathbf{0}$ is required to satisfy QoS constraint C2 for each user. 
Thus, considering K3, $\boldsymbol{\Lambda}^*_{k,{\rm{C6}}}$ cannot be full-rank.
Hence, $\mathrm{Rank}(\boldsymbol{\Lambda}^*_{k,{\rm{C6}}}) = N_{\mathrm{t}}-1$ and $\mathrm{Rank}\left(\mathbf{W}^*_k\right) = 1$, which completes the proof\footnote{Note that the above proof does not imply the uniqueness of the optimal solution of the relaxed problem in \eqref{RAProblemFormulationIII}.
	There may be multiple optimal solutions that all satisfy the KKT conditions in \eqref{kktConditions} and achieve the same objective value.
	Nevertheless, as shown above, all optimal solutions are rank-one with probability one.}.
}
	
\bibliographystyle{IEEEtran}
\bibliography{WPT}

\begin{thebibliography}{100}
\providecommand{\url}[1]{#1}
\csname url@samestyle\endcsname
\providecommand{\newblock}{\relax}
\providecommand{\bibinfo}[2]{#2}
\providecommand{\BIBentrySTDinterwordspacing}{\spaceskip=0pt\relax}
\providecommand{\BIBentryALTinterwordstretchfactor}{4}
\providecommand{\BIBentryALTinterwordspacing}{\spaceskip=\fontdimen2\font plus
\BIBentryALTinterwordstretchfactor\fontdimen3\font minus
  \fontdimen4\font\relax}
\providecommand{\BIBforeignlanguage}[2]{{%
\expandafter\ifx\csname l@#1\endcsname\relax
\typeout{** WARNING: IEEEtran.bst: No hyphenation pattern has been}%
\typeout{** loaded for the language `#1'. Using the pattern for}%
\typeout{** the default language instead.}%
\else
\language=\csname l@#1\endcsname
\fi
#2}}
\providecommand{\BIBdecl}{\relax}
\BIBdecl

\bibitem{7123563}
A.~{Al-Fuqaha}, M.~{Guizani}, M.~{Mohammadi}, M.~{Aledhari}, and M.~{Ayyash},
  ``Internet of things: A survey on enabling technologies, protocols, and
  applications,'' \emph{IEEE Commun. Surveys Tuts.}, vol.~17, no.~4, pp.
  2347--2376, Fourthquater 2015.

\bibitem{7565189}
C.~Bockelmann, N.~Pratas, H.~Nikopour, K.~Au, T.~Svensson, C.~Stefanovic,
  P.~Popovski, and A.~Dekorsy, ``Massive machine-type communications in {5G}:
  {Physical} and {MAC}-layer solutions,'' \emph{IEEE Commun. Mag.}, vol.~54,
  no.~9, pp. 59--65, Sep. 2016.

\bibitem{Bruno2019Review}
B.~Clerckx, R.~Zhang, R.~Schober, D.~W.~K. Ng, D.~I. Kim, and H.~V. Poor,
  ``Fundamentals of wireless information and power transfer: From {RF} energy
  harvester models to signal and system designs,'' \emph{IEEE J. Select. Areas
  Commun.}, vol.~37, no.~1, pp. 4--33, Jan. 2019.

\bibitem{DerrickOFDMAHybridBS}
D.~W.~K. Ng, E.~S. Lo, and R.~Schober, ``Energy-efficient resource allocation
  in {OFDMA} systems with hybrid energy harvesting base station,'' \emph{IEEE
  Trans. Wireless Commun.}, vol.~12, no.~7, pp. 3412--3427, Jul. 2013.

\bibitem{KimInductiveCoupling}
C.-G. Kim, D.-H. Seo, J.-S. You, J.-H. Park, and B.~Cho, ``Design of a
  contactless battery charger for cellular phone,'' \emph{IEEE Trans. on
  Industrial Electronics}, vol.~48, no.~6, pp. 1238--1247, Dec. 2001.

\bibitem{Kurs83}
A.~Kurs, A.~Karalis, R.~Moffatt, J.~D. Joannopoulos, P.~Fisher, and M.~Solja{\v
  c}i{\'c}, ``Wireless power transfer via strongly coupled magnetic
  resonances,'' \emph{Science}, vol. 317, no. 5834, pp. 83--86, Jul. 2007.

\bibitem{summerer2009concepts}
L.~Summerer and O.~Purcell, ``Concepts for wireless energy transmission via
  laser,'' \emph{Europeans Space Agency (ESA)-Advanced Concepts Team}, 2009.

\bibitem{ZhangRuiModel}
R.~Zhang and C.~K. Ho, ``{MIMO} broadcasting for simultaneous wireless
  information and power transfer,'' \emph{IEEE Trans. Wireless Commun.},
  vol.~12, no.~5, pp. 1989--2001, May 2013.

\bibitem{SuzhiMag}
S.~Bi, C.~K. Ho, and R.~Zhang, ``Wireless powered communication: opportunities
  and challenges,'' \emph{IEEE Commun. Mag.}, vol.~53, no.~4, pp. 117--125,
  Apr. 2015.

\bibitem{QingQingEEWPC}
Q.~Wu, M.~Tao, D.~W. Kwan~Ng, W.~Chen, and R.~Schober, ``Energy-efficient
  resource allocation for wireless powered communication networks,'' \emph{IEEE
  Trans. Wireless Commun.}, vol.~15, no.~3, pp. 2312--2327, Mar. 2016.

\bibitem{NguyenWPR}
K.-G. Nguyen, Q.-D. Vu, L.-N. Tran, and M.~Juntti, ``Energy efficiency fairness
  for multi-pair wireless-powered relaying systems,'' \emph{IEEE J. Select.
  Areas Commun.}, vol.~37, no.~2, pp. 357--373, Feb. 2019.

\bibitem{YinghuiWPBC}
Y.~Ye, L.~Shi, Q.~R. Hu, and G.~Lu, ``Energy-efficient resource allocation for
  wirelessly powered backscatter communications,'' \emph{IEEE Commun. Lett.},
  vol.~23, no.~8, pp. 1418--1422, Aug. 2019.

\bibitem{FengWPMEC}
J.~Feng, Q.~Pei, F.~R. Yu, X.~Chu, and B.~Shang, ``Computation offloading and
  resource allocation for wireless powered mobile edge computing with latency
  constraint,'' \emph{IEEE Wireless Commun. Lett.}, vol.~8, no.~5, pp.
  1320--1323, Oct. 2019.

\bibitem{PsomasMEC}
C.~Psomas and I.~Krikidis, ``Wireless powered mobile edge computing: Offloading
  or local computation?'' \emph{IEEE Commun. Lett.}, vol.~24, no.~11, pp.
  2642--2646, Nov. 2020.

\bibitem{XiaoyanMEC}
X.~Hu, K.-K. Wong, and K.~Yang, ``Wireless powered cooperation-assisted mobile
  edge computing,'' \emph{IEEE Trans. Wireless Commun.}, vol.~17, no.~4, pp.
  2375--2388, Apr. 2018.

\bibitem{LuyueMEC}
L.~Ji and S.~Guo, ``Energy-efficient cooperative resource allocation in
  wireless powered mobile edge computing,'' \emph{IEEE Internet Things J.},
  vol.~6, no.~3, pp. 4744--4754, Nov. 2019.

\bibitem{Morsi2020}
R.~Morsi, ``Analysis and design of communication systems with wireless power
  transfer,'' Doctoral Thesis, Friedrich-Alexander-Universit{\"a}t
  Erlangen-N{\"u}rnberg (FAU), 2020.

\bibitem{RogerOFDM1999}
C.~Y. Wong, R.~Cheng, K.~Lataief, and R.~Murch, ``Multiuser {OFDM} with
  adaptive subcarrier, bit, and power allocation,'' \emph{IEEE J. Select. Areas
  Commun.}, vol.~17, no.~10, pp. 1747--1758, Oct. 1999.

\bibitem{DerrickOFDMA2012}
D.~W.~K. Ng, E.~S. Lo, and R.~Schober, ``Energy-efficient resource allocation
  in {OFDMA} systems with large numbers of base station antennas,'' \emph{IEEE
  Trans. Wireless Commun.}, vol.~11, no.~9, pp. 3292--3304, Sep. 2012.

\bibitem{WeiNOMA7934461}
Z.~Wei, D.~W.~K. Ng, J.~Yuan, and H.-M. Wang, ``Optimal resource allocation for
  power-efficient {MC-NOMA} with imperfect channel state information,''
  \emph{IEEE Trans. Commun.}, vol.~65, no.~9, pp. 3944--3961, Sep. 2017.

\bibitem{LiangTS}
L.~Liu, R.~Zhang, and K.-C. Chua, ``Wireless information transfer with
  opportunistic energy harvesting,'' \emph{IEEE Trans. Wireless Commun.},
  vol.~12, no.~1, pp. 288--300, Jul. 2013.

\bibitem{LiangPS}
------, ``Wireless information and power transfer: A dynamic power splitting
  approach,'' \emph{IEEE Trans. Commun.}, vol.~61, no.~9, pp. 3990--4001, Sep.
  2013.

\bibitem{ZhouOPS}
X.~Zhou, R.~Zhang, and C.~K. Ho, ``Wireless information and power transfer:
  Architecture design and rate-energy tradeoff,'' \emph{IEEE Trans. Commun.},
  vol.~61, no.~11, pp. 4754--4767, Nov. 2013.

\bibitem{XuSWIPT}
J.~Xu, L.~Liu, and R.~Zhang, ``Multiuser {MISO} beamforming for simultaneous
  wireless information and power transfer,'' \emph{IEEE Trans. Signal
  Process.}, vol.~62, no.~18, pp. 4798--4810, Sep. 2014.

\bibitem{ShiQingtwc}
Q.~Shi, L.~Liu, W.~Xu, and R.~Zhang, ``Joint transmit beamforming and receive
  power splitting for {MISO SWIPT} systems,'' \emph{IEEE Trans. Wireless
  Commun.}, vol.~13, no.~6, pp. 3269--3280, Jun. 2014.

\bibitem{KaibinTSP}
K.~Huang and E.~Larsson, ``Simultaneous information and power transfer for
  broadband wireless systems,'' \emph{IEEE Trans. Signal Process.}, vol.~61,
  no.~23, pp. 5972--5986, Dec. 2013.

\bibitem{JR:WIPT_fullpaper_OFDMA}
D.~W.~K. Ng, E.~S. Lo, and R.~Schober, ``Wireless information and power
  transfer: Energy efficiency optimization in {OFDMA} systems,'' \emph{IEEE
  Trans. Wireless Commun.}, vol.~12, pp. 6352--6370, Dec. 2013.

\bibitem{ZhouTWC}
X.~Zhou, R.~Zhang, and C.~K. Ho, ``Wireless information and power transfer in
  multiuser {OFDM} systems,'' \emph{IEEE Trans. Wireless Commun.}, vol.~13,
  no.~4, pp. 2282--2294, Apr. 2014.

\bibitem{JR:MOOP_SWIPT}
D.~W.~K. Ng, E.~S. Lo, and R.~Schober, ``Multiobjective resource allocation for
  secure communication in cognitive radio networks with wireless information
  and power transfer,'' \emph{IEEE Trans. Veh. Technol.}, vol.~65, no.~5, pp.
  3166--3184, May 2016.

\bibitem{MengLiSWIPTMOO}
M.~Li, X.~Tao, N.~Li, and H.~Wu, ``Multi-objective optimization for full-duplex
  {SWIPT} systems,'' \emph{IEEE Access}, vol.~8, pp. 30\,838--30\,853, Feb.
  2020.

\bibitem{GaofengAverage}
G.~Pan, H.~Lei, Y.~Deng, L.~Fan, J.~Yang, Y.~Chen, and Z.~Ding, ``On secrecy
  performance of {MISO SWIPT} systems with {TAS} and imperfect {CSI},''
  \emph{IEEE Trans. Commun.}, vol.~64, no.~9, pp. 3831--3843, Sep. 2016.

\bibitem{KhandakeOutage}
M.~R.~A. Khandaker, K.-K. Wong, Y.~Zhang, and Z.~Zheng, ``Probabilistically
  robust {SWIPT} for secrecy {MISOME} systems,'' \emph{IEEE Trans. on Inf.
  Forensics and Security}, vol.~12, no.~1, pp. 211--226, Jan. 2017.

\bibitem{BinbinOutageWorst}
B.~Su, Q.~Ni, and W.~Yu, ``Robust transmit beamforming for {SWIPT}-enabled
  cooperative {NOMA} with channel uncertainties,'' \emph{IEEE Trans. Commun.},
  vol.~67, no.~6, pp. 4381--4392, Jun. 2019.

\bibitem{XiangRobust}
Z.~Xiang and M.~Tao, ``Robust beamforming for wireless information and power
  transmission,'' \emph{IEEE Wireless Commun. Lett.}, vol.~1, no.~4, pp.
  372--375, Aug. 2012.

\bibitem{Xin_WorstCase}
X.~Su, L.~Li, H.~Yin, and P.~Zhang, ``Robust power- and rate-splitting-based
  transceiver design in $k$-user {MISO SWIPT} interference channel under
  imperfect {CSIT},'' \emph{IEEE Commun. Lett.}, vol.~23, no.~3, pp. 514--517,
  Mar. 2019.

\bibitem{XinruiWorst}
X.~Li, W.~Wang, M.~Zhang, F.~Zhou, and N.~Al-Dhahir, ``Robust secure
  beamforming for {SWIPT}-aided relay systems with full-duplex receiver and
  imperfect {CSI},'' \emph{IEEE Trans. Veh. Technol.}, vol.~69, no.~2, pp.
  1867--1878, Feb. 2020.

\bibitem{LengFDSWIPT}
S.~Leng, D.~W.~K. Ng, N.~Zlatanov, and R.~Schober, ``Multi-objective resource
  allocation in full-duplex {SWIPT} systems,'' in \emph{Proc. IEEE Intern.
  Commun. Conf.}, 2016, pp. 1--7.

\bibitem{HuFDswipt}
Z.~Hu, C.~Yuan, and F.~Gao, ``Maximizing harvested energy for full-duplex
  {SWIPT} system with power splitting,'' \emph{IEEE Access}, vol.~5, pp.
  24\,975--24\,987, Oct. 2017.

\bibitem{5706317}
R.~{Irmer}, H.~{Droste}, P.~{Marsch}, M.~{Grieger}, G.~{Fettweis}, S.~{Brueck},
  H.~{Mayer}, L.~{Thiele}, and V.~{Jungnickel}, ``Coordinated multipoint:
  Concepts, performance, and field trial results,'' \emph{IEEE Commun. Mag.},
  vol.~49, no.~2, pp. 102--111, Feb. 2011.

\bibitem{JR:rui_zhang_secrecy}
L.~Liu, R.~Zhang, and K.-C. Chua, ``Secrecy wireless information and power
  transfer with {MISO} beamforming,'' \emph{IEEE Trans. Signal Process.},
  vol.~62, pp. 1850--1863, Apr. 2014.

\bibitem{NgSecureSWIPT}
D.~W.~K. Ng, E.~S. Lo, and R.~Schober, ``Robust beamforming for secure
  communication in systems with wireless information and power transfer,''
  \emph{IEEE Trans. Wireless Commun.}, vol.~13, no.~8, pp. 4599--4615, Aug.
  2014.

\bibitem{IoannisSWIPT}
I.~Krikidis, ``Simultaneous information and energy transfer in large-scale
  networks with/without relaying,'' \emph{IEEE Trans. Commun.}, vol.~62, no.~3,
  pp. 900--912, Mar. 2014.

\bibitem{HimalJSAC}
D.~S. Michalopoulos, H.~A. Suraweera, and R.~Schober, ``Relay selection for
  simultaneous information transmission and wireless energy transfer: A
  tradeoff perspective,'' \emph{IEEE J. Select. Areas Commun.}, vol.~33, no.~8,
  pp. 1578--1594, Aug. 2015.

\bibitem{TangCoMPSWIPT}
J.~Tang, A.~Shojaeifard, D.~K.~C. So, K.-K. Wong, and N.~Zhao, ``Energy
  efficiency optimization for {CoMP-SWIPT} heterogeneous networks,'' \emph{IEEE
  Trans. Commun.}, vol.~66, no.~12, pp. 6368--6383, Dec. 2018.

\bibitem{7037480}
D.~W.~K. {Ng} and R.~{Schober}, ``Resource allocation for coordinated
  multipoint networks with wireless information and power transfer,'' in
  \emph{Proc. IEEE Global Commun. Conf.}, Dec. 2014, pp. 4281--4287.

\bibitem{CN:EH_measurement_2}
J.~Guo and X.~Zhu, ``An improved analytical model for {RF-DC} conversion
  efficiency in microwave rectifiers,'' in \emph{Proc. IEEE MTT-S Intern.
  Microwave Sympos.}, Jun. 2012, pp. 1--3.

\bibitem{JR:EH_measurement_1}
T.~Le, K.~Mayaram, and T.~Fiez, ``Efficient far-field radio frequency energy
  harvesting for passively powered sensor networks,'' \emph{IEEE J. Solid-State
  Circuits}, vol.~43, pp. 1287--1302, May 2008.

\bibitem{JR:Rania_nonlinear_circuit_based_TCOM}
R.~{Morsi}, V.~{Jamali}, A.~{Hagelauer}, D.~W.~K. {Ng}, and R.~{Schober},
  ``Conditional capacity and transmit signal design for {SWIPT} systems with
  multiple nonlinear energy harvesting receivers,'' \emph{IEEE Trans. Commun.},
  vol.~68, no.~1, pp. 582--601, Jan. 2020.

\bibitem{JR:non_linear_model}
E.~Boshkovska, D.~Ng, N.~Zlatanov, and R.~Schober, ``Practical non-linear
  energy harvesting model and resource allocation for {SWIPT} systems,''
  \emph{IEEE Commun. Lett.}, vol.~19, pp. 2082--2085, Dec. 2015.

\bibitem{JR:Elena_TCOM}
E.~Boshkovska, D.~W.~K. Ng, N.~Zlatanov, A.~Koelpin, and R.~Schober, ``Robust
  resource allocation for {MIMO} wireless powered communication networks based
  on a non-linear {EH} model,'' \emph{IEEE Trans. Commun.}, vol.~65, no.~5, pp.
  1984--1999, May 2017.

\bibitem{book:Kwan_power_transfer}
D.~W.~K. Ng, T.~Q. Duong, C.~Zhong, and R.~Schober, \emph{Wireless Information
  and Power Transfer: Theory and Practice}.\hskip 1em plus 0.5em minus
  0.4em\relax {Wiley}, 2019.

\bibitem{BrunoTSP}
B.~Clerckx, ``Wireless information and power transfer: Nonlinearity, waveform
  design, and rate-energy tradeoff,'' \emph{IEEE Trans. Signal Process.},
  vol.~66, no.~4, pp. 847--862, Nov. 2018.

\bibitem{KrikidisSWIPT}
I.~{Krikidis}, S.~{Timotheou}, S.~{Nikolaou}, G.~{Zheng}, D.~W.~K. {Ng}, and
  R.~{Schober}, ``Simultaneous wireless information and power transfer in
  modern communication systems,'' \emph{IEEE Commun. Mag.}, vol.~52, no.~11,
  pp. 104--110, Nov. 2014.

\bibitem{NiyatoEH}
D.~Niyato, D.~I. Kim, M.~Maso, and Z.~Han, ``Wireless powered communication
  networks: Research directions and technological approaches,'' \emph{IEEE
  Wireless Commun.}, vol.~24, no.~6, pp. 88--97, Dec. 2017.

\bibitem{ZengWPT}
Y.~Zeng, B.~Clerckx, and R.~Zhang, ``Communications and signals design for
  wireless power transmission,'' \emph{IEEE Trans. Commun.}, vol.~65, no.~5,
  pp. 2264--2290, May 2017.

\bibitem{clerckx2021wireless}
B.~Clerckx, K.~Huang, L.~R. Varshney, S.~Ulukus, and M.-S. Alouini, ``Wireless
  power transfer for future networks: Signal processing, machine learning,
  computing, and sensing,'' \emph{IEEE J. Select. Topics Signal Process.},
  early access, 2021.

\bibitem{JR:Energy_harvesting_circuit}
C.~Valenta and G.~Durgin, ``Harvesting wireless power: Survey of
  energy-harvester conversion efficiency in far-field, wireless power transfer
  systems,'' \emph{IEEE Microw. Mag.}, vol.~15, pp. 108--120, Jun. 2014.

\bibitem{JR:SWIPT_antennas}
X.~Chen, C.~Yuen, and Z.~Zhang, ``Wireless energy and information transfer
  tradeoff for limited-feedback multiantenna systems with energy beamforming,''
  \emph{IEEE Trans. Veh. Technol.}, vol.~63, pp. 407--412, Jan. 2014.

\bibitem{CN:OFDM_Kwan}
D.~W.~K. Ng, E.~S. Lo, and R.~Schober, ``Energy-efficient resource allocation
  in multiuser {OFDM} systems with wireless information and power transfer,''
  in \emph{Proc. IEEE Wireless Commun. and Networking Conf.}, Apr. 2013, pp.
  1--6.

\bibitem{JR:EE_SWIPT_Massive_MIMO}
X.~Chen, X.~Wang, and X.~Chen, ``Energy-efficient optimization for wireless
  information and power transfer in large-scale {MIMO} systems employing energy
  beamforming,'' \emph{IEEE Wireless Commun. Lett.}, vol.~2, pp. 667--670, Dec.
  2013.

\bibitem{Ding2014}
Z.~Ding, C.~Zhong, D.~W.~K. Ng, M.~Peng, H.~A. Suraweera, R.~Schober, and H.~V.
  Poor, ``Application of smart antenna technologies in simultaneous wireless
  information and power transfer,'' \emph{IEEE Commun. Mag.}, vol.~53, no.~4,
  pp. 86--93, Apr. 2015.

\bibitem{corless1996lambertw}
R.~M. Corless, G.~H. Gonnet, D.~E. Hare, D.~J. Jeffrey, and D.~E. Knuth, ``On
  the {LambertW} function,'' \emph{Advances in Computational Mathematics},
  vol.~5, no.~1, pp. 329--359, 1996.

\bibitem{ADS}
{The Keysight Technologies, Inc.}, ``{Electronic Design Automation (EDA)
  Software, Advanced Design System (ADS), Version 2017}.''

\bibitem{Georgiadis_WPT_book_2016}
S.~Nikoletseas, Y.~Yang, and A.~Georgiadis, Eds., \emph{Wireless Power Transfer
  Algorithms, Technologies and Applications in Ad Hoc Communication
  Networks}.\hskip 1em plus 0.5em minus 0.4em\relax Springer International
  Publishing, 2016.

\bibitem{JR:WPCN_nonlinear_Elena}
E.~{Boshkovska}, D.~W.~K. {Ng}, L.~{Dai}, and R.~{Schober}, ``Power-efficient
  and secure {WPCNs} with hardware impairments and non-linear {EH} circuit,''
  \emph{IEEE Trans. Commun.}, vol.~66, no.~6, pp. 2642--2657, Jun. 2018.

\bibitem{CN:Elena_non_linear_scheduling}
E.~{Boshkovska}, R.~{Morsi}, D.~W.~K. {Ng}, and R.~{Schober}, ``Power
  allocation and scheduling for {SWIPT} systems with non-linear energy
  harvesting model,'' in \emph{Proc. IEEE Intern. Commun. Conf.}, 2016, pp.
  1--6.

\bibitem{VarastehWPT}
M.~Varasteh, J.~Hoydis, and B.~Clerckx, ``Learning to communicate and energize:
  Modulation, coding, and multiple access designs for wireless
  information-power transmission,'' \emph{IEEE Trans. Commun.}, vol.~68,
  no.~11, pp. 6822--6839, Nov. 2020.

\bibitem{BrunoBeneficial}
B.~Clerckx and J.~Kim, ``On the beneficial roles of fading and transmit
  diversity in wireless power transfer with nonlinear energy harvesting,''
  \emph{IEEE Trans. Wireless Commun.}, vol.~17, no.~11, pp. 7731--7743, Nov.
  2018.

\bibitem{Waveform_design_WPT_Clerckx_2016}
B.~Clerckx and E.~Bayguzina, ``Waveform design for wireless power transfer,''
  \emph{IEEE Trans. Signal Process.}, vol.~64, no.~23, pp. 6313--6328, Dec.
  2016.

\bibitem{PaulART}
P.~Horowitz and W.~Hill, \emph{The Art of Electronics}.\hskip 1em plus 0.5em
  minus 0.4em\relax Cambridge University Press, 1989.

\bibitem{JR:Energy_harvesting_circiut_memory_Robert}
N.~Shanin, L.~Cottatellucci, and R.~Schober, ``Markov decision process based
  design of {SWIPT} systems: Non-linear {EH} circuits, memory, and impedance
  mismatch,'' \emph{IEEE Trans. Commun.}, vol.~69, no.~2, pp. 1259--1274, Feb.
  2021.

\bibitem{ShiTSP}
Q.~Shi, C.~Peng, W.~Xu, M.~Hong, and Y.~Cai, ``Energy efficiency optimization
  for {MISO SWIPT} systems with zero-forcing beamforming,'' \emph{IEEE Trans.
  Signal Process.}, vol.~64, no.~4, pp. 842--854, Feb. 2016.

\bibitem{ZhiqiangChapter}
Z.~Wei, Y.~Cai, D.~W.~K. Ng, and J.~Yuan, \emph{Energy-Efficient Radio Resource
  Management}.\hskip 1em plus 0.5em minus 0.4em\relax In Wiley 5G Ref (eds R.
  Tafazolli, C.-L. Wang and P. Chatzimisios), American Cancer Society, 2019,
  pp. 1--23.

\bibitem{JR:WPC_Rui_Zhang}
H.~Ju and R.~Zhang, ``Throughput maximization in wireless powered communication
  networks,'' \emph{IEEE Trans. Wireless Commun.}, vol.~13, pp. 418--428, Jan.
  2014.

\bibitem{QingqingWuSWIPT}
Q.~Wu and R.~Zhang, ``Weighted sum power maximization for intelligent
  reflecting surface aided {SWIPT},'' \emph{IEEE Wireless Commun. Lett.},
  vol.~9, no.~5, pp. 586--590, May 2020.

\bibitem{BoaventuraDC}
A.~S. Boaventura and N.~B. Carvalho, ``Maximizing {DC} power in energy
  harvesting circuits using multisine excitation,'' in \emph{Proc. IEEE MTT-S
  Intern. Microwave Sympos.}, 2011, pp. 1--4.

\bibitem{JR:Yan_MOOP}
Y.~Sun, D.~W.~K. Ng, J.~Zhu, and R.~Schober, ``Multi-objective optimization for
  robust power efficient and secure full-duplex wireless communication
  systems,'' \emph{IEEE Trans. Wireless Commun.}, vol.~15, no.~8, pp.
  5511--5526, Aug. 2016.

\bibitem{BinSWIPTmmWave}
B.~Li, Y.~Dai, Z.~Dong, E.~Panayirci, H.~Jiang, and H.~Jiang,
  ``Energy-efficient resources allocation with millimeter-wave massive {MIMO}
  in ultra dense hetnets by {SWIPT} and {CoMP},'' \emph{IEEE Trans. Wireless
  Commun.}, vol.~20, no.~7, pp. 4435--4451, Jul. 2021.

\bibitem{XieUAVSWIPT}
L.~Xie, X.~Cao, J.~Xu, and R.~Zhang, ``{UAV}-enabled wireless power transfer:
  {A }tutorial overview,'' \emph{IEEE Trans. on Green Commun. Netw.}, early
  access, 2021.

\bibitem{ChunshengGIOT}
C.~Zhu, V.~C.~M. Leung, L.~Shu, and E.~C.-H. Ngai, ``Green internet of things
  for smart world,'' \emph{IEEE Access}, vol.~3, pp. 2151--2162, Nov. 2015.

\bibitem{ShuaifeiIoE}
S.~Chen, J.~Zhang, Y.~Jin, and B.~Ai, ``Wireless powered {IoE} for {6G}:
  Massive access meets scalable cell-free massive {MIMO},'' \emph{China
  Communications}, vol.~17, no.~12, pp. 92--109, Dec. 2020.

\bibitem{ZhiQuanNPHard}
Z.-Q. Luo and S.~Zhang, ``Dynamic spectrum management: Complexity and
  duality,'' \emph{IEEE J. Select. Topics Signal Process.}, vol.~2, no.~1, pp.
  57--73, Feb. 2008.

\bibitem{TuyMonotonic}
H.~Tuy, ``Monotonic optimization: Problems and solution approaches,''
  \emph{SIAM J. on Optim.}, vol.~11, no.~2, pp. 464--494, 2000.

\bibitem{MatthiesenTSP}
B.~Matthiesen, C.~Hellings, E.~A. Jorswieck, and W.~Utschick, ``Mixed monotonic
  programming for fast global optimization,'' \emph{IEEE Trans. Signal
  Process.}, vol.~68, pp. 2529--2544, Mar. 2020.

\bibitem{7812683}
Y.~{Sun}, D.~W.~K. {Ng}, Z.~{Ding}, and R.~{Schober}, ``Optimal joint power and
  subcarrier allocation for full-duplex multicarrier non-orthogonal multiple
  access systems,'' \emph{IEEE Trans. Commun.}, vol.~65, no.~3, pp. 1077--1091,
  Mar. 2017.

\bibitem{ZapponeEEMonotonic}
A.~Zappone, E.~Björnson, L.~Sanguinetti, and E.~Jorswieck, ``Globally optimal
  energy-efficient power control and receiver design in wireless networks,''
  \emph{IEEE Trans. Signal Process.}, vol.~65, no.~11, pp. 2844--2859, Jun.
  2017.

\bibitem{konno2000branch}
H.~Konno and K.~Fukaishi, ``A branch and bound algorithm for solving low rank
  linear multiplicative and fractional programming problems,'' \emph{J. of
  Global Optim.}, vol.~18, no.~3, pp. 283--299, 2000.

\bibitem{Tuy2005Chapter}
H.~Tuy, F.~Al-Khayyal, and P.~T. Thach, \emph{Monotonic Optimization: Branch
  and Cut Methods}.\hskip 1em plus 0.5em minus 0.4em\relax Springer US, 2005,
  pp. 39--78.

\bibitem{dinkelbach1967nonlinear}
W.~Dinkelbach, ``On nonlinear fractional programming,'' \emph{Management
  Science}, vol.~13, no.~7, pp. 492--498, Mar. 1967.

\bibitem{CaiEEUAV}
Y.~Cai, Z.~Wei, R.~Li, D.~W.~K. Ng, and J.~Yuan, ``Joint trajectory and
  resource allocation design for energy-efficient secure {UAV} communication
  systems,'' \emph{IEEE Trans. Commun.}, vol.~68, no.~7, pp. 4536--4553, Jul.
  2020.

\bibitem{QingjiangWSMSE}
Q.~Shi, M.~Razaviyayn, Z.-Q. Luo, and C.~He, ``An iteratively weighted {MMSE}
  approach to distributed sum-utility maximization for a {MIMO} interfering
  broadcast channel,'' \emph{IEEE Trans. Signal Process.}, vol.~59, no.~9, pp.
  4331--4340, Sep. 2011.

\bibitem{ShenFP}
K.~Shen and W.~Yu, ``Fractional programming for communication systems - part
  {I}: Power control and beamforming,'' \emph{IEEE Trans. Signal Process.},
  vol.~66, no.~10, pp. 2616--2630, May 2018.

\bibitem{ZhiqiangEENOMA}
Z.~Wei, D.~W.~K. Ng, and J.~Yuan, ``{NOMA} for hybrid mmwave communication
  systems with beamwidth control,'' \emph{IEEE J. Select. Topics Signal
  Process.}, vol.~13, no.~3, pp. 567--583, Jun. 2019.

\bibitem{ZhiQuanTSPM}
Z.-Q. Luo, W.-K. Ma, A.~M.-C. So, Y.~Ye, and S.~Zhang, ``Semidefinite
  relaxation of quadratic optimization problems,'' \emph{IEEE Signal Process.
  Mag.}, vol.~27, no.~3, pp. 20--34, May 2010.

\bibitem{book:convex}
S.~Boyd and L.~Vandenberghe, \emph{Convex Optimization}.\hskip 1em plus 0.5em
  minus 0.4em\relax {Cambridge University Press}, 2004.

\bibitem{Birge2011introduction}
J.~R. Birge and F.~Louveaux, \emph{Introduction to stochastic
  programming}.\hskip 1em plus 0.5em minus 0.4em\relax Springer Science \&
  Business Media, 2011.

\bibitem{Kun-Yu2014}
K.-Y. Wang, A.~M.-C. So, T.-H. Chang, W.-K. Ma, and C.-Y. Chi, ``Outage
  constrained robust transmit optimization for multiuser {MISO} downlinks:
  Tractable approximations by conic optimization,'' \emph{IEEE Trans. Signal
  Process.}, vol.~62, no.~21, pp. 5690--5705, Nov. 2014.

\bibitem{VandenbergheSDP}
L.~Vandenberghe and S.~Boyd, ``Semidefinite programming,'' \emph{SIAM Rev.},
  vol.~38, no.~1, pp. 49--95, Mar. 1996.

\bibitem{HaotianIPM}
H.~Jiang, T.~Kathuria, Y.~T. Lee, S.~Padmanabhan, and Z.~Song, ``A faster
  interior point method for semidefinite programming,'' in \emph{IEEE 61st
  Annual Sympos. on Foundations of Comput. Sci.}, 2020, pp. 910--918.

\bibitem{HongxiangCSI}
H.~Xie, F.~Gao, S.~Jin, J.~Fang, and Y.-C. Liang, ``Channel estimation for
  {TDD/FDD} massive {MIMO} systems with channel covariance computing,''
  \emph{IEEE Trans. Wireless Commun.}, vol.~17, no.~6, pp. 4206--4218, Jun.
  2018.

\bibitem{book:david_wirelss_com}
D.~Tse and P.~Viswanath, \emph{{Fundamentals of Wireless Communication}},
  1st~ed.\hskip 1em plus 0.5em minus 0.4em\relax {Cambridge University Pres},
  2005.

\bibitem{ShanpuMIMOWPT}
S.~Shen and B.~Clerckx, ``Beamforming optimization for {MIMO} wireless power
  transfer with nonlinear energy harvesting: {RF} combining versus {DC}
  combining,'' \emph{IEEE Trans. Wireless Commun.}, vol.~20, no.~1, pp.
  199--213, Jan. 2021.

\bibitem{ShanpuWaveformMIMO}
------, ``Joint waveform and beamforming optimization for {MIMO} wireless power
  transfer,'' \emph{IEEE Trans. Commun.}, vol.~69, no.~8, pp. 5441--5455, Aug.
  2021.

\bibitem{di2019smart}
M.~Di~Renzo \emph{et~al.}, ``Smart radio environments empowered by
  reconfigurable {AI} meta-surfaces: An idea whose time has come,''
  \emph{EURASIP J. Wireless Commun. Netw.}, vol. 129, no.~1, pp. 1--20, May
  2019.

\bibitem{yu2021smart}
X.~Yu, V.~Jamali, D.~Xu, D.~W.~K. Ng, and R.~Schober, ``Smart and
  reconfigurable wireless communications: From {IRS} modeling to algorithm
  design,'' \emph{arXiv:2103.07046}, Mar. 2021.

\bibitem{hu2021robust}
S.~Hu, Z.~Wei, Y.~Cai, C.~Liu, D.~W.~K. Ng, and J.~Yuan, ``Robust and secure
  sum-rate maximization for multiuser {MISO} downlink systems with
  self-sustainable {IRS},'' \emph{IEEE Trans. Commun.}, early access, 2021.

\bibitem{9133130}
X.~{Yu}, D.~{Xu}, Y.~{Sun}, D.~W.~K. {Ng}, and R.~{Schober}, ``Robust and
  secure wireless communications via intelligent reflecting surfaces,''
  \emph{IEEE J. Sel. Areas Commun.}, vol.~38, no.~11, pp. 2637--2652, Nov.
  2020.

\bibitem{9136595}
W.~{Feng}, J.~{Tang}, Y.~{Yu}, J.~{Song}, N.~{Zhao}, G.~{Chen}, K.~K. {Wong},
  and J.~{Chambers}, ``{UAV}-enabled {SWIPT} in {IoT} networks for emergency
  communications,'' \emph{IEEE Wireless Commun.}, vol.~27, no.~5, pp. 140--147,
  Oct. 2020.

\bibitem{9174765}
Z.~Sun, Z.~Wei, N.~Yang, and X.~Zhou, ``Two-tier communication for
  {UAV}-enabled massive {IoT} systems: Performance analysis and joint design of
  trajectory and resource allocation,'' \emph{IEEE J. Select. Areas Commun.},
  vol.~39, no.~4, pp. 1132--1146, Apr. 2021.

\bibitem{ZhiqiangUAV}
Z.~Wei, Y.~Cai, Z.~Sun, D.~W.~K. Ng, J.~Yuan, M.~Zhou, and L.~Sun, ``Sum-rate
  maximization for {IRS}-assisted {UAV OFDMA} communication systems,''
  \emph{IEEE Trans. Wireless Commun.}, vol.~20, no.~4, pp. 2530--2550, Apr.
  2021.

\bibitem{8642372}
H.~{Zheng}, K.~{Xiong}, P.~{Fan}, Z.~{Zhong}, and K.~B. {Letaief},
  ``Fog-assisted multiuser {SWIPT} networks: Local computing or offloading,''
  \emph{IEEE Internet of Things J.}, vol.~6, no.~3, pp. 5246--5264, Jun. 2019.

\bibitem{9140412}
J.~Liu, K.~Xiong, D.~W.~K. Ng, P.~Fan, Z.~Zhong, and K.~B. Letaief, ``Max-min
  energy balance in wireless-powered hierarchical fog-cloud computing
  networks,'' \emph{IEEE Trans. Wireless Commun.}, vol.~19, no.~11, pp.
  7064--7080, Nov. 2020.

\bibitem{pmlr-v54-mcmahan17a}
B.~McMahan, E.~Moore, D.~Ramage, S.~Hampson, and B.~A. y~Arcas,
  ``Communication-efficient learning of deep networks from decentralized
  data,'' in \emph{Proc. 20th Int. Conf. Artificial Intelligence Statistics},
  Fort Lauderdale, FL, USA, Apr. 2017, pp. 1273--1282.

\bibitem{9374105}
Q.-V. Pham, M.~Zeng, R.~Ruby, T.~Huynh-The, and W.-J. Hwang, ``{UAV}
  communications for sustainable federated learning,'' \emph{IEEE Trans. Veh.
  Technol.}, vol.~70, no.~4, pp. 3944--3948, Apr. 2021.

\bibitem{qiao2021communication}
Z.~Qiao, X.~Yu, J.~Zhang, and K.~B. Letaief, ``Communication-efficient
  federated learning with dual-side low-rank compression,''
  \emph{arXiv:2104.12416}, Apr. 2021.

\bibitem{7792374}
C.~{Jiang}, H.~{Zhang}, Y.~{Ren}, Z.~{Han}, K.~{Chen}, and L.~{Hanzo},
  ``Machine learning paradigms for next-generation wireless networks,''
  \emph{IEEE Wireless Commun.}, vol.~24, no.~2, pp. 98--105, Apr. 2017.

\bibitem{MengyuanLearning}
M.~Lee, G.~Yu, and G.~Y. Li, ``Learning to branch: Accelerating resource
  allocation in wireless networks,'' \emph{IEEE Trans. Veh. Technol.}, vol.~69,
  no.~1, pp. 958--970, Jan. 2020.

\bibitem{MatthiesenLearning}
B.~Matthiesen, A.~Zappone, K.-L. Besser, E.~A. Jorswieck, and M.~Debbah, ``A
  globally optimal energy-efficient power control framework and its efficient
  implementation in wireless interference networks,'' \emph{IEEE Trans. Signal
  Process.}, vol.~68, pp. 3887--3902, Jun. 2020.

\bibitem{LiangLearning}
F.~Liang, C.~Shen, W.~Yu, and F.~Wu, ``Towards optimal power control via
  ensembling deep neural networks,'' \emph{IEEE Trans. Commun.}, vol.~68,
  no.~3, pp. 1760--1776, Mar. 2020.

\bibitem{8626195}
J.~{Luo}, J.~{Tang}, D.~K.~C. {So}, G.~{Chen}, K.~{Cumanan}, and J.~A.
  {Chambers}, ``A deep learning-based approach to power minimization in
  multi-carrier {NOMA} with {SWIPT},'' \emph{IEEE Access}, vol.~7, pp.
  17\,450--17\,460, Jan. 2019.

\bibitem{6485022}
P.~{Blasco}, D.~{Gunduz}, and M.~{Dohler}, ``A learning theoretic approach to
  energy harvesting communication system optimization,'' \emph{IEEE Trans.
  Wireless Commun.}, vol.~12, no.~4, pp. 1872--1882, Apr. 2013.

\bibitem{Hochreiter1997long}
S.~Hochreiter and J.~Schmidhuber, ``Long short-term memory,'' \emph{Neural
  Computation}, vol.~9, no.~8, pp. 1735--1780, 1997.

\bibitem{AndreaLearning}
A.~Ortiz, H.~Al-Shatri, X.~Li, T.~Weber, and A.~Klein, ``Reinforcement learning
  for energy harvesting point-to-point communications,'' in \emph{Proc. IEEE
  Intern. Commun. Conf.}, 2016, pp. 1--6.

\bibitem{ManLearning}
M.~Chu, X.~Liao, H.~Li, and S.~Cui, ``Power control in energy harvesting
  multiple access system with reinforcement learning,'' \emph{IEEE Internet of
  Things J.}, vol.~6, no.~5, pp. 9175--9186, Oct. 2019.

\bibitem{SharmaDRL}
M.~K. Sharma, A.~Zappone, M.~Assaad, M.~Debbah, and S.~Vassilaras,
  ``Distributed power control for large energy harvesting networks: A
  multi-agent deep reinforcement learning approach,'' \emph{IEEE Trans. on
  Cognitive Commun. and Networking}, vol.~5, no.~4, pp. 1140--1154, Dec. 2019.

\end{thebibliography}

\begin{IEEEbiography}[{\includegraphics[width=1.1in,height=1.25in,clip,keepaspectratio]{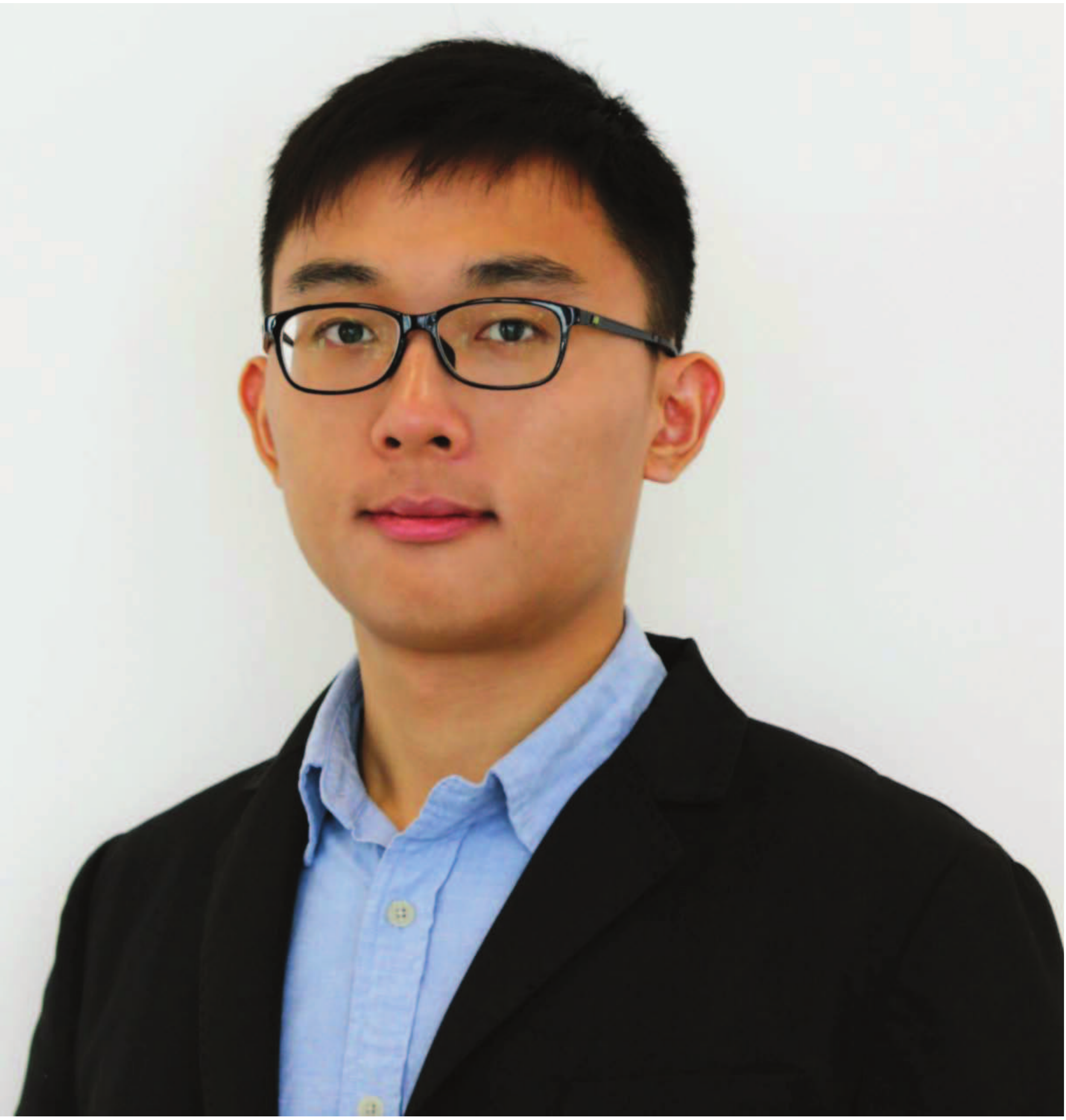}}]{Zhiqiang Wei} (Member, IEEE) received the B.E. degree in information engineering from Northwestern Polytechnical University (NPU), Xi'an, China, in 2012, and the Ph.D. degree in electrical engineering and telecommunications from the University of New South Wales (UNSW), Sydney, Australia, in 2019. From 2019 to 2020, he was a Postdoctoral Research Fellow with UNSW. He is currently a Humboldt Postdoctoral Research Fellow with the Institute for Digital Communications, Friedrich-Alexander University Erlangen-Nuremberg (FAU), Erlangen, Germany. He received the Best Paper Awards at the IEEE International Conference on Communications (ICC), 2018. He has been serving as the TPC Co-Chair of the IEEE ICC 2021 Workshop on orthogonal time frequency space (OTFS) and IEEE International Conference on Communications in China (ICCC) 2021 Workshop on OTFS. He was recognized as an Exemplary Reviewer of IEEE Transactions on Communications in 2017-2020 and IEEE Transactions on Wireless Communications in 2017 and 2018.
\end{IEEEbiography}

\begin{IEEEbiography}[{\includegraphics[width=1.1in,height=1.25in,clip,keepaspectratio]{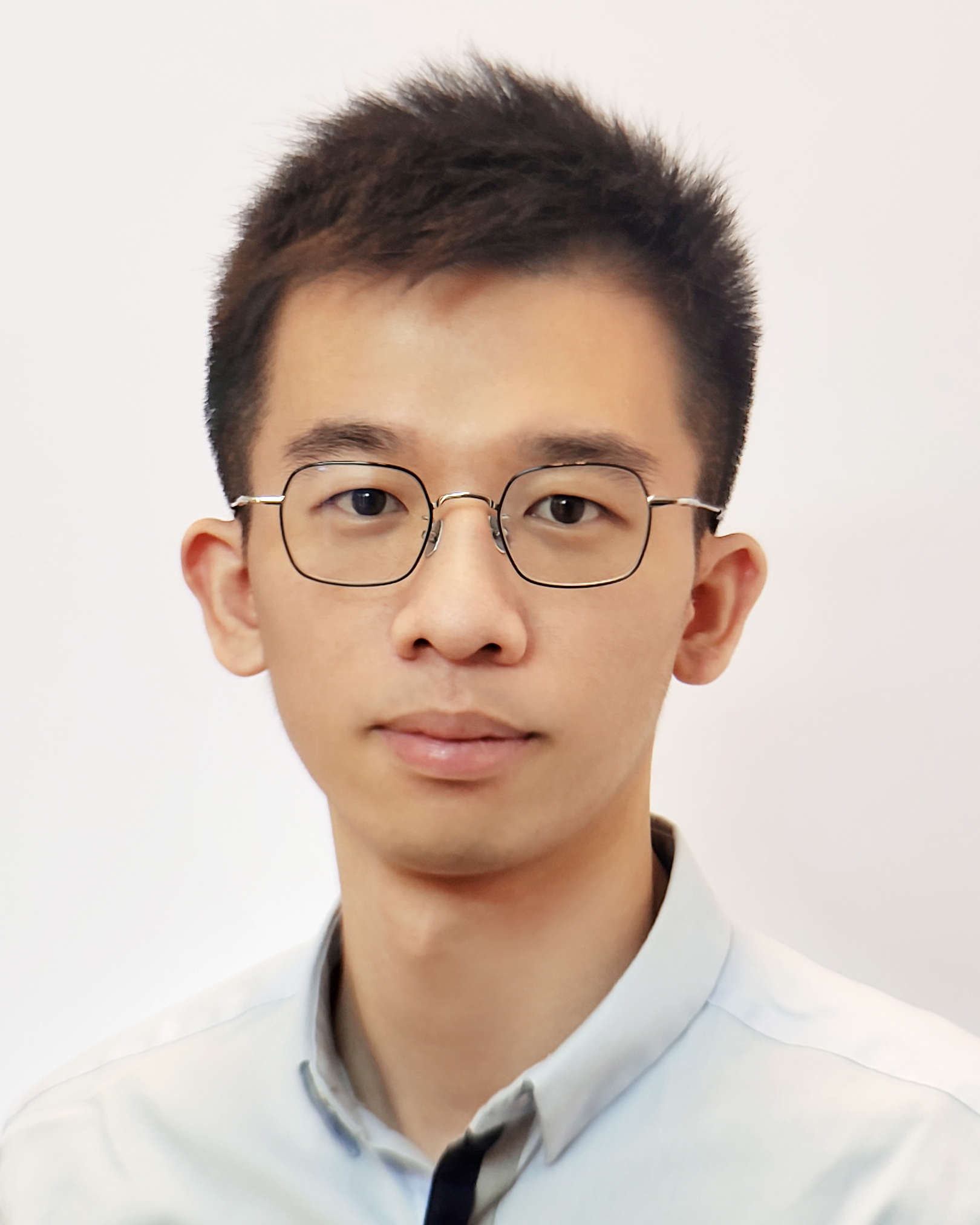}}]{Xianghao Yu} (Member, IEEE) received the B.Eng. degree in information engineering from Southeast University, Nanjing, China, in 2014, and the Ph.D. degree in electronic and computer engineering from The Hong Kong University of Science and Technology (HKUST), Hong Kong, China, in 2018.
		
He was a Humboldt Post-Doctoral Research Fellow with the Institute for Digital Communications, Friedrich-Alexander University of Erlangen-Nuremberg (FAU), Erlangen, Germany. He is currently a Research Assistant Professor with the Department of Electronic and Computer Engineering, HKUST. He has coauthored the book \textit{Stochastic Geometry Analysis of Multi-Antenna Wireless Networks} (Springer, 2019). His research interests include millimeter wave communications, intelligent reflecting surface-assisted communications, and wireless artificial intelligence. He received the IEEE Global Communications Conference (GLOBECOM) 2017 Best Paper Award, the 2018 IEEE Signal Processing Society Young Author Best Paper Award, and the IEEE GLOBECOM 2019 Best Paper Award. He was also recognized as an Exemplary Reviewer of IEEE Transactions on Wireless Communications in 2017 and 2018.
\end{IEEEbiography}

\begin{IEEEbiography}[{\includegraphics[width=1.1in,height=1.25in,clip,keepaspectratio]{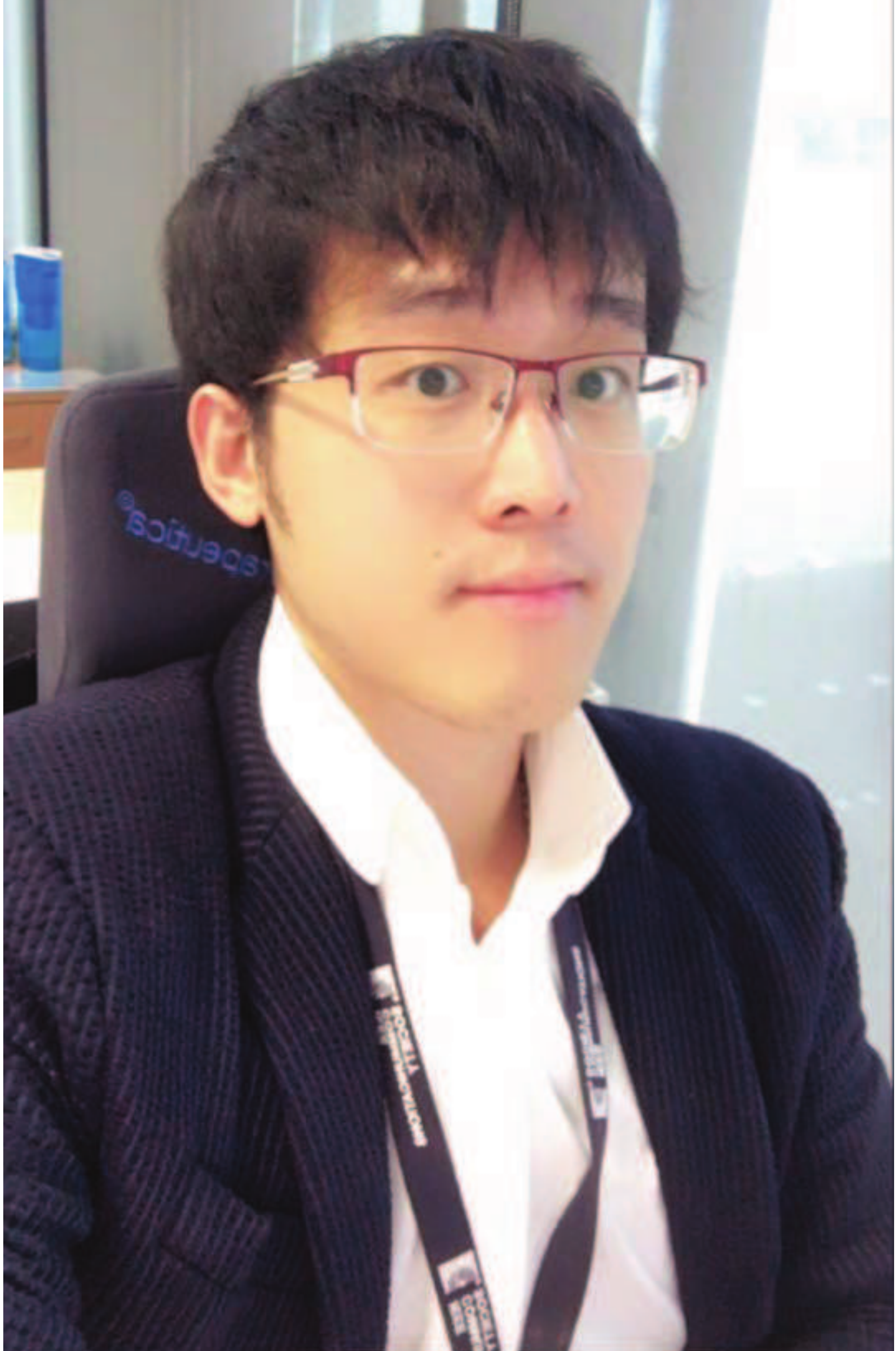}}]{Derrick
		Wing Kwan Ng} (Fellow, IEEE) received the bachelor degree with first-class honors and the Master of Philosophy (M.Phil.) degree in electronic engineering from the Hong Kong University of Science and Technology (HKUST) in 2006 and 2008, respectively. He received his Ph.D. degree from the University of British Columbia (UBC) in Nov. 2012. He was a senior postdoctoral fellow at the Institute for Digital Communications, Friedrich-Alexander University of Erlangen-Nuremberg (FAU), Germany. He is now working as Scientia  Senior Lecturer at the University of New South Wales, Sydney, Australia.  His research interests include convex and non-convex optimization, physical layer security, IRS-assisted communication, UAV-assisted communication, wireless information and power transfer, and green (energy-efficient) wireless communications. 
	
	Dr. Ng has been listed as a Highly Cited Researcher by Clarivate Analytics since 2018.  He received the Australian Research Council (ARC) Discovery Early Career Researcher Award 2017, the Best Paper Awards at the WCSP 2020,  IEEE TCGCC Best Journal Paper Award 2018, INISCOM 2018, IEEE International Conference on Communications (ICC) 2018, 2021,  IEEE International Conference on Computing, Networking and Communications (ICNC) 2016,  IEEE Wireless Communications and Networking Conference (WCNC) 2012, the IEEE Global Telecommunication Conference (Globecom) 2011, and the IEEE Third International Conference on Communications and Networking in China 2008. He has been serving as an editorial assistant to the Editor-in-Chief of the IEEE Transactions on Communications from Jan. 2012 to Dec. 2019. He is now serving as an editor for the IEEE Transactions on Communications,  the IEEE Transactions on Wireless Communications, and an area editor for the IEEE Open Journal of the Communications Society. 
\end{IEEEbiography}

\begin{IEEEbiography}[{\includegraphics[width=1.25in,height=1.25in,clip,keepaspectratio,angle=90]{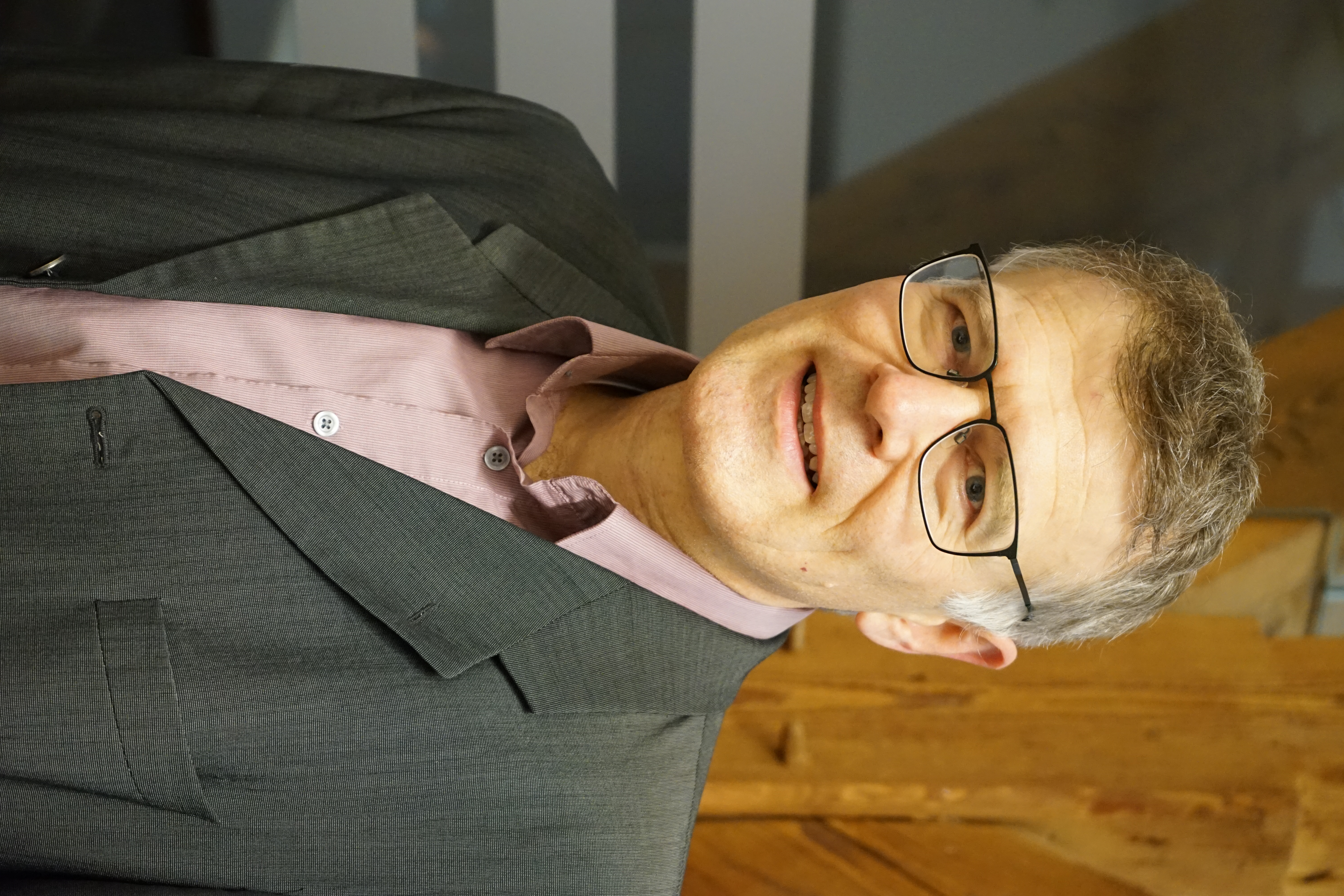}}]{Robert Schober} (Fellow, IEEE) received the Diplom (Univ.) and the Ph.D. degrees in electrical engineering from Friedrich-Alexander
University of Erlangen-Nuremberg (FAU), Germany, in 1997 and 2000, respectively. From 2002 to 2011, he was a Professor and Canada Research
Chair at the University of British Columbia (UBC), Vancouver, Canada. Since January 2012 he is an Alexander von Humboldt Professor and the Chair for Digital Communication at FAU. His research interests fall into the broad areas of Communication Theory, Wireless and Molecular Communications, and Statistical Signal Processing.

Robert received several awards for his work including the 2002 Heinz Maier Leibnitz Award of the German Science Foundation (DFG), the 2004
Innovations Award of the Vodafone Foundation for Research in Mobile Communications, a 2006 UBC Killam Research Prize, a 2007 Wilhelm Friedrich Bessel Research Award of the Alexander von Humboldt Foundation, the 2008 Charles McDowell Award for Excellence in Research from UBC, a 2011 Alexander von Humboldt Professorship, a 2012 NSERC E.W.R. Stacie Fellowship, and a 2017 Wireless Communications Recognition Award by the IEEE Wireless Communications Technical Committee. Since 2017, he has been listed as a Highly Cited Researcher by the Web of Science. Robert is a Fellow of the Canadian Academy of Engineering, a Fellow of the Engineering Institute of Canada, and a Member of the German National Academy of Science and Engineering. From 2012 to 2015, he served as Editor-in-Chief of the IEEE Transactions on Communications. Currently, he serves as Member of the Editorial Board of the Proceedings of the IEEE and as VP Publications for the
IEEE Communication Society (ComSoc).
\end{IEEEbiography}

\end{document}